%% file: topreview.tex
\documentclass[aps,prd,superscriptaddress,onecolumn,floatfix,nofootinbib,preprintnumbers]{revtex4}

\usepackage{amsmath,graphicx,tabularx,url}

\begin{document}

\include{declare}
\title{Top Tagging}

\author{Tilman Plehn}
\affiliation{Institut f\"ur Theoretische Physik, 
             Universit\"at Heidelberg, Germany}

\author{Michael Spannowsky}
\affiliation{IPPPP, Department of Physics, 
             Durham University, United Kingdom}

\begin{abstract}
 Top tagging is a recent approach to identifying boosted hadronic top
 quarks. It avoids reconstructing individual top decay products and
 instead uses a jet algorithm to reconstruct the entire top
 decay. Quite generally, geometrically large jets including heavy
 particles (fat jets) can be analyzed on the level of their subjet
 constituents. LHC data will soon allow us to establish this new
 analysis method. We discuss different tagging algorithms, their
 critical QCD aspects, and currently available experimental
 results. For the development of taggers and their different
 applications this review should provide a firm theoretical and
 algorithmic background.
\end{abstract}

\maketitle
\tableofcontents

\newpage

\section{Physics Case}
\label{sec:case}

The top quark~\cite{topdiscovery} is the only observed fermion with a
weak-scale mass and the only quark which decays before it
hadronizes~\cite{top_dec}. At the LHC it should be a perfect
laboratory to study the mechanism of electroweak symmetry breaking,
orthogonal to searches for a scalar Higgs boson.  Because the top
quark also induces the largest quantum corrections to the bare Higgs
mass it lives in the center of all new physics models motivated by the
hierarchy problem~\cite{review}.  Many such extensions of the Standard
Model, like supersymmetry or little Higgs models, predict top partners
which naturally decay to top quarks. Other models, like extra
dimensions with geometries linked to the mass hierarchy in the
Standard Model, predict large couplings of new states to top quarks,
again giving rise to new physics decays to top quarks.\bigskip

The problem is that in a high-multiplicity QCD environment the
reconstruction of top quarks is difficult. Obviously, two leptonically
decaying top quarks can be observed well through their decay leptons
and missing transverse energy. Because of the two invisible neutrinos
a full reconstruction of leptonic top pairs is not possible. A
semi-leptonically decaying top pair can be reconstructed
approximately. For this we replace the missing 4-momentum of the
neutrino with the assumption that we know its mass, the
two-dimensional measured missing transverse momentum vector, and at
least one of the two $W$ and top on-shell mass constraints. The
problem is that these assumptions make it hard to disentangle Standard
Model top quarks for example from top partners decaying to a top quark
and missing energy. Finally, purely hadronically decaying top pairs
produce six decay jets with two $b$-tags plus any number of QCD jets.
The mass scale of QCD radiation and the $W$ decay mass scale $m_W/2$
are very similar. Reliably reconstructing such a top pair event will
be a serious challenge at the LHC. This list of channels and their
individual challenges supports the claim that top identification is
one of the hardest tasks in LHC experiments. This is the reason why to
date there exist essentially no published search limits of top
partners~\cite{desai}. On the other hand, following theory arguments
we really want to study top pairs in signal or background events,
including a full reconstruction. This is only possible for hadronic
top decays.\bigskip

Two main motivations to study such hadronic top quarks in a boosted
regime exist, one theoretical and one experimental. From a theory
perspective heavy $s$-channel resonances preferably decaying to top
quarks will produce highly boosted top quarks. They require a
dedicated analysis strategy because the large boost makes traditional
analysis methods
hard~\cite{semi_resonances,had_resonances}. Similarly, models with top
partners and a weakly interacting dark matter agent naturally predict
scenarios where top partners are pair-produced at the LHC and then
decay into a top quark pair plus missing energy~\cite{meade}. The
higher the typical new physics mass scale is pushed by other LHC
searches, the more boosted also these top quarks will become.

Secondly, a very general experimental concern with LHC events
involving top quarks is the high jet multiplicity. The associated
combinatorial backgrounds are a problem for many LHC searches; the
best known victim of such combinatorics is the Higgs search based on
$t\bar{t}H$ events with hadronic $H \to b\bar{b}$
decays~\cite{tth_exp,tth}. An identification algorithm for hadronic
top quarks which automatically takes care of jet combinatorics should
have the potential to revive a sizeable number of LHC analyses. One
phase space region where this seems feasible are boosted top quarks,
because the different top decay jets and QCD radiation are well
separated from one another.\bigskip

The idea of studying the substructure of jets is already a
classic~\cite{mike}. It is closely tied to the development of
recombination jet algorithms. Its impact in solving the two problems
laid out above has only been recognized recently. The LHC will for the
first time produce enough boosted top quarks (or other heavy Standard
Model particles) to systematically study these analysis ideas. As an
illustration, in the left panel of Fig.~\ref{fig:kinematics} we show
the geometric size of a top decay as a function of the transverse
momentum of the decaying top. Values of $\Delta R_{bjj} \sim \pi$
imply that in the transverse plane the top decay fills in the central region an entire hemisphere. In the right panel of Fig.~\ref{fig:kinematics} we show
the number of top quarks we can tag inside a fat jet\footnote{The name fat jet refers to geometrically large jet
  which include a heavy particle decay.} of size
$R=1.5$. The minimum transverse
momentum we can probe is around 200~GeV. At the LHC, this corresponds
to $\mathcal{O}(5-10\%)$ of all top pairs. If we start with 60000 top
pairs in the currently available $5~\ifb$ sample, several thousand of
them should have a sufficient boost, so we can test top taggers on
them. The recent FastJet update~\cite{fastjet} allows larger jet
sizes, so increasing the fat jet size to $R=1.8$ can double the number
of available top quarks in ATLAS and CMS. Now is the right time for
detailed studies of top taggers in particular and subjet methods in
general on LHC data.\bigskip

\begin{figure}[t]
  \includegraphics[width=0.35\textwidth]{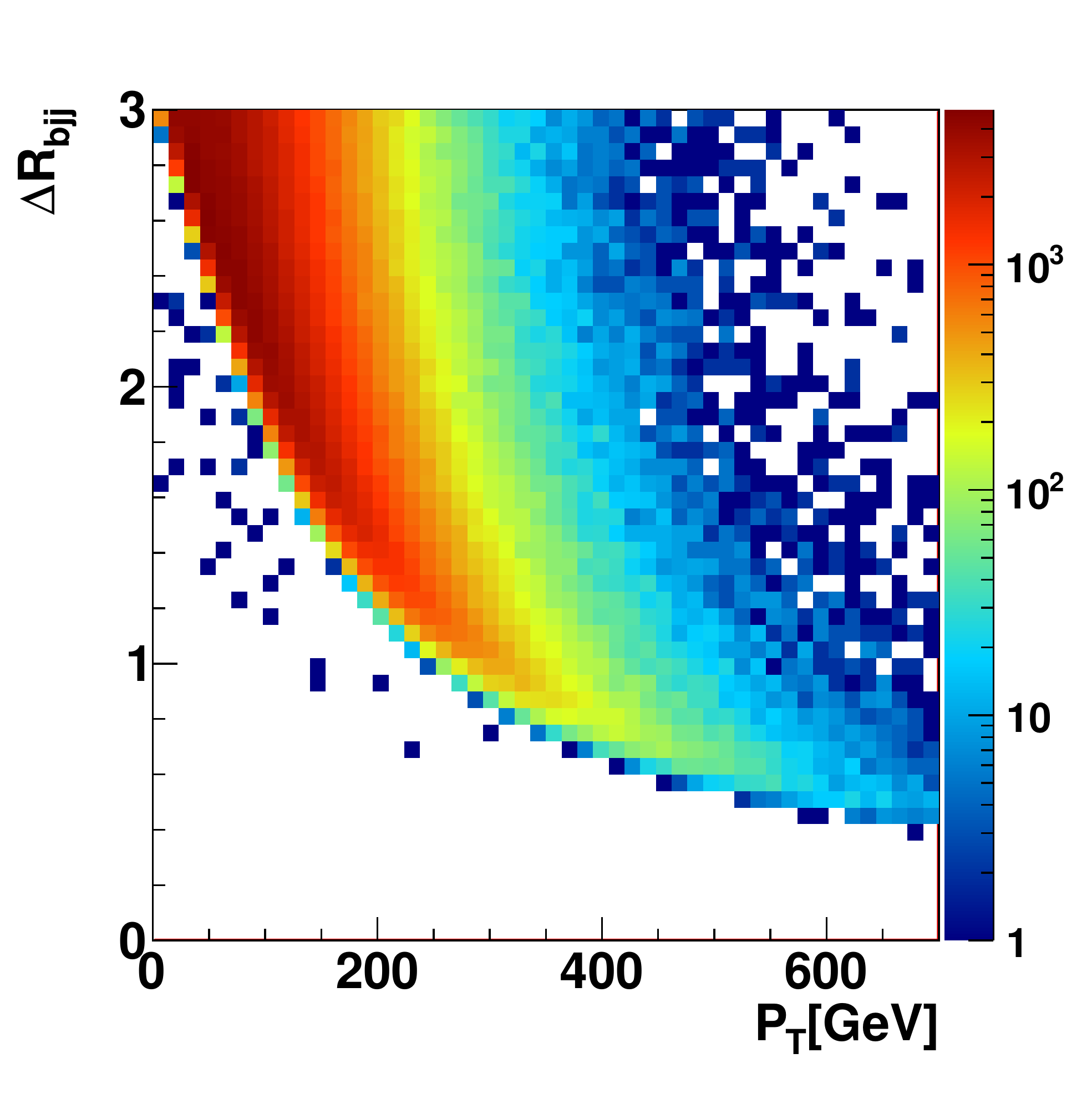}
  \hspace*{0.1\textwidth}
  \includegraphics[width=0.35\textwidth]{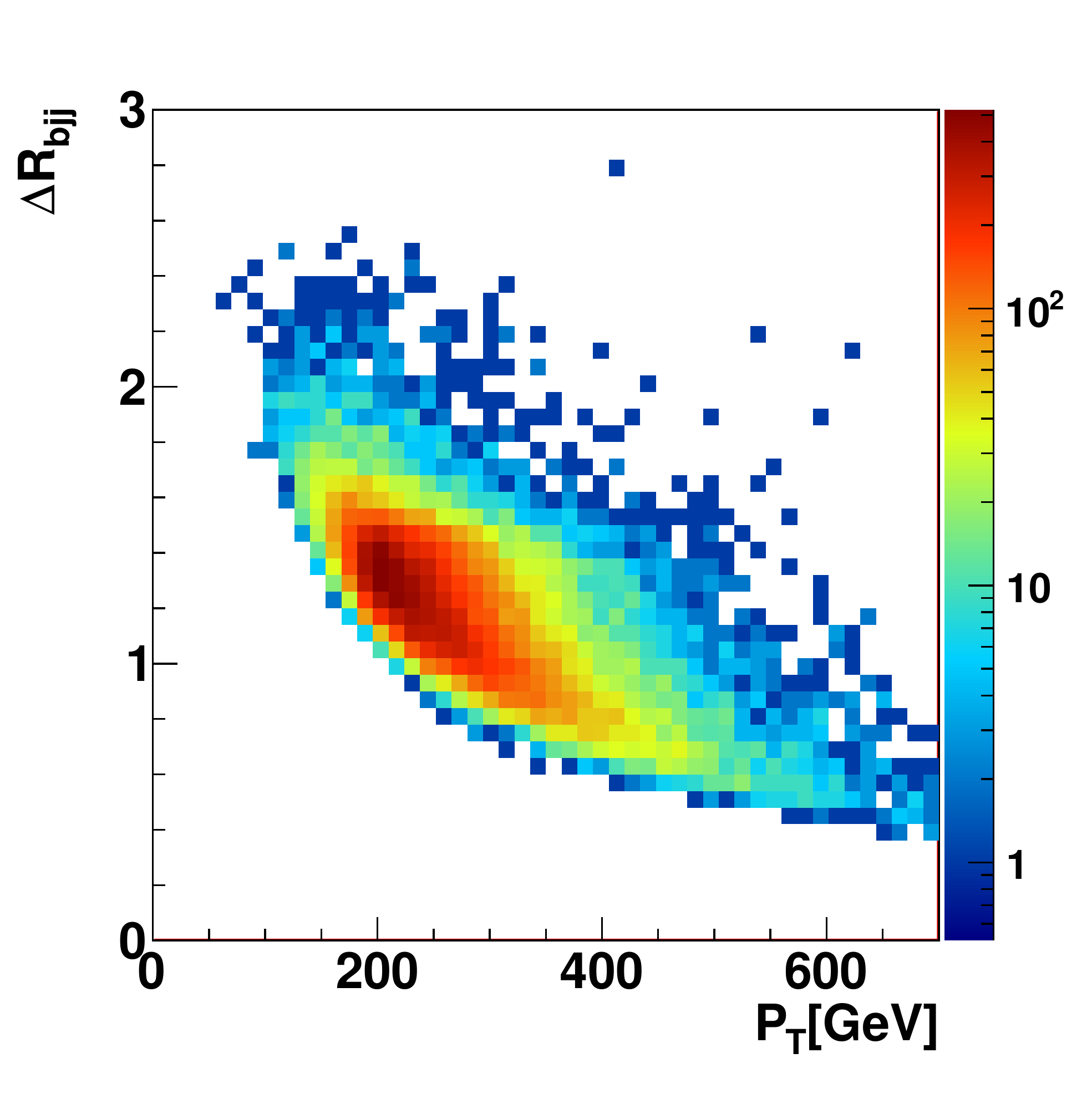}
  \caption{Approximate C/A distance of all three top decay products
    for Standard Model $t\bar{t}$ events as a function of
    $p_{T,t}$. The left panel shows all events while the right panel
    shows only tagged events using C/A fat jets of size $R=1.5$.
    Figure from Ref.~\cite{heptop1}.}
  \label{fig:kinematics}
\end{figure}

In this review we summarize the recent developments on top
taggers. This will not include a comprehensive comparison of the
performance of different taggers, which has to be left to data
analyses. Neither do we spend much effort describing possible analysis
ideas which fat jet analyses will eventually allow. The task is simply
to describe different approaches to top tagging, including a wide
variety of algorithms with individual advantages and challenges. For
more detailed comparisons of the performance of different taggers we
refer to the frequent BOOST proceedings~\cite{boost2010,boost2011} which
present the state of the art. What we provide are the underlying
concepts and algorithms.

As an audience for this review we envision graduate students or
researchers who would like to enter this field or who are looking for
a comprehensive discussion of the relevant ideas and their
realizations. We limit ourselves to top tagging only and neglect for
example (most) Higgs taggers or other related tools. For alternative
reviews about jet substructure we refer for example to
Refs~\cite{Salam:2009jx,boost2010,boost2011,subreview}. Going beyond
the top tagging focus would have made this review considerably less
coherent, and as far as we can tell the concepts used and realized in
top taggers cover the entire substructure field well.

In its current form we recommend reading through each of the main
parts of this review from the beginning to the end. The
discussions of the taggers as well as of the QCD effects have a
pedagogical setup, and even though the individual sections are self
contained enough to maybe serve as a reference book, some information
will always be hidden in introductory discussions.\bigskip

This review consists of three major parts: first, in
Sec.~\ref{sec:algos} we will introduce all major top tagging
algorithms. Major in this context is defined as `likely to be tested
on LHC data soon'. All described taggers include a set of tunable
parameters, however, explicit numbers for kinematic cuts are
only illustrations and should be taken with a grain of salt. On the
other hand, the different algorithms have very different physics
backgrounds, so it is useful to discuss each of them in some
detail. In the second part~\ref{sec:qcd} we discuss the related QCD
issues, most notably different ways of dealing with soft QCD effects,
underlying event, and pile-up. The different approaches are
historically developed for individual taggers and often have major
impact on their performance. On the other hand, as far as they
contribute to the definition of the physics objects of the actual top
tagging algorithms they can be relatively easily exchanged between and
added to taggers.  Finally, in Sec.~\ref{sec:performance} we discuss
some of the published experimental results. This part will suffer from
the fact that most ATLAS and CMS results on top tagging are at best
published in internal notes, so we will be very brief.

\section{Top Tagging Algorithms}
\label{sec:algos}

Top tagging algorithms are typically based on two classes of
observables. On the one hand, we can generalize the well established
event shapes to jet shapes, \ie observables defined on calorimeter
clusters of the energy flow inside a geometrically large fat jet. Such
jet shapes are directly accessible by the LHC detectors. For our
purpose the most relevant jet shape is the jet mass, on which all top
tagging algorithms are based. The second class of observables is 
the clustering history of all jet constituents. This history cannot
be observed directly. Instead, we have to rely on our understanding of
QCD to simulate it, based on the energy depositions we observe in the
calorimeters (and trackers).\bigskip

To backwards engineer the splitting history of a jet we can use our
picture of collinear quark and gluon splittings predicted by first
principles QCD.  The successive splitting of quarks and gluons
radiated off an $n$-particle hard process ($\sigma_n$) factorizes in
the soft or collinear limits into the simple form
\begin{alignat}{5}
\sigma_{n+1}
 = \int \sigma_n \; 
   \frac{d p_j^2}{p_j^2} dz \; 
   \frac{\alpha_s}{2 \pi} \; \hat{P}_{j_1 \leftarrow j}(z) \; ,
\label{eq:factorize}
\end{alignat}
where $p_j$ is the momentum of the splitting parton and $z$ is the
energy fraction of one of the splitting products $j \to j_1 j_2$. The
different splitting kernels $ \hat{P}(z)$ depend on the partonic quark
or gluon process and are known. They often diverge in the soft limit
$z \to 0$, so we will encounter an overlapping enhancement and
eventually divergence for soft and for collinear
radiation~\cite{qcd,gavin_review}. The factorization shown in
Eq.\eqref{eq:factorize} describes the splitting of parton radiation
off incoming as well as off outgoing hard partons until the radiated
partons become soft enough to hadronize. The numerical implementation
of Eq.\eqref{eq:factorize} is the parton shower, and it describes the
transition from hard partons to a large number of hadrons which
eventually decay and appear in the calorimeters of the LHC
experiments.\bigskip

Inverting this successive splitting and hence extracting a hard parton
momentum from a measured jet is what jet algorithms do. Historically,
an important issue is the infrared safety of observables and
algorithms; a soft or collinear splitting of any parton momentum
cannot impact the macroscopic observables. While some cone algorithms
are not collinear save, recombination algorithms are.
Such recombination algorithms iteratively determine which of the
observed calorimeter towers should be merged into subjets and which of
these subjets should then be merged together step by step, such that
finally we arrive at few hard jets per event. The end of this
successive splitting can be defined in terms of a given minimum jet
separation or a given maximum number of jets. Different recombination
algorithms are based on different subjet distance measures:
\begin{alignat}{8}
&k_T &
d_{j_1 j_2} &= \frac{\Delta R_{j_1 j_2}^2}{D^2} \;
          \min \left( p_{T,j_1}^2, p_{T,j_2}^2 \right) \qquad  \qquad  \qquad &
d_{j_1 B} &= p_{T,j_1}^2 \notag \\
&\text{Cambridge/Aachen} \qquad \qquad &
d_{j_1 j_2} &=\frac{\Delta R_{j_1 j_2}^2}{D^2} &
y_{j_1 B} &= 1 \notag \\
&\text{anti-}k_T &
d_{j_1 j_2} &=\frac{\Delta R_{j_1 j_2}^2}{D^2} 
          \min \left( \frac{1}{p_{T,j_1}^2}, \frac{1}{p_{T,j_2}^2} \right) &
d_{j_1 B} &= \frac{1}{p_{T,j_1}^2} \; .
\label{eq:jetalgos}
\end{alignat}
These measures can be generalized to $d_{j_1 j_2} = \Delta R_{j_1
  j_2}^2/D^2 \times \min (p_{T,j_1}^{2n} p_{T,j_2}^{2n})$ for
$n=-1,0,1$. The $k_T$-algorithm~\cite{kt_algo} mimics the soft and
collinear enhancement of the QCD splitting kernels in
Eq.\eqref{eq:factorize}. For the top tagging application it should
best reconstruct the QCD splitting history. The Cambridge/Aachen (C/A)
algorithm~\cite{ca_algo} always combines the two closest (most
collinear) subjets. It is sensitive to collinear but not to soft
splittings, but as we will see later it has some advantages in fat jet
searches.  The anti-$k_T$~\cite{akt_algo} algorithm first combines the
hardest subjets, to define a particularly stable jet recombination
with clean geometric jet boundaries.  Intermediate subjets based on
the anti-$k_T$-algorithm have not resemblance with what we would
expect from QCD.  All three algorithms are available through the
FastJet package~\cite{fastjet}.\bigskip

Closely related to the $k_T$ and C/A measures is the JADE
distance~\cite{jade} which essentially is a transverse subjet mass:
\begin{alignat}{5}
d_{j_1 j_2} = p_{T,j_1} p_{T,j_2} \; \Delta R_{j_1 j_2}^2 \sim m_{T,j_1 j_2}^2 \; .
\label{eq:jade}
\end{alignat}
In this notation we label the splitting partons as well as the
reconstructed subjets in the recombination algorithms as $j_i$. In the
remainder of the paper we will only use subjets, so this notation does
not pose any problems. Moreover, we will refer to all intermediate
clusterings inside all recombination algorithms as subjets. More
stable objects, like filtered subjets we will introduce in
Sec.~\ref{sec:qcd}.\bigskip

Independent of the choice of subjets to be merged by the jet algorithm
we also have to define a scheme for the combination of the two
4-momenta. In particular when looking for massive jets we should not
assume anything about the mass of the partons. Instead, we can simply
add the two 4-vectors $p_j = p_{j_1} + p_{j_2}$ in the E-scheme. The
subjet mass is defined as $m_j^2 = p_j^2$. In most (soft or collinear)
QCD splittings it should not exceed the $B$ meson mass, and even
including detector effects we usually find $m_j \lesssim 30$~GeV in
the absence of massive weak-scale splittings.\bigskip

In contrast to the dynamic clustering history which we can think of
as a time evolution, jet shapes are observables based on the final jet
constituents. A priory, it is not clear that these two approaches
include the same information. Therefore, the comparison of different
top taggers is first of all an interesting QCD experiment.

Because different top taggers rely on very different jet shapes we
will not introduce them in general here. The definitions are often
inherited from event shapes, like most noticeably
thrust~\cite{thrust}. Unlike jet clustering histories, which depending
on the underlying jet algorithm are either theoretically well defined
(\ie infrared save) or not, jet shapes have to be classified one by
one. Much work has for example gone into appropriate definitions of
the jet mass, introduced above~\cite{jet_mass}.

The kinematics underlying this jet mass, assuming widely separated
jets with a good 4-momentum reconstruction, is fairly
simple. Following our QCD picture, it is based on successive ($1 \to
2$) splittings. If one of these splittings corresponds to the $t \to
Wb$, $W \to jj$, or even $H \to b\bar{b}$ decay, the corresponding jet
mass should be around the electroweak scale.  In the leading
logarithmic approximation we can describe a massive jet composed out
of two subjets using~\cite{pruning2}
\begin{alignat}{5}
\frac{m_j^2}{p_{T,j}^2} \sim z (1-z) \; \Delta R_{j_1 j_2}^2
\qquad \text{with} \quad 
z = \frac{\min p_{T,j_i}}{p_{T,j}} \; .
\label{eq:basics}
\end{alignat}
\bigskip

As mentioned above, all top taggers include at least one jet shape,
namely the jet mass. The early subjet tools combine the jet mass with
a clustering history. This includes the first $W$ and top taggers by
Mike Seymour (Sec.~\ref{sec:early}), the $W$ predecessor to YSplitter
(Sec.~\ref{sec:early}) and the BDRS Higgs tagger
(Sec.~\ref{sec:early}).

More advanced tools like YSplitter (Sec.~\ref{sec:ysplitter}), the
Seattle or pruning tagger (Sec.~\ref{sec:seattle}), the Johns Hopkins
tagger (Sec.~\ref{sec:hopkins}), the HEPTopTagger
(Sec.~\ref{sec:heptop}), or the Thaler-Wang tagger
(Sec.~\ref{sec:thaler_wang}) supplement the jet mass with a detailed
analysis of the clustering history. Differences between them arise
because of different jet algorithms and different selection criteria
to extract the massive $t \to Wb$ and $W \to jj$ splittings.

Following the success of event shapes at LEP, the $N$-subjettiness
tagger (Sec.~\ref{sec:scet}), the template method
(Sec.~\ref{sec:alternatives}), or the tree-less algorithm
(Sec.~\ref{sec:alternatives}) are exclusively based on (sub-)jet
shapes. The choice of jet algorithms in this approach does not play
any role, except for removing underlying event and pile-up, as we will
discuss in Sec.~\ref{sec:qcd}.\bigskip

Testing which family of taggers is best suited for studies of the
inside of jets will shed light on experimental QCD issues way beyond
the identification of top jets. For example at LEP, event shapes
became the standard tools for any kind of precision QCD measurements,
like for example the $\alpha_s$ measurement. At the Tevatron, simple
cone jets were used most of the time because they were deemed to be
most stable in the hadron collider environment.  At the LHC we are
already observing serious problems with pile-up, even though the
collider energy and luminosity are still small.  In this massively
complex QCD environment it is not at all clear how we will analyze QCD
effects in the coming years. Ongoing jet substructure and top tagging
studies will significantly contribute to answering this very
fundamental question.

\subsection{Early developments}
\label{sec:early}

The first tagger for $W$ bosons and top quarks was developed in
1994~\cite{mike}. According to its author it was meant to illustrate
the power of new jet algorithms and to replace the at that time common
cone algorithms with recombination algorithms. The basic idea is that
step-wise combining calorimeter towers to jets by following geometric
and energy flow patterns includes more information than simply
collecting all towers within a certain $R$ distance. Very early
clustering algorithms were developed and used by the JADE
collaboration~\cite{jade}, based on the distance measure
Eq.\eqref{eq:jade} which unfortunately does not reflect the
soft-collinear splitting kernels which we can derive from
QCD~\cite{qcd}. A more appropriate clustering history includes
valuable information on the content and on the origin of a jet. Note
that such an additional source of information bypasses for example the
definitions of optimal use of information using matrix element
methods, because it acts on the objects of our usual analyses, not on
correlations between known objects.\bigskip

The original $W$-tagger~\cite{mike} is based on the $k_T$-clustering
algorithm. It starts with a fat jet of size $R=1.0$, extracts the two
hardest subjets, and then cuts on the $R$ distance between the two
subjets $R_{j_1, j_2} > 0.25$ and between the fat jet and each of the
subjets $R_{j,j_i} < 0.81$. Subjets with an energy below 17~GeV are
excluded from the $W$ reconstruction. The two subjets finally have to
reconstruct the $W$ mass to $\pm 10$~GeV. As we will see later, this
strategy defines most relevant ingredients of the corresponding modern
Higgs and top taggers.

\begin{figure}[b]
  \includegraphics[width=0.40\textwidth]{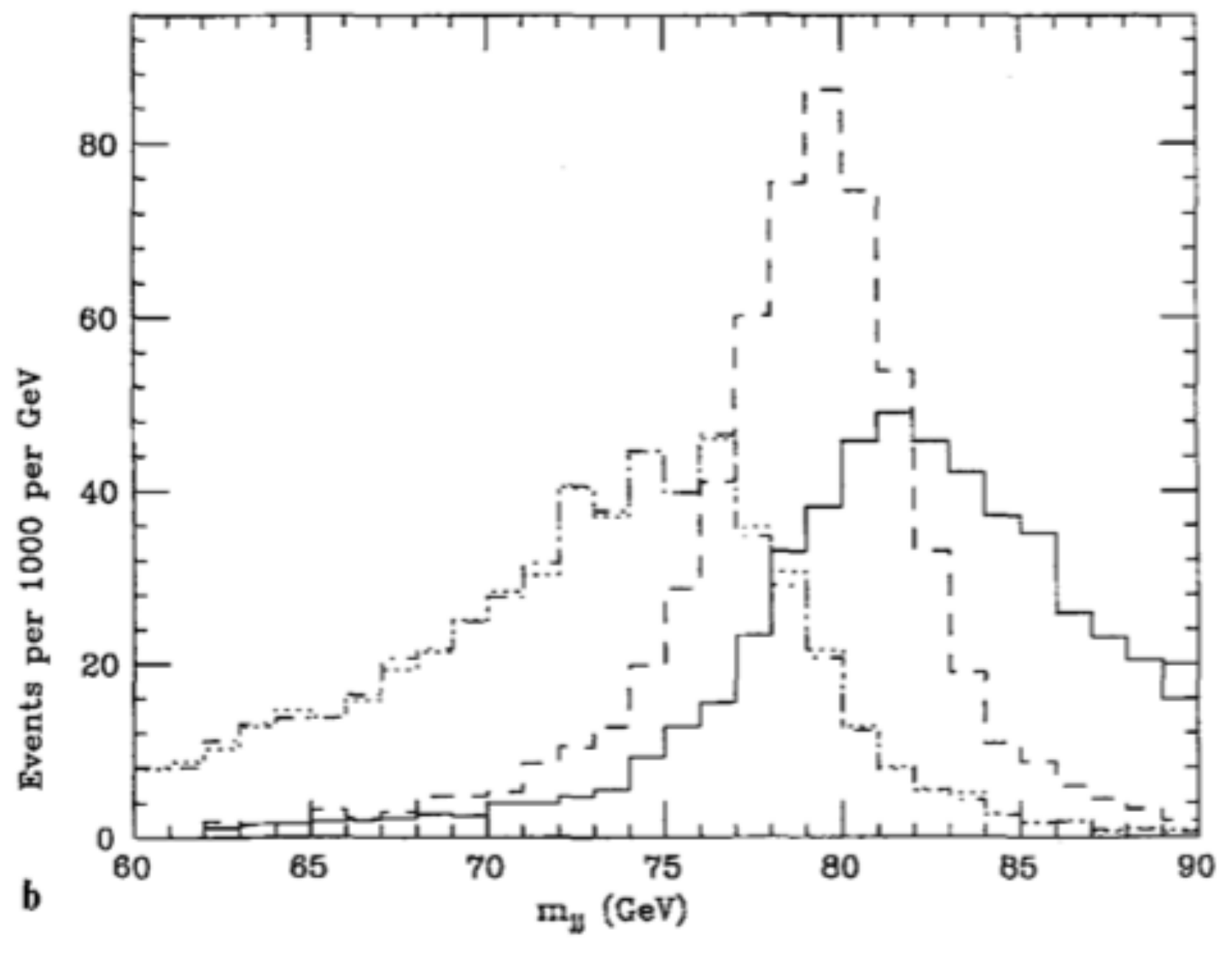}
  \hspace*{0.1\textwidth}
  \includegraphics[width=0.40\textwidth]{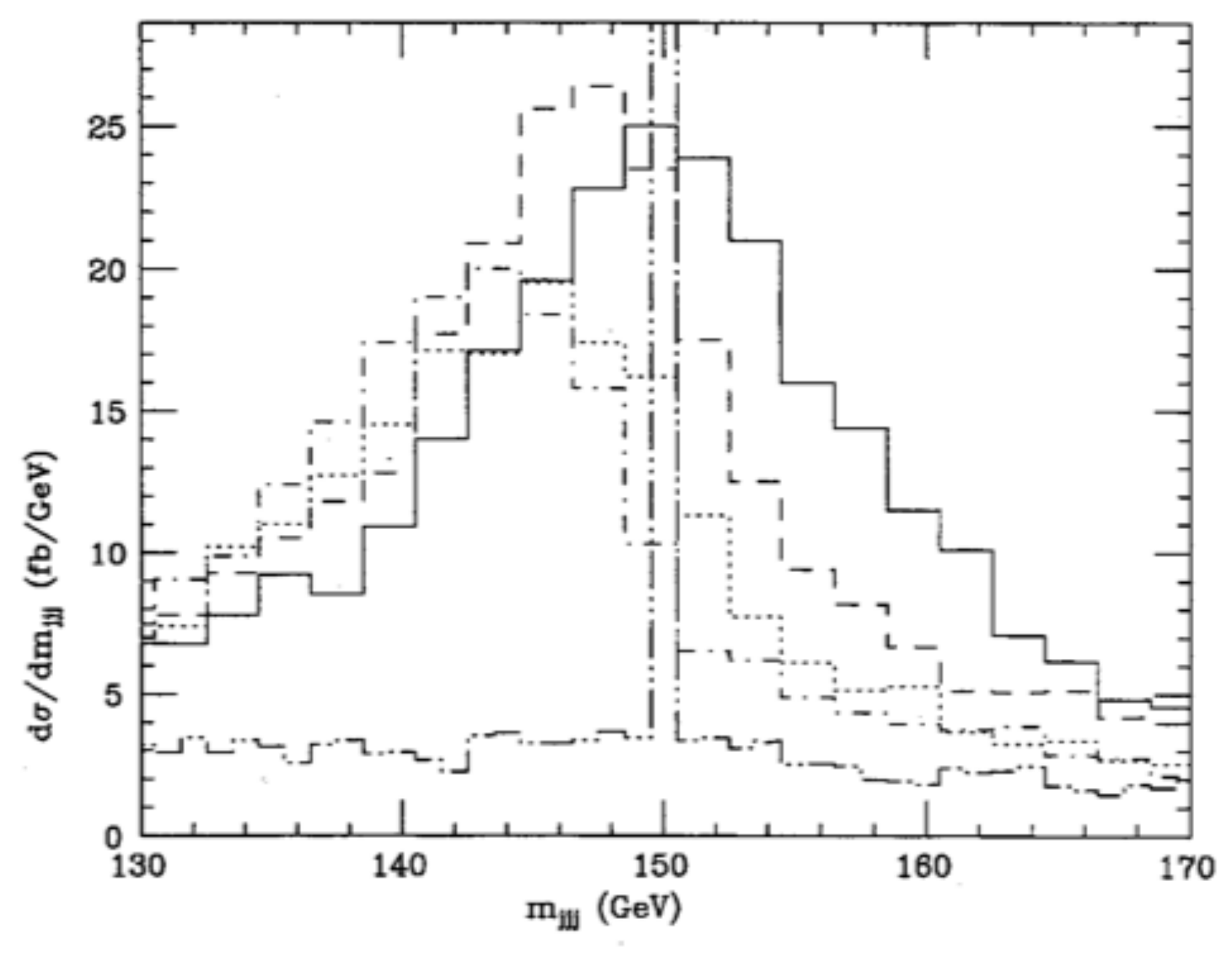}
  \caption{Reconstructed $W$ mass (left) and top mass (right) from
    cluster (solid and dashed) and cone (dotted and dot-dashed)
    algorithms. The solid and dotted curves include underlying
    event. For the top mass we also show the mass reconstructed from
    partons instead of subjets as the double-dot-dashed line. Figures
    from Ref.~\cite{mike}.}
  \label{fig:mike}
\end{figure}

For comparison, a similar tagger based on a cone algorithm~\cite{mike}
uses the same two cone radii of 0.81 and 0.25 to look for a large jet
which includes two well separated smaller jets with transverse energy above
20~GeV. Those two subjets also have to reconstruct $m_W$ within $\pm
10$~GeV. In the left panel of Fig.~\ref{fig:mike} the dashed and
dot-dashed curves show the results from the clustering and cone
algorithms. The clustering algorithm gives a significantly narrower
distribution centered well around the true $m_W = 80.4$~GeV.

Serious differences in the performance of these two taggers arise once
we include hadronization and underlying event. Hadronization will
clearly be better described by the clustering algorithm. On the other
hand, underlying event and eventually pile-up will compromise the
clustering result because soft QCD activity inside the fat jet will
eventually be included in the jet mass, cut off only by the global
minimum energy requirement for example from individual detector
cells. The double cone algorithm will ignore a large part of the
original large cone when it reconstructs the $W$ mass from the two
small cones. A drop by a factor $0.81/0.25 \sim 3.24$ in the jet
radius corresponds to an order of magnitude in area, which is the
relevant measure for any kind of uncorrelated hadronic
activity.\bigskip

Based on this observation, in Sec.~\ref{sec:qcd} we introduce
different ways of removing underlying event and pile-up from
recombination algorithms. The main idea~\cite{mike} is to re-cluster
all particles which we find inside the $R=1.0$ fat jet with an adapted
smaller size, \eg $R = 0.6\times R_{j_1 j_2}$, and reconstruct the $W$
mass from these cleaner subjets.  The solid and dotted curves in the
left panel of Fig.~\ref{fig:mike} again show the $m_W$ distributions
for both algorithms, including underlying event and cleaned subjets
for the mass reconstruction. The advantage of the clustering algorithm
is still visible, but it is significantly reduced. For the remaining
discussions in this review that implies that the power of subjet
techniques at the LHC will be largely decided by our shielding of the
clustering history from uncorrelated hadronic activity.\bigskip

The top tagger presented in the same work is optimized for a top mass
`less than about 200~GeV' in semi-leptonically decaying top
pairs at the Tevatron (before they were actually discovered in
1995). The reconstructed lepton guarantees the triggering of
the event. Additional $b$-tagging is not employed, so we should see
four top decay jets plus any number of QCD radiation jets per
event. Three mass combinations $m_{jjj}$, $m_{jj}$ and $m_{\ell \nu
  j}$ should reproduce $m_W$ and $m_t$, so a $\chi^2$ test is well
suited to solve the combinatorics in the decay jet assignment. The
minimum $p_T$ value for subjets to be considered is 15~GeV.

In the right panel of Fig.~\ref{fig:mike} we show the top mass
distributions for both jet algorithms with and without underlying
event. The true value $m_t = 150$~GeV can be seen in the parton-level
reconstruction (double-dot-dashed). The cluster algorithm in the
presence of underlying event at the Tevatron should reconstruct the
top mass to better than 10\%. Compared to newer developments in this
direction there are two aspects missing: first, there is no
information on the reconstruction of the entire 4-momentum of the
heavy particles; second, there is no explicit mentioning of the
advantageous boosted phase space regions as compared to the
overwhelming QCD backgrounds. Nevertheless, many of the taggers
discussed later in this section closely follow the ideas of
Ref.\cite{mike}.\bigskip

A second early study on the use of taggers for heavy particles in the
Standard Model applies a $W$-tagger to high energy $WW$ scattering at
a 14~TeV LHC~\cite{butterbrot}, \ie the regime where in the absence of
a fundamental Higgs scalar the Standard Model description breaks down
because it breaks unitarity. Of a semi-leptonically decaying $W$ pair
we again use the decay lepton to guarantee triggering. Two forward
tagging jets~\cite{tagging} are not part of the central event, so we
can ignore them in the reconstruction of the $WW$ system. The analysis
is driven towards subjet techniques because for $p_{T,W} > 320$~GeV
the standard $k_T$-algorithm in the vast majority of all events (98\%)
cannot resolve the two $W$ decay jets.

\begin{figure}[t]
  \includegraphics[width=0.40\textwidth]{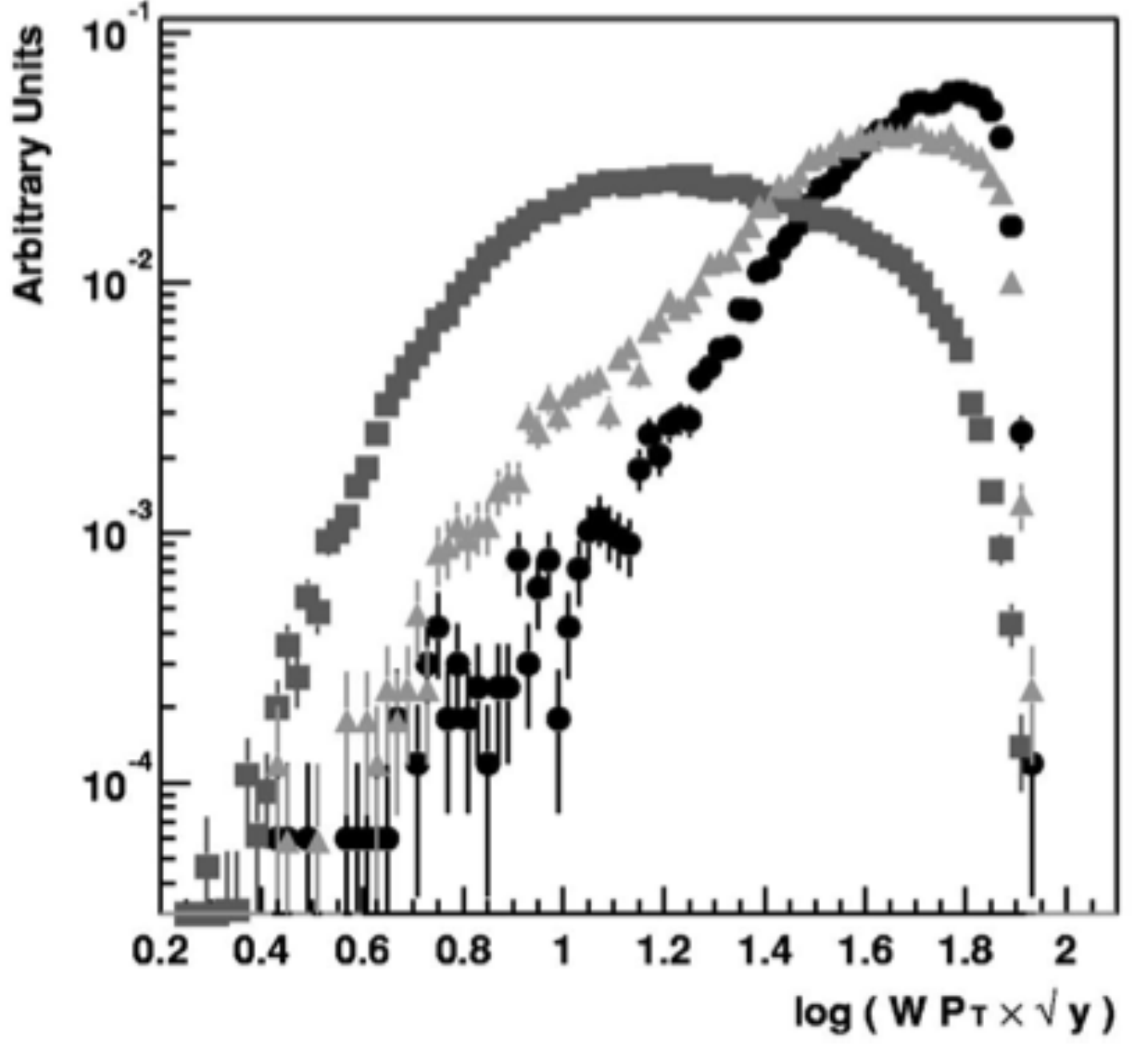}
  \caption{Left: Normalized $p_T \sqrt{y}$ distributions in the $W$
    candidate jet for a massive $WW$ system decaying to two
    relativistic semi-leptonic $W$ bosons (dots), $t\bar{t}$ events
    (triangles), and $W$+jet production (squares). Figure from
    Ref.~\cite{butterbrot}.}
  \label{fig:ysplit1}
\end{figure}

This analysis first studies the scale $y$ at which the $k_T$-algorithm
clusters subjets. In a slightly modified way from
Eq.\eqref{eq:jetalgos} a dimensionless separation measure can be
defined as
\begin{alignat}{5}
y_{j_1 j_2}= 2 \left( 1- \cos \theta_{j_1 j_2} \right) 
            \frac{\min E_{j_i}}{m_j^2} \; .
\label{eq:kt_y}
\end{alignat}
If we search for a hadronic $W$ decay inside a jet we can force the
jet algorithm to produce exactly two subjets, so a general clustering
history is reduced to one $y$ value.  For jets originating from a
heavy $W$ boson we expect this relevant splitting to occur at $y p_T^2
\sim m_W^2$, see Eq.\eqref{eq:basics}. For QCD events with no hard
scale above $\mathcal{O}(m_b)$ appearing in shower and hadronization,
all $y$ values should be much smaller. In Fig.~\ref{fig:ysplit1} we
see that a cut $1.6 < \log (p_T \sqrt{y}) < 2.0$ extracts the signal
well and efficiently removes the $W$+jets continuum
background. Numerically, $(p_T \sqrt{y}) = 10^{1.9} \sim 79$~GeV
indeed corresponds to the $W$ mass. Correlated to this cut on the
splitting history, an additional cut on the jet mass of the $W$
candidate of $70...90$~GeV helps reject the QCD backgrounds
further. The direct generalization of this approach to top tagging at
the LHC leads to the YSplitter or ATLAS top tagger, which we will
discuss in Sec.~\ref{sec:ysplitter}.\bigskip

The third of the early and ground-breaking jet substructure papers
applies a combination of the two approaches presented above and
constructs the BDRS Higgs tagger~\cite{bdrs}. The motivation for such
a Higgs tagger is easy to see: for Higgs masses around 120~GeV two
thirds of all Higgs bosons decay to a pair of bottom quarks. If we
want to study and measure properties of the Higgs sector at the LHC, a
measurement of this branching ratio is
mandatory~\cite{sfitter}. Without such a measurement, the denominator
of any LHC counting measurement $\sigma \times \text{BR}$ and with it
any link between the measured rate and the Higgs couplings is
undetermined.

Inclusive Higgs production as well as Higgs production in weak
boson fusion with a hadronic Higgs decay will not be observed at the
LHC, most likely not even triggered. Associated Higgs production
either with a $V=W,Z$ boson or with a top quark pair have been studied
in detail over many years, but the continuum QCD backgrounds
$Wb\bar{b}$ and $t\bar{t}b\bar{b}$ might well be impossible to
suppress using the standard techniques.

The BDRS Higgs tagger~\cite{bdrs} in its optimized setup is based
on the C/A jet algorithm in combination with a mass drop criterion in
the jet un-clustering. It individually analyzes clusterings where the
original jet algorithm combines $j_1$ and $j_2$ into $j$. The
iterative un-clustering and selection goes through the following
steps:
\begin{enumerate}
\item Un-do the last step of the fat jet clustering where the (parent)
  subjet $j$ breaks into two (daughter) subjets $j_{1,2}$. For the
  BDRS Higgs tagger we use the C/A jet algorithm.
\item Test three conditions: first, the drop in jet mass has to be
  large for a heavy particle decay; second, the splitting should then
  be symmetric; finally the subjets $j_{1.2}$ have to be sufficiently
  hard:
\begin{alignat}{5}
\frac{\min m_{j_i}}{m_j} < 0.67
\qqquad 
\frac{\min(p_{T,j_i}^2) \Delta R_{j_1 j_2}^2}{m_j^2} \sim 
\frac{\min p_{T,j_i}}{\max p_{T,j_i}} > 0.09
\qqquad 
p_{T,j_i}> 30~\gev  \; ,
\label{eq:bdrs}
\end{alignat}
\item For all other splittings, identify the more massive subjet of
  the $j_{1.2}$ with $j$ and remove the less massive one. This
  splitting is then removed from the relevant splitting history of the
  fat jet.
\item Go to the next splittings with the parent subjet $j \equiv j_1$ and,
  if applicable, $j \equiv j_2$. The un-clustering loop will stop once
  the last condition in Eq.\eqref{eq:bdrs} cannot be met anymore.
\item Reconstruct the Higgs mass from the jet mass $m_j$ of the
  relevant splitting(s). To remove effects from soft radiation,
  underlying event, and pile-up the BDRS tagger employs a filtering
  stage described in Sec.~\ref{sec:filtering}. The Higgs mass will
  then be reconstructed from filtered subjets.
\end{enumerate}
The three conditions in Eq.\eqref{eq:bdrs} effectively reject typical
soft and collinear QCD splitting, even though the C/A algorithm only
takes into account soft structures. Differences between jet algorithms
we will discuss for example in Sec.~\ref{sec:seattle} or in
Sec.~\ref{sec:heptop}.

If the jet $j$ is a candidate for the Higgs boson, the two immediate
daughters $j_i$ should be bottom jets. This means we can eventually
apply two $b$-tags, one on each of the subjets inside the fat
jet. This increases the light-flavor QCD rejection by up to four
orders of magnitude. It turns out that $b$-tagging of reasonably
boosted objects with a geometric separation of the order of $R_{bb}
\sim 0.4$ shows an even better performance than $b$-tagging of
continuum jets because the boost aligns the $B$ decay
products~\cite{giacinto}.  The relatively small cut on $m_{j_i}/m_j$
in Eq.\eqref{eq:bdrs} allows configurations where one of the two
bottom decay jets still includes a non-collinear gluon radiation, so
in some situations a softer threshold of 0.9 might improve the
signal-to-background ratio~\cite{tth}.\bigskip

As mentioned above, the results shown in Fig.~\ref{fig:mike} indicate
that the success of any recombination algorithm in fat jet analyses
crucially depends on the treatment of underlying event and
pile-up. The BDRS Higgs tagger includes a filtering strategy which we
will discuss in detail in Sec.~\ref{sec:filtering}. Inspired by
Ref.~\cite{mike} it defines a finer geometric resolution
$R_\text{filter} \sim R_{bb}/2$ and re-clusters the leading Higgs
decay product on this scale. The three leading subjets, corresponding
to two bottom subjets and one, possibly hard, gluon subjet it
identifies with the Higgs mass and momentum. If the original fat jet
is of size $R=1.2$ and the filtering scale is around
$R_\text{filter}=0.3$ the effective area included after filtering is
only $3 \times (0.3/1.2)^2 = 0.2$ of the original fat jet.

One slight complication can arise when we use the BDRS Higgs tagger in
a busy jet environment, like for $t\bar{t}H$ searches~\cite{tth}.  Due
to the high jet multiplicity in the event there might be more than one
candidate for a massive splitting inside the Higgs fat jet. To limit
the combinatorial background we cannot simply include all possible
pairings. Instead, we only keep the (three) leading pairings ordered
by the modified JADE distance $p_{T,j_1} p_{T,j_2} (\Delta R_{j_1
  j_2})^4$ with a bias towards large geometric separation. For the
general cutoff on all subjets considered a relatively large value
$p_T>40$~GeV will add to the QCD rejection~\cite{tth}.\bigskip

It is important to notice that this Higgs tagger does not require a
knowledge of the Higgs mass, which means that we can show
$m_H^\text{rec}$ distributions for signal and background and search
for a mass peak with proper side-bins and all the associated analysis
benefits.  A detailed comparison between the BDRS results and the full
ATLAS detector simulation reveals very few potential problems in Higgs
tagging. The C/A un-clustering combined with the mass drop criterion
and the $b$-tagging work at least as well as expected. Potential
problems are only charm-induced mis-tagging, which will not affect the
top taggers discussed below, and the reconstructed Higgs mass window,
which should be of the order of $\pm 10\%$.

\subsection{YSplitter}
\label{sec:ysplitter}

Early on, ATLAS developed a top tagger based on the structure of the
splitting history $y_i$, as originally suggested in
Ref.~\cite{butterbrot}. It is usually referred to as YSplitter of
`ATLAS default tagger'~\cite{ysplitter}. The physics case for this
tagger is best illustrated by the signal process in the original
publication~\cite{ysplitter}, namely a heavy $Z'$ boson decaying to a
semi-leptonic top pair. As usual, the lepton guarantees that the
events are efficiently triggered and reduces the jet combinatorics.
As suggested in Fig.~\ref{fig:kinematics}, YSplitter requires one hard
fat jet with $p_T > 300$~GeV. However, varying the transverse momentum
range the cuts proposed in Ref.~\cite{ysplitter} give a broad plateau
in the top tagging efficiency for $p_{T,t} = 700...1400$~GeV.\bigskip

\begin{figure}[b]
  \includegraphics[width=0.325\textwidth]{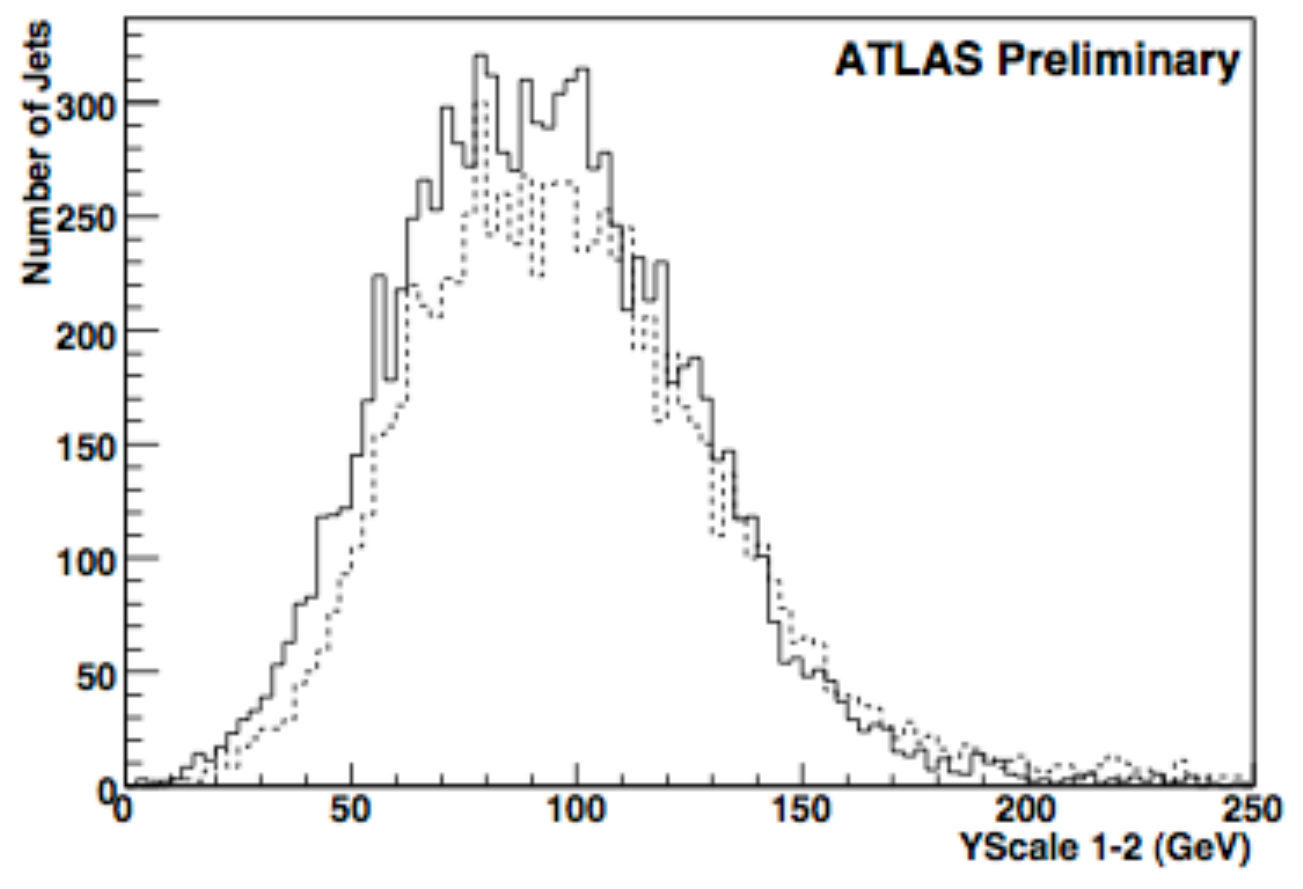}
  \hspace*{0.00\textwidth}
  \includegraphics[width=0.325\textwidth]{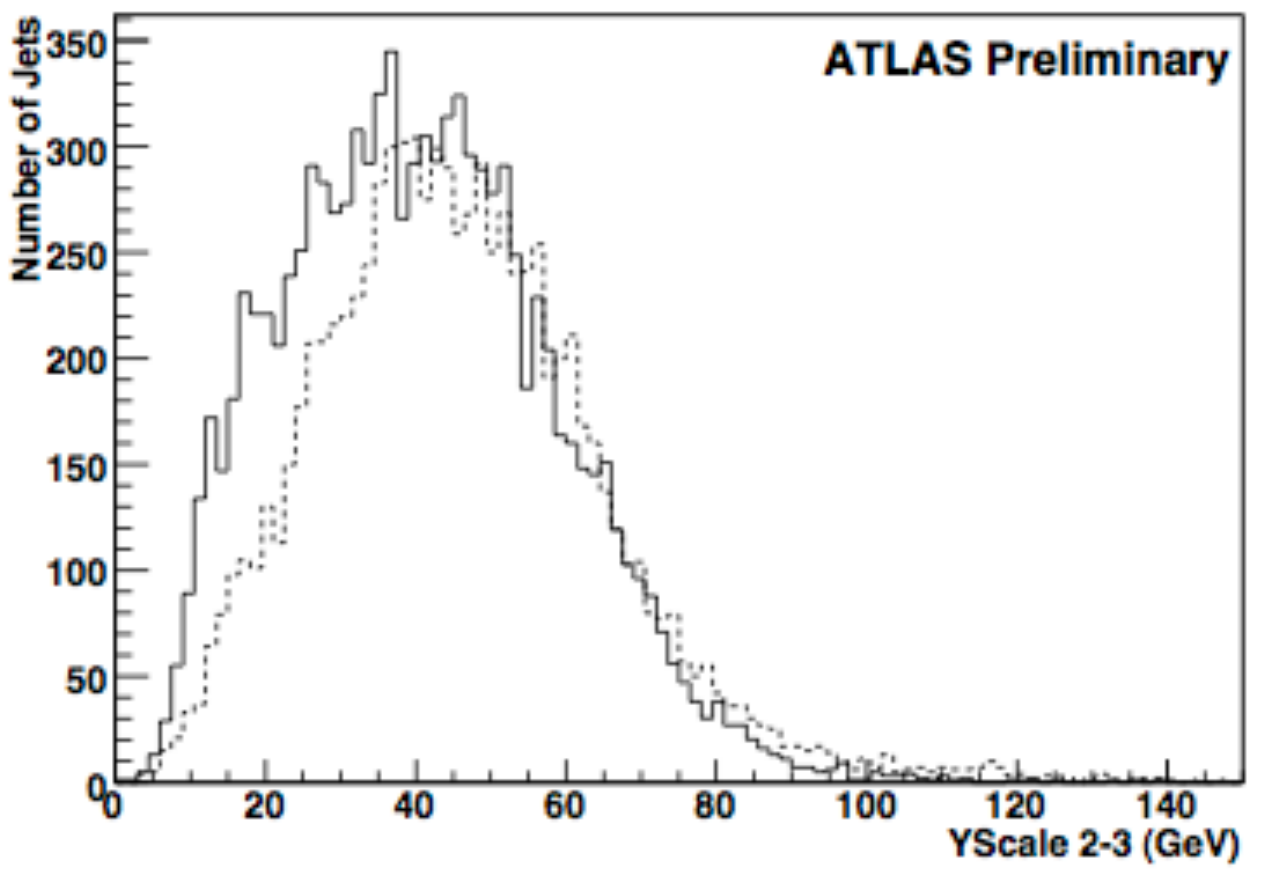}
  \hspace*{0.00\textwidth}
  \includegraphics[width=0.325\textwidth]{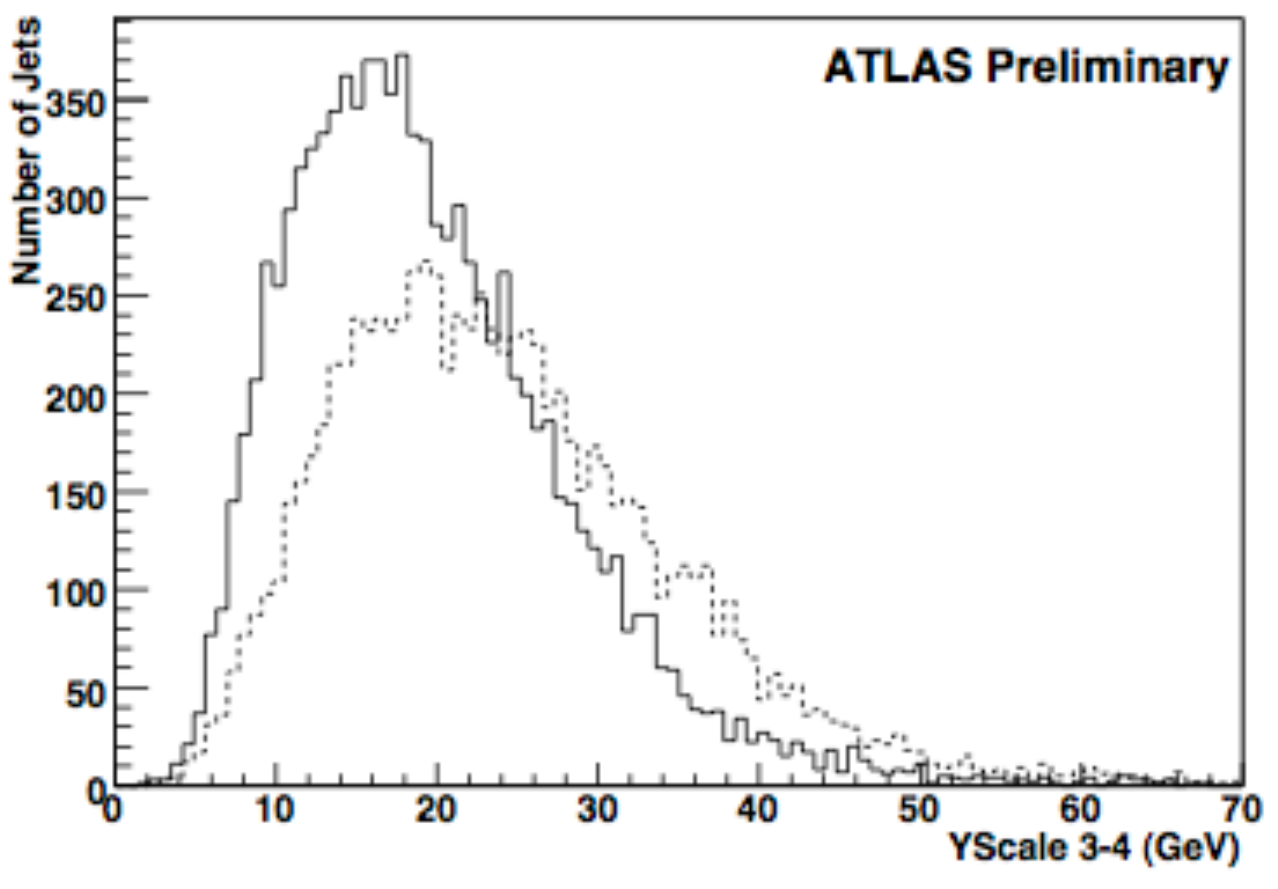}
  \caption{Three splitting points $y_i$ for fat jets from $Z'$ decays
    with $m_{Z'} = 2$~TeV (solid) and 3~TeV (dashed). The $x$-axis
    shows the three leading splitting points of the $k_T$-algorithm as
    $p_T \sqrt{y}/2$.  Figures from Ref.~\cite{ysplitter}.}
  \label{fig:ysplit2}
\end{figure}

The relevant parameter to identify heavy particles in the $k_T$
splitting history of the fat jet are for example three $y$ values, as
defined in Eq.\eqref{eq:kt_y}, corresponding to the
decay jets and one QCD jet or the $b$ jet from the leptonically
decaying top.  In Fig.~\ref{fig:ysplit2} we show the leading three
splitting points for the signal sample. For the first two splittings
which correspond to the $t \to Wb$ and the $W \to jj$ decay steps the
mass of the heavy resonance has a negligible effect.  From the
original $W$ analysis we know that $p_T \sqrt{y}$ scales well with the
mass of the decaying massive particle. This mass variable $p_T
\sqrt{y}/2$ is shown on the $x$-axis of Fig.~\ref{fig:ysplit2} and
indeed reproduces $m_t/2$ and $m_W/2$ for the top decay steps. The
problem with using these one-dimensional distributions for a tagging
algorithm is that they are very broad. For example the top mass
reconstruction only works to roughly $m_t/2 \sim (90 \pm 40)$~GeV.  In
addition, the correlation between the three extracted values for $p_T
\sqrt{y}$ is minor.

For the fourth splitting which arises because of the additional $b$
jet from the leptonic top decay and continuum QCD jet radiation we
find typical values $p_T \sqrt{y} \sim 30$~GeV. This is much harder
than expected from QCD. In part this is due to the very heavy particle
in the $s$-channel and the associated very hard collinear radiation
scale. This means this value increases even more for heavier $Z'$
scenarios. While such an effect of a generic hard process is
understood from appropriate QCD simulations~\cite{skands}, it does not
probe the top content of the signal and is likely to be mimicked by
the backgrounds once all other cuts have moved them into a signal-like
phase space region. Generally, the last splitting is in the range
where jet substructure algorithms typically have a soft cutoff on all
subjets considered.\bigskip

The original $W$ tagger~\cite{butterbrot} applies a second cut on the
$W$ jet mass without any study of the correlation between the jet mass
and the splitting history cuts. For well isolated jets the
reconstructed top jet mass is much more narrow than the $p_T \sqrt{y}$
distributions, typically reproducing the input top mass to
$\mathcal{O}(10\%)$. Implementing cuts on two-dimensional correlations
of $p_T \sqrt{y}$, $m_j$, and $p_{T,j}$ improves the signal efficiency
as well as the QCD background rejection. Therefore, in the most recent
versions of the ATLAS default tagger a neural net is employed for all
measured parameters from the $k_T$ splitting history.  It is worth
noting that a very important initial step in experimentally
establishing subjet techniques has been achieved with this tagger: in
an ATLAS study it was shown that for signals with a heavy $s$-channel
resonance the $Z'$ mass reconstruction from boosted top quarks is
better than from the known reconstruction techniques of semi-leptonic
top quarks.

\subsection{Seattle Tagger}
\label{sec:seattle}

A second tagging algorithm based on the $k_T$-algorithm follows a
different approach. Instead of studying the splitting history the way
YSplitter~\cite{butterbrot,ysplitter} does it, it aims at removing all
soft and collinear splittings as expected from pure massless
QCD~\cite{pruning1} and then studying the remaining massive splitting
candidates~\cite{pruning2}.

In the $k_T$-algorithm, two cases of preferred jet merging appear. On
the one hand, if one subjet is much softer than others, it should be
merged into one of the harder subjets. This can be identified through
a $\min (p_{T,j_1},p_{T,j_2})/p_{T,j}$ measurement. On the other hand,
collinear splitting in QCD does not require one of the jets to be
soft, so also two jets with small $\Delta R_{j_1,j_2}$ should be
merged. Any splitting in the $k_T$-history which does not satisfy 
these two conditions is a candidate for a non-QCD
splitting. The details of this `pruning' procedure we will discuss in
Sec.~\ref{sec:pruning}. Originally, it was meant to improve tagging
algorithms based on jet mass measurements both using the $k_T$ or the
C/A jet algorithm~\cite{pruning1}. In a second step, it can be used to
tag heavy particles inside of fat jets~\cite{pruning2}.\bigskip

The Seattle or pruning top tagger includes three parameters: the jet
mass $m_j$, the transverse momentum drop $\min p_{T,j_i}/p_{T,j}$ as
defined in Eq.\eqref{eq:basics}, and the angular spread of the
daughter subjets $\Delta R_{j_1 j_2}$. The details of the pruning
algorithm we discuss in Sec.~\ref{sec:pruning}.  A comprehensive study
first shows that the jet mass from the C/A algorithm is more stable
than from the $k_T$-algorithm. The reason is that according to
Eq.\eqref{eq:basics} we can write the (parent) jet mass in terms of
the distance measure of the $k_T$-algorithm
\begin{alignat}{5}
  m_j^2 = \left( p_{T,j} \, z \, \Delta R_{j_1 j_2} \right)^2 \; \frac{1-z}{z} 
        = d_{j_1 j_2} \; \frac{1-z}{z} \qquad (k_T~\text{algorithm}) \; .
\end{alignat}
This formula shows that soft or small-$z$ splitting will contribute
disproportionally to the jet mass, leaving it vulnerable to many
problems with soft radiation detection and identification.

Looking at the second observable, we find that the $p_T$ drop for the
C/A algorithm prefers small values of $\min p_{T,j_i}/p_{T,j} < 0.04$
even for boosted top decay jets. Unless we do something about it, this
makes top tagging in this regime difficult. Finally, the $R$
separation of the last splitting in the case of the C/A algorithm has
to correspond to $\Delta R_{bW} > \Delta R_{jj}$ because of the
geometric ordering of the recombination. For the $k_T$-algorithm the
$R$ separation of the last splitting can be somewhat smaller. These
basic differences can be blurred for example by pruning or similar
procedures discussed in Sec.~\ref{sec:qcd}, but they nevertheless
explain why the BDRS Higgs tagger, based on the C/A jet algorithm,
requires a drop in the jet mass and not in the transverse momentum of
the splittings~\cite{bdrs}. In Sec.~\ref{sec:heptop} we will discuss
the application of the C/A algorithm in combination with a mass drop
criterion for top tagging.\bigskip

Following a detailed analysis the critical observables for the Seattle
tagger are chosen to be jet masses, \ie the reconstructed top and $W$
jet masses. A fat jet for example of size $R=1.0$ is un-clustered
iteratively, like described for the BDRS Higgs tagger in
Sec.~\ref{sec:early}. An additional pruning step ensures that only
hard and well separated splittings survive. For these candidate
splittings the jet masses are compared to their known values. Because
of the problems with the jet mass reconstruction described above, the
allowed deviations for a top tag are of the order of $m_t \pm 14$~GeV
and $m_W \pm 12$~GeV for the $k_T$-algorithm while for the C/A
algorithm they can be smaller, $m_t \pm 11$~GeV and $m_W \pm
8$~GeV. Nevertheless, due to the other issues with the C/A-algorithms
Ref.~\cite{pruning2} finds that a pruning-based top tagger shows a
better performance when combined with the $k_T $ algorithm, in
particular towards larger $p_{T,t}$ values.\bigskip

When comparing different top taggers it should be noted that the
pruning or Seattle tagger can be viewed as a more general analysis
tool for fat jets with suspected massive splittings. In addition, the
pruning stage can be used in combination with any specialized
Higgs~\cite{spanno_soper} and top tagger~\cite{heptop2}. 

\subsection{Johns Hopkins Tagger}
\label{sec:hopkins}

The Johns Hopkins top tagger~\cite{hopkins} was the first public top
tagger applying the successful BDRS setup to the two-step top
decay. Because the additional kinematic conditions on now three decay
jets should allow for a more effective QCD and $W$+jets background
rejection, an additional $b$-tag is not foreseen. However, it can be
added to it, as we discuss towards the end of
Sec.~\ref{sec:heptop}.\bigskip

In its original setup the tagger is optimized for relatively highly
boosted top quarks, so for fat jet transverse energies $E_{T,j} >
1$~TeV is starts with a C/A jet of size $R=0.8$. Translated into
transverse momentum a second basic requirement is $p_{T,j} > 0.35 \;
E_{T,j}$. From Fig.~\ref{fig:kinematics} we see that boosted top
quarks with $p_{T,t} > 350$~GeV indeed require at least $R = 0.8$ to be
fully included inside the fat jet. Following the BDRS recipe described
in Sec.~\ref{sec:early} the fat jet is then iteratively de-clustered. A
splitting is kept as the candidate for one of the two top decay steps
if
\begin{alignat}{5}
\frac{\min p_{T,j_i}}{p_T^\text{hard}} > 0.1
\qqqquad 
\Delta R_{j_1 j_2} > 0.19 \; .
\label{eq:hopkins}
\end{alignat}
A slight change with respect to the BDRS algorithm is that
$p_T^\text{hard}$ is the same for all splittings. It is fixed by the
parent $p_{T,j}$ in the first relevant splitting and stays at this
value for the remaining un-clustering sequence. This way, the first
condition in Eq.\eqref{eq:hopkins} also terminates the un-clustering
loop and no additional soft cutoff is required. Once two successive
clusterings corresponding to the top and $W$ decays are identified,
the algorithm also terminates.

The numerical values for the tunable parameters shown in
Eq.\eqref{eq:hopkins} we quote for the softest fat jets considered in
Ref.~\cite{hopkins}, \ie $p_{T,t} > 350$~GeV. Comparing this condition to
Eq.\eqref{eq:bdrs} we see that the mass drop criterion in the BDRS
algorithm is replaced by a $p_T$ drop, as for example discussed in
Sec.~\ref{sec:seattle}.

\begin{figure}[t]
  \includegraphics[width=0.45\textwidth]{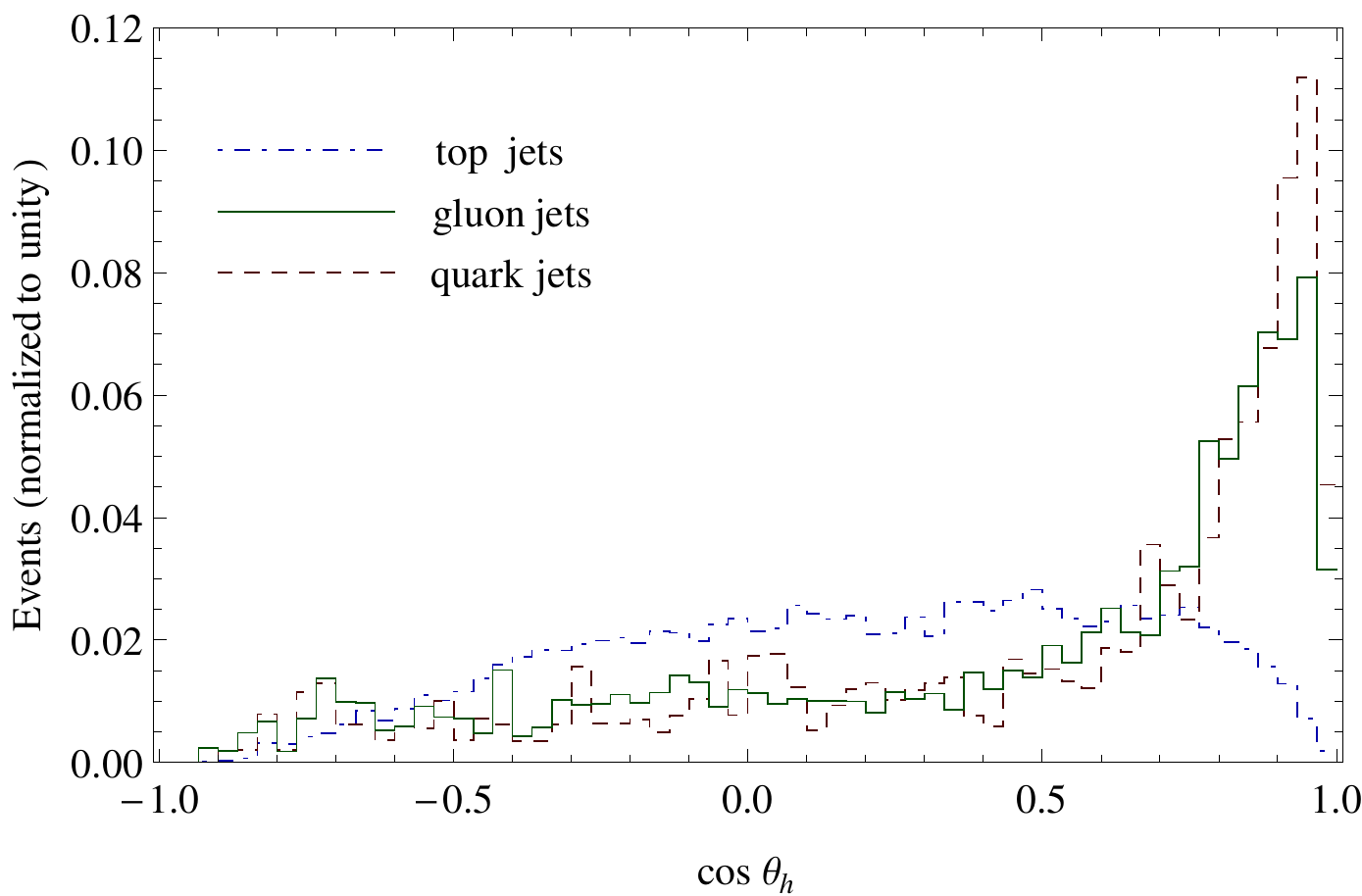}
  \hspace*{0.05\textwidth}
  \includegraphics[width=0.45\textwidth]{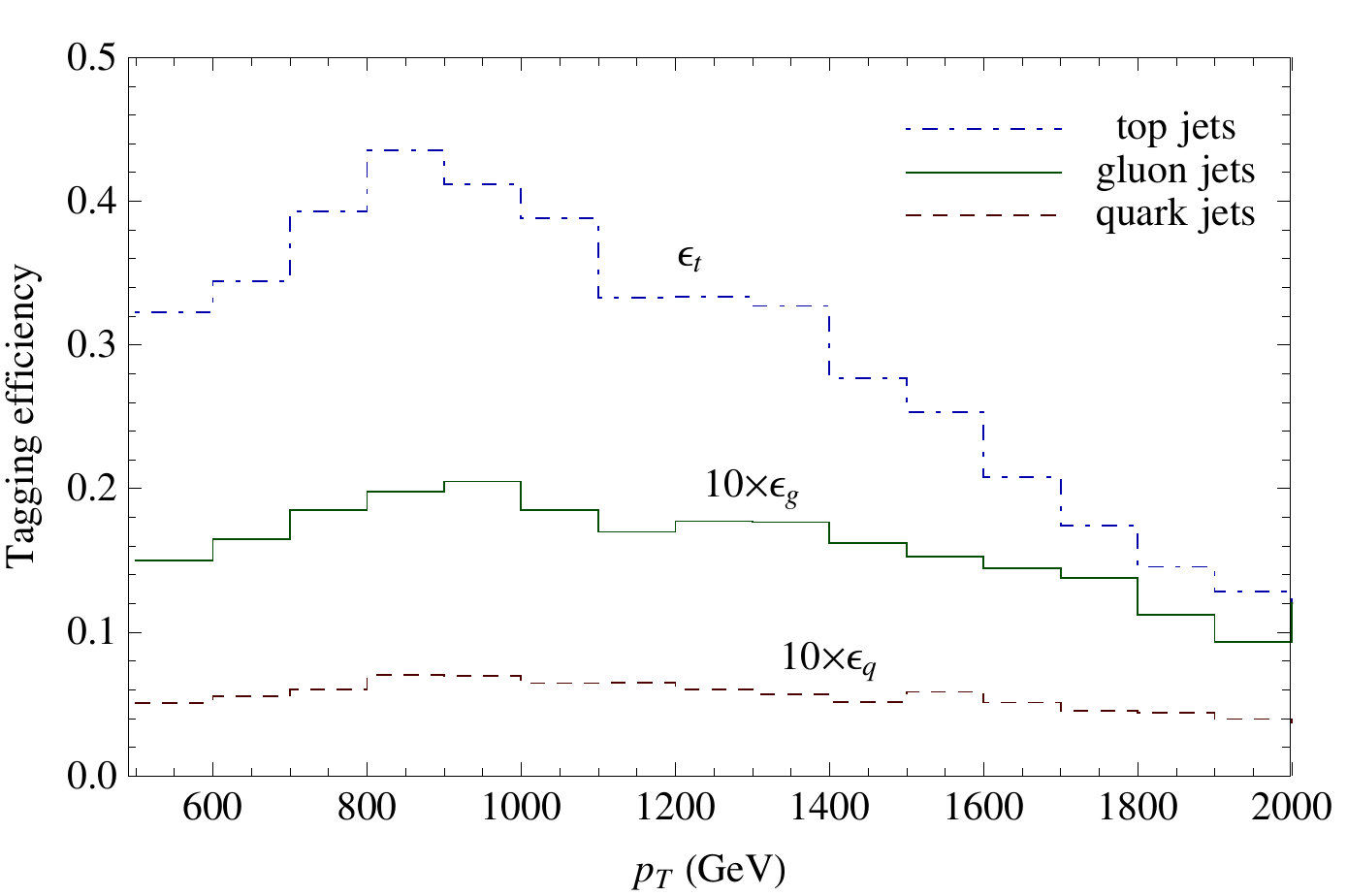}
  \caption{Left: Helicity angle distribution for top, gluon, and
    light-flavor quark jets fulfilling $p_T > 700$~GeV. All other
    tagging cuts have been already imposed. Right: original tagging
    and mis-tagging efficiency estimates for top and QCD jets as a
    function of the fat jet $p_T$.  Figures from Ref.~\cite{hopkins}.}
  \label{fig:hopkins}
\end{figure}

For signal events the iterative algorithm should stop after
identifying three or four decay subjets, allowing for one final state
gluon in the top decay. At this stage, three conditions have to be met
by one combination of the jet momenta
\begin{alignat}{5}
 m_{jjj} = m_t \pm 30~\gev
\qqquad 
 m_{jj} = m_W \pm 15~\gev
\qqquad 
 \cos \theta_h < 0.7 \; .
\label{eq:3kin}
\end{alignat}
The angle $\theta_h$ is the top helicity angle, measured in the rest
frame of the reconstructed $W$ boson. It is defined as the opening
angle of the incoming top momentum and the softer of the two $W$ decay
subjets. In leptonic top decays the corresponding $W$ decay product is
usually chosen as the lepton momentum, but for hadronic $W$ decays the
two jets are indistinguishable. The helicity angle distributions for
signal and backgrounds we show in Fig.~\ref{fig:hopkins}. For top jets
it is essentially flat, while for massless QCD splittings it is
strongly peaked towards small angles. This corresponds to the
collinear divergence in the splitting kernels which runs into a $1/(1-
\cos \theta_h) \sim 1/\theta_h^2$ pole. In other words, the top
helicity angle does not act to reject top decays with wrong angular
correlations; it is simply another independent observable of the three
top decay jets in addition to the two mass constraints.

Experimental QCD effects like underlying event or pile-up will
eventually require some kind of filtering stage in the algorithm, as
we will describe in Sec.~\ref{sec:qcd} and as already included in the
HEPTopTagger discussed below. This modification is straightforward.
In addition, the identification of the top helicity angle as one of
the kinematic observables means that the tagging algorithm identifies
the $b$-jet from the top decay. This would allow us to immediately add
a $b$-tag to the Johns Hopkins tagger~\cite{heptop2}, if we are
interested in additional background rejection.\bigskip

In the right panel of Fig.~\ref{fig:hopkins} we show the original
estimates for the tagging efficiencies. While the final numbers will
most likely change significantly after including pile-up, some kind of
filtering, and detector effects, we do observe a clear structure in
the signal efficiency: for low transverse momenta the tagging
efficiency decreases mostly due to a limited number of top jets fully
included in the original $R=0.8$ fat jet. To some degree this can be
improved by allowing larger $R$ values as one of the tuning parameters
of the tagger~\cite{heptop2}. The drop towards large transverse
momenta has a two-fold reason; for strongly boosted top quarks
eventually the calorimeter resolution will limit the tagging
performance. Even before this, once we arrive at $m_W \ll p_{T,t}$
some of the fat jet splitting history will be dominated by soft
kinematics. In that case the C/A re-clustering will fail to
reconstruct the correct top decay products~\cite{heptop2}. In this
case we should switch to the $k_T$-algorithm for the top
reconstruction stage.\bigskip

The CMS tagger~\cite{cmstagger} is essentially a Johns Hopkins tagger
with very few modifications. The algorithm is exactly the same, but
the three kinematic conditions defined in Eq.\eqref{eq:hopkins} are
replaced by
\begin{alignat}{5}
m_{jjj(j)} = 100...250~\gev
\qqqquad 
\min_{j_1,j_2,j_3} m_{jj} > 50~\gev \; .
\end{alignat}
The second condition means that we first identify the three hardest
subjets inside the fat jet, then combine them into six two-jet pairs,
and finally require that all of those have a sufficiently large jet
mass. The CMS study tests that indeed the C/A jet algorithm leads to
the best results. An obvious and critical check of the CMS tagger
would be an observation of tops in Standard Model top pair events with
a top mass peak in the $m_{jjj(j)}$.\bigskip

\subsection{HEPTopTagger}
\label{sec:heptop}

Similar to the Johns Hopkins tagger the
HEPTopTagger~\cite{heptop1,heptop2} (Heidelberg-Eugene-Paris) makes
use of the BDRS setup and generalizes it to the multi-step top decay
structure.  It is originally used to study $t\bar{t}H$ searches with
two fat jets in a high-multiplicity environment, one from the Higgs
and the other one from the top~\cite{tth}. Its public version is in
detail described in the Appendix of Ref.~\cite{heptop1}. Additional
improvements, discussed in Ref.~\cite{heptop2}, we will briefly
discuss at the end of this section. The reference analyses for the
HEPTopTagger are associated top-Higgs production~\cite{heptop1} and
supersymmetric scalar top pairs decaying to top quarks and missing
energy~\cite{heptop2}. Hence, this tagger needs to aim at considerably
lower $p_{T,t}$ values. This makes it a promising tool to establish
subjet techniques and top tagging on Standard Model top samples at the
7~TeV LHC.\bigskip

From Fig.~\ref{fig:kinematics} we can see that starting a top tagger
with C/A jets of size $R=1.5$ should allow us to access top quarks
down to $p_{T,t} \sim 200$~GeV. For example looking at Standard Model
top pairs an increase from $R=0.9$ to $R=1.5$ means that twice as many
top quarks are accessible even for relatively large $p_{T,t} \sim
400$~GeV.  On the negative side, this increase in the jet area poses two
problems for the tagging algorithm. First, subjet combinatorics will
increase and it will get harder to identify the individual top decay
products. Second, pile-up will become a huge problem, so the
HEPTopTagger always includes a filtering stage as described in
Sec.~\ref{sec:filtering}.

The tagging algorithm proceeds in steps similar to the BDRS Higgs
tagger with its mass drop criterion Eq.\eqref{eq:bdrs}:
\begin{enumerate}
\item Un-doing the last clustering of the jet $j$ the mass drop
  criterion $\min m_{j_i} < 0.8 \, m_j$ determines if we keep $j_1$
  and $j_2$. Subjets with $m_{j_i} < 30~\gev$ are not considered,
  which eventually ends the iterative un-clustering. A symmetry
  requirement, as in the BDRS algorithm, is not included.
\item Apply a filtering stage to construct one three-subjet
  combination with a jet mass within $m_t \pm 25$~GeV.
\item Order these three subjets by $p_T$. If their jet masses
  $(m_{12}, m_{13},m_{23})$ satisfy one of the following three
  criteria, accept them as a top candidate:
\begin{alignat}{5}
&0.2 <\arctan \frac{m_{13}}{m_{12}} < 1.3
\qquad \text{and} \quad
R_{\min}< \frac{m_{23}}{m_{123}} < R_{\max}
\notag \\
&R_{\min}^2 \left(1+\left(\frac{m_{13}}{m_{12}}\right)^2 \right) 
< 1-\left(\frac{m_{23}}{m_{123}} \right)^2
< R_{\max}^2 \left(1+\left(\frac{m_{13}}{m_{12}}\right)^2 \right)       
\quad \text{and} \quad 
\frac{m_{23}}{m_{123}} > R_\text{soft}
\notag \\
&R_{\min}^2\left(1+\left(\frac{m_{12}}{m_{13}}\right)^2 \right) 
< 1-\left(\frac{m_{23}}{m_{123}} \right)^2
< R_{\max}^2\left(1+\left(\frac{m_{12}}{m_{13}}\right)^2 \right)        
\quad \text{and} \quad 
\frac{m_{23}}{m_{123}}> R_\text{soft}
\label{eq:heptop}
\end{alignat} 
\item For consistency, require the combined $p_T$ of the three subjets
  to be above 200~GeV.
\end{enumerate}
The dimensionless mass windows $R_{\min}=85\% \times m_W/m_t$ and
$R_{\max}=115\% \times m_W/m_t$ are tunable and will be optimized by
the experimental collaborations. The soft cutoff $R_\text{soft} =
0.35$ removes QCD events which the C/A algorithm cannot correctly
identify as soft radiation.

\begin{figure}[t]
\includegraphics[width=0.30\textwidth]{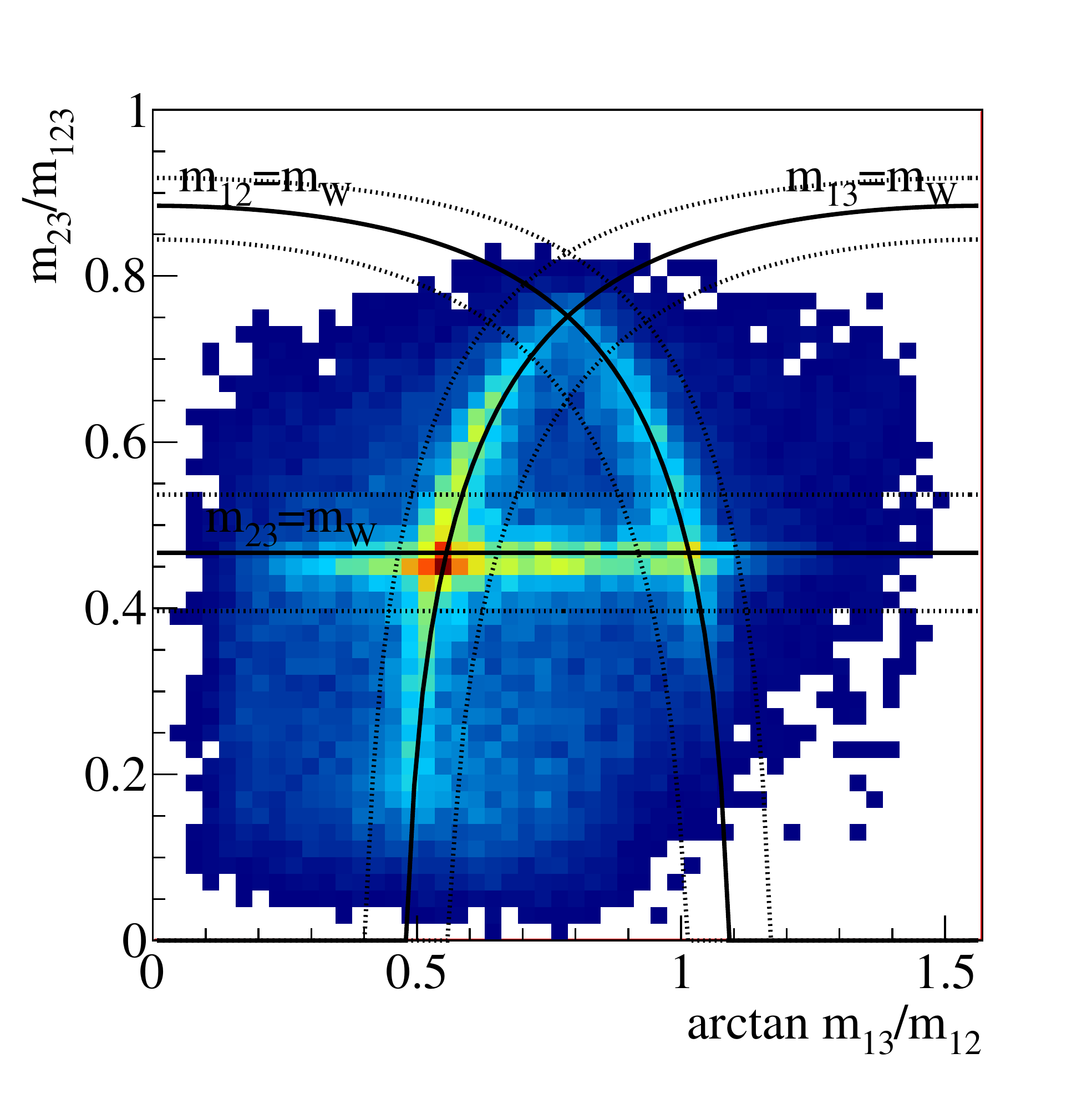}
\includegraphics[width=0.30\textwidth]{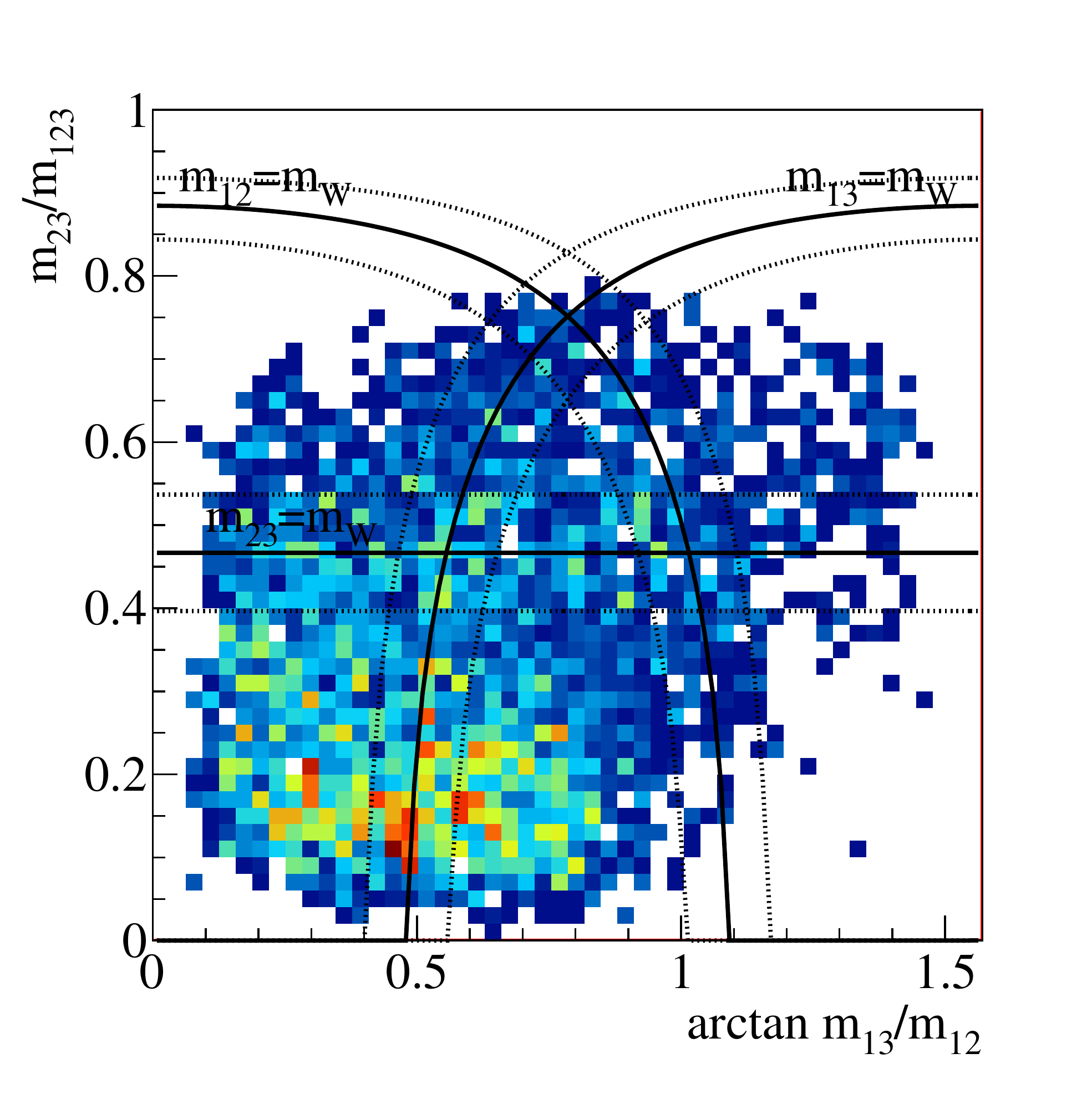}
\includegraphics[width=0.30\textwidth]{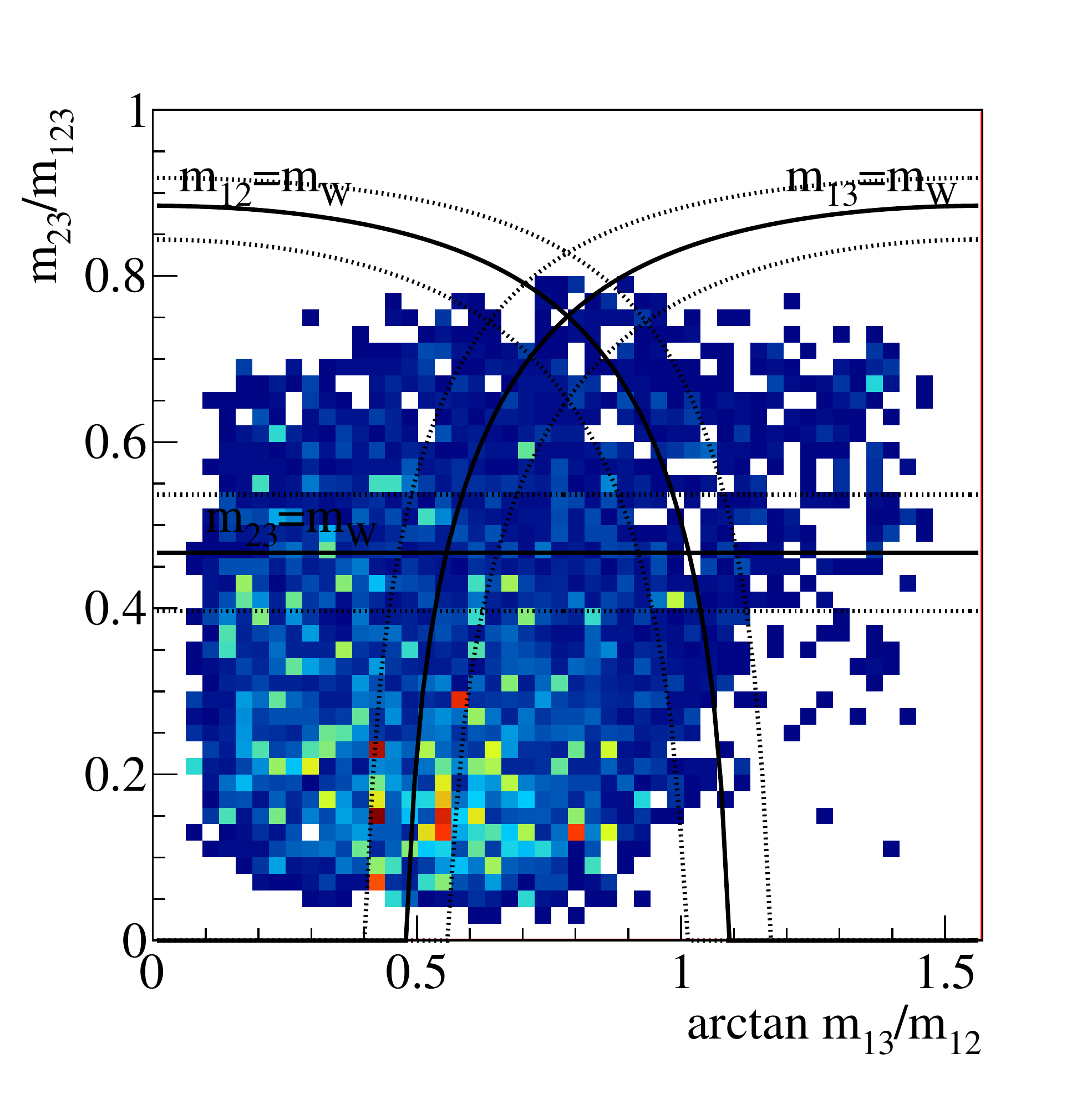}
\caption{Events in the $\arctan m_{13}/m_{12}$ vs $m_{23}/m_{123}$
  plane for $t\bar{t}$ (left). $W$+jets (center) and pure QCD jets
  (right) samples. More densely populated regions of the phase space
  appear in red. Figures from Ref.~\cite{heptop1}.}
\label{fig:heptop1}
\end{figure}

Compared to the Johns Hopkins or CMS taggers there are four main
differences. First, the HEPTopTagger determines the entire clustering
history and does not enforce exactly two hard splittings. Second, it
is based on a mass drop criterion instead of a transverse momentum
drop. As a matter of fact, the entire HEPTopTagger algorithm is only
based on jet masses. The C/A algorithm should reconstruct those very
well after filtering. Third, in the left panel of
Fig.~\ref{fig:heptop1} we see that it is quite likely that two
invariant mass combinations of the three top decay products will be in
the $m_W$ range.  This is due to the specific top kinematics with its
endpoint $\max m_{bj_i}^2 < m_t^2 - m_W^2$, which is numerically close
to $m_W^2$~\cite{wjets}. Consequently, we avoid assigning two subjets
to the $W$ decay and instead impose a symmetric condition, like the
one given in Eq.\eqref{eq:heptop}. Finally, similar to the BDRS
approach the HEPTopTagger uses filtered subjets to reconstruct the top
decay products. This has significant impact on the performance in
particular on data, as discussed in Sec~\ref{sec:qcd}.\bigskip

\begin{figure}[t]
\includegraphics[width=0.30\textwidth]{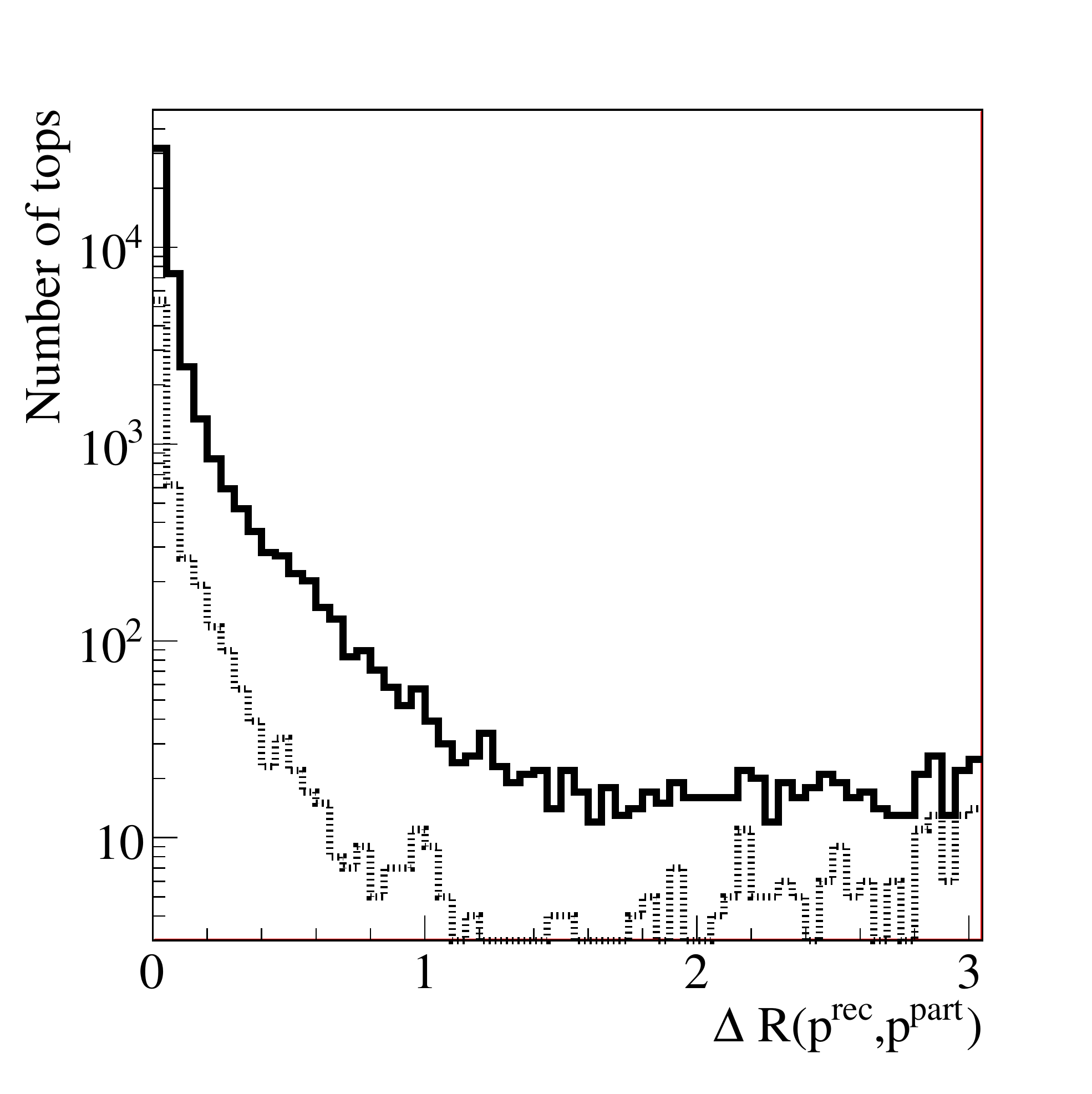}
\includegraphics[width=0.30\textwidth]{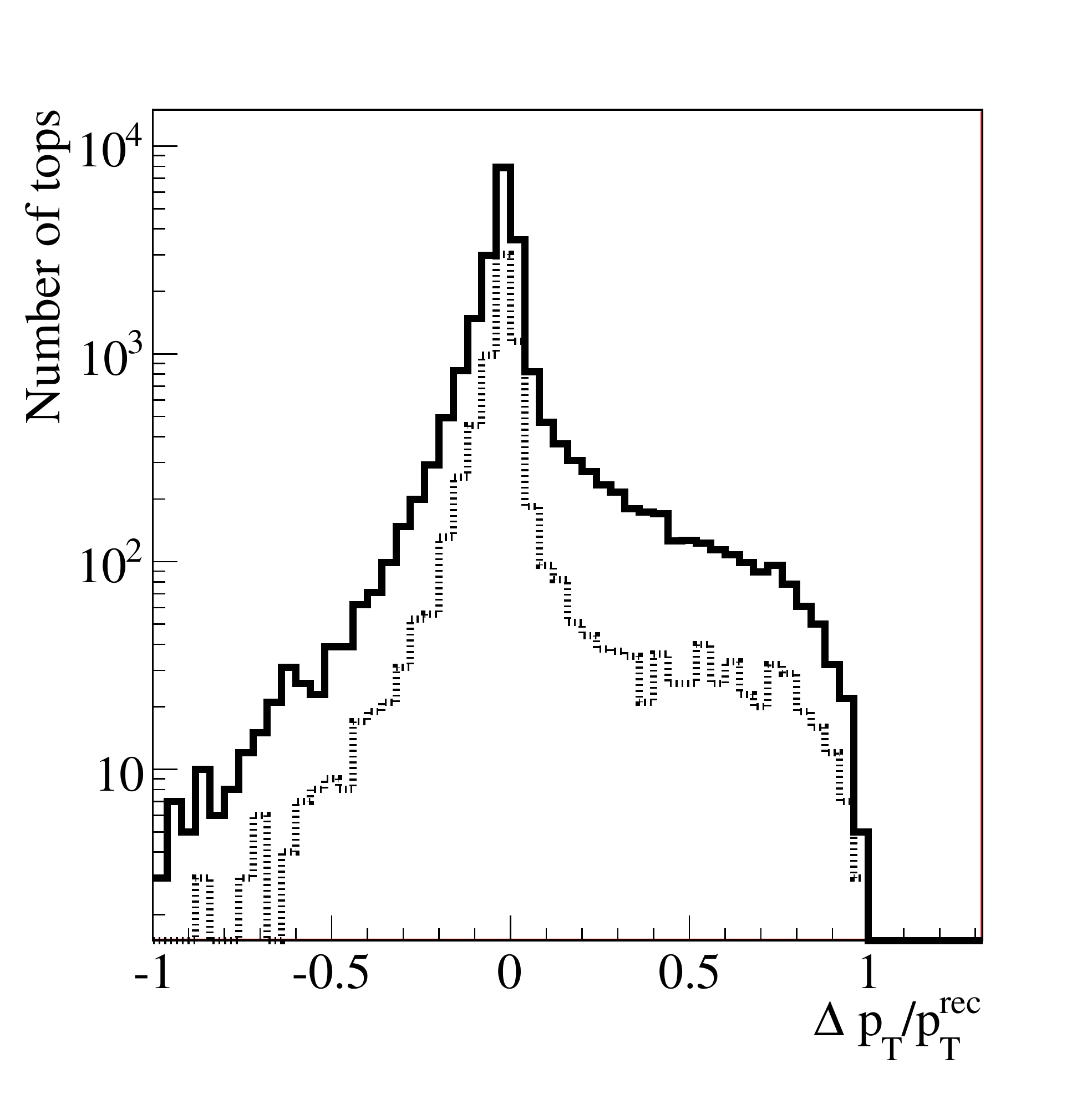}
\includegraphics[width=0.30\textwidth]{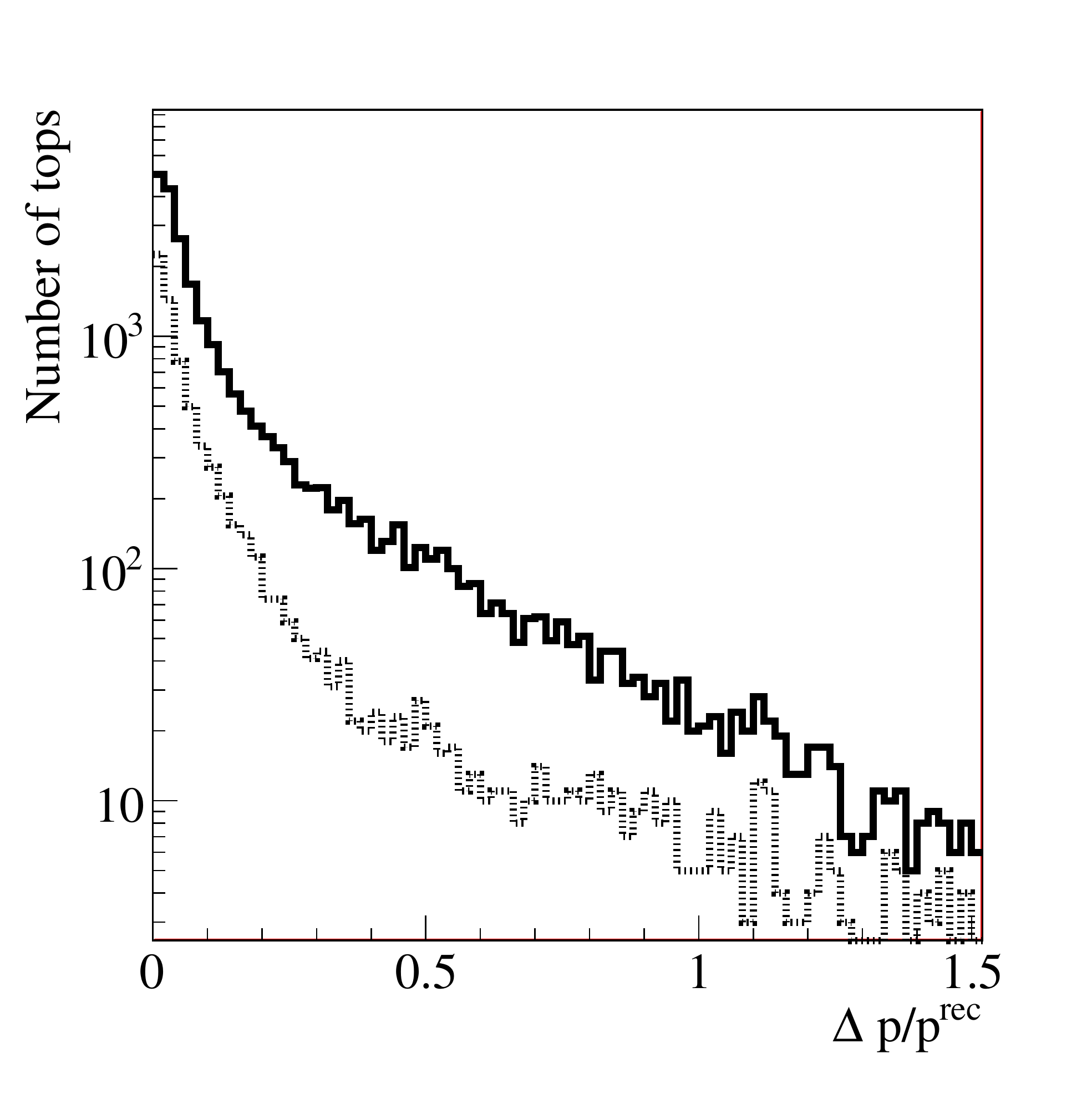}
\caption{$\Delta R$ between the reconstructed and the parton-level top
  quark in $t\bar{t}$ events (left), $\Delta p_T/p_T^\text{rec}$ for
  the same sample (center) and $\Delta |\vec{p}|/|\vec{p}|^\text{rec}$
  (right). For the solid curves we only apply the default cut
  $p_{T,t}^\text{rec} > 200$~GeV while the dashed curves require
  $p_{T,t}^\text{rec} > 300$~GeV. Figures from Ref.~\cite{heptop1}.}
\label{fig:heptop2}
\end{figure}

No matter if the new physics signature is a massive $s$-channel
resonance or a decaying supersymmetric top squark we need to rely on
the top tagger to not only identify top quarks but also reconstruct
their 4-momenta. For the HEPTopTagger the quality of this momentum
reconstruction has been studied in detail. The question if the top
tagger really reconstructs all top decay products is surprisingly
irrelevant for this test --- a generic tagger will always be fairly
likely to correctly assign the hardest two top decay products, while
the softer $W$ decay subjet will contribute little to the
reconstructed top momentum. This is the reason why even for moderately
boosted tops 95\% of the tagged events show a correctly reconstructed
direction within $\Delta R = 0.5$; for more than 80\% of the tops the
momentum is reconstructed within 20\% of the Monte Carlo truth. For
stop pairs decaying to fully reconstructed hadronic tops and missing
energy this means we can apply the usual $m_{T,2}$ methods to measure
the supersymmetric masses in the process.\bigskip

In a subsequent study, possible improvements to the HEPTopTagger are
tested~\cite{heptop2}. Similar results should apply to other top
taggers, like the Johns Hopkins tagger. As a first improvement, we can
use the $k_T$-algorithm in the filtering and the re-clustering
stages. For the re-clustering this stabilizes the HEPTopTagger tagging
efficiency on a plateau between $p_{T,t} = 200$~GeV to 600~GeV, where
the C/A version shows a significant drop in performance. Second, as
discussed in Sec.~\ref{sec:filtripru} we can add a pruning stage
parallel to the filtering and then include both reconstructed top
masses in the selection. Compared to the critical pure QCD background
this can improve the signal-to-background ratio by a factor of two. In
contrast, increasing the size of the fat jet to $R=1.8$ has little
benefit even for Standard Model top pairs because subjet combinatorics
compensate the possible benefits.\footnote{It does increase the
  absolute number of signal events with constant $S/B$, so it might be
  useful in the first stage of an experimental
  test~\cite{gregor_private}.} Finally, simply adding a $b$-tag at the
end of the top tagging algorithm can improve the background
rejection. Including the $b$-tagging information inside the
HEPTopTagger algorithm does not appear promising.\bigskip

A combination of the HEPTopTagger and a (sub-)jet count, as
we will describe in Sec.~\ref{sec:scet} is presented in
Ref.~\cite{vernon}. It first defines an anti-$k_T$ fat jet of size
$R=1.5$, reclusters it as a C/A jet with a mass drop criterion,
applies the usual top and $W$ mass constraints and then adds a
$b$-tag. As an additional criterion it uses an $R=0.6$ anti-$k_T$ jet
algorithm on the same event and compares the number of jets per event.
Only events with two fat jets of size $R=1.5$ and at least three
smaller jets of size $R=0.6$ are kept as top pair candidates. The
reasoning behind this is that QCD dijet events with two hard jets will
show two hard jet structures independent of the jet size while
high-multiplicity top jets will show an increased number of jets.

\subsection{Thaler-Wang Tagger}
\label{sec:thaler_wang}

Around the same time when the Johns Hopkins tagger adopted the BDRS
approach for boosted top quarks, the Thaler-Wang tagger took a
different approach. It describes a subjet splitting in terms of the
jet mass $m_j$ of the parent subjet and the subjet energy drop in the
splitting. The definition of the jet energy drop is not unique, so it
can be implemented into a tagging algorithm in different ways, all
equivalent in the massless and collinear limit
\begin{alignat}{5}
 z = \frac{\min E_{j_i}}{E_j} 
   \sim \frac{d_{j_1 j_2}}{d_{j_1 j_2}+m_j^2}
   \sim \frac{\min (p_{j_i} \cdot p_\text{ref})}{(p_j \cdot p_\text{ref})}
\label{eq:thaler_wang1}
\end{alignat}
where $d_{j_1 j_2} = \min (p_{T,j_i}^2) \, \Delta R_{j_1 j_2}^2$ is the
distance measure in the $k_T$-algorithm introduced in
Eq.\eqref{eq:jetalgos} and $p_\text{ref}$ is a free reference
4-vector, for example the direction of one of the incoming
protons. This energy drop is expected to only be weakly correlated
with the jet mass, which simplifies the tagging algorithm.

Because all definitions in Eq.\eqref{eq:thaler_wang1} coincide in the
collinear limit it is unlikely that we will be able to compare their
performance inside a tagger on Monte-Carlo data, which is generated
with a parton shower.  For top jets simple simulations show that the
second two definitions are essentially equivalent while the actual
energy ratio has a significantly softer $z$ spectrum. For QCD jets
all definitions are equally strongly peaked towards small $z$ values,
but the energy drop has much smaller tails for $z >
0.3$~\cite{thaler_wang}.\bigskip

To extract massive splittings the Thaler-Wang tagger starts with an
anti-$k_T$ jet of size $R=1$. Of this fat jet only the regions with
the hardest jets are labelled and re-clustered with a
$k_T$-algorithm. To apply the tagger to LHC data in the presence of
underlying event and pile-up it needs to be supplemented with a
trimming stage to remove soft calorimeter activity, as described in
Sec.~\ref{sec:trimming}.

\begin{figure}[t]
  \includegraphics[width=0.32\textwidth]{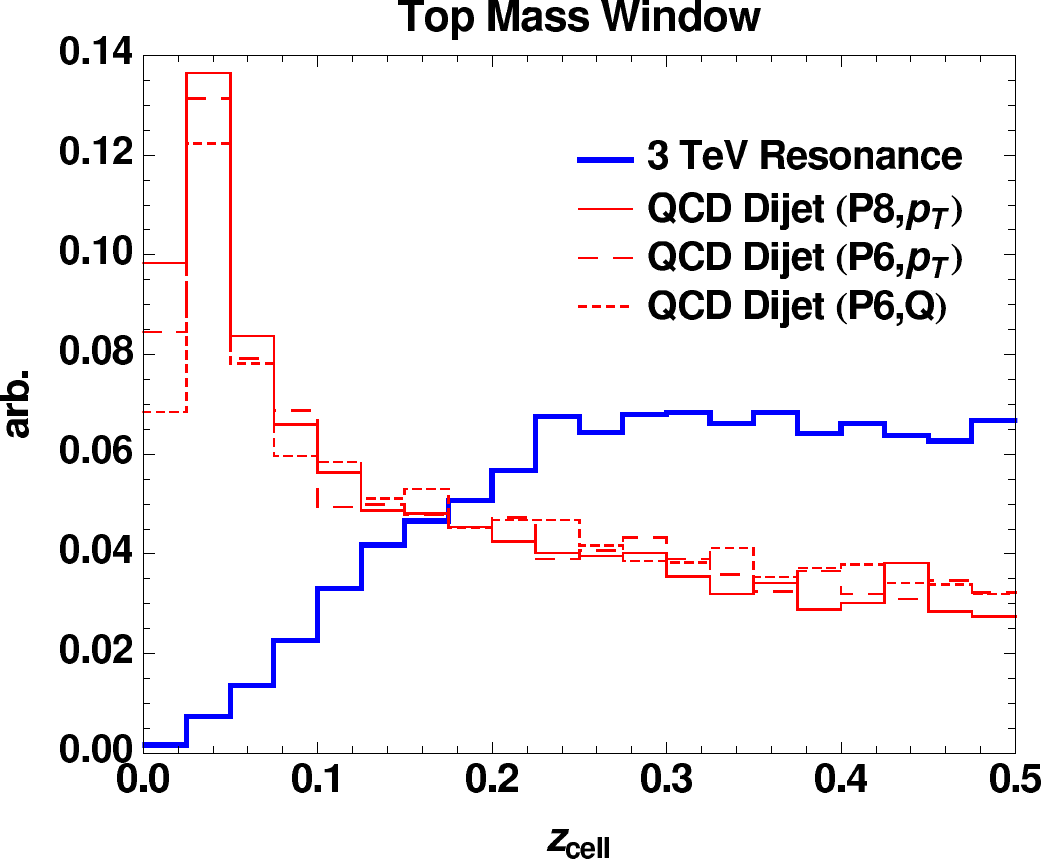}
  \includegraphics[width=0.32\textwidth]{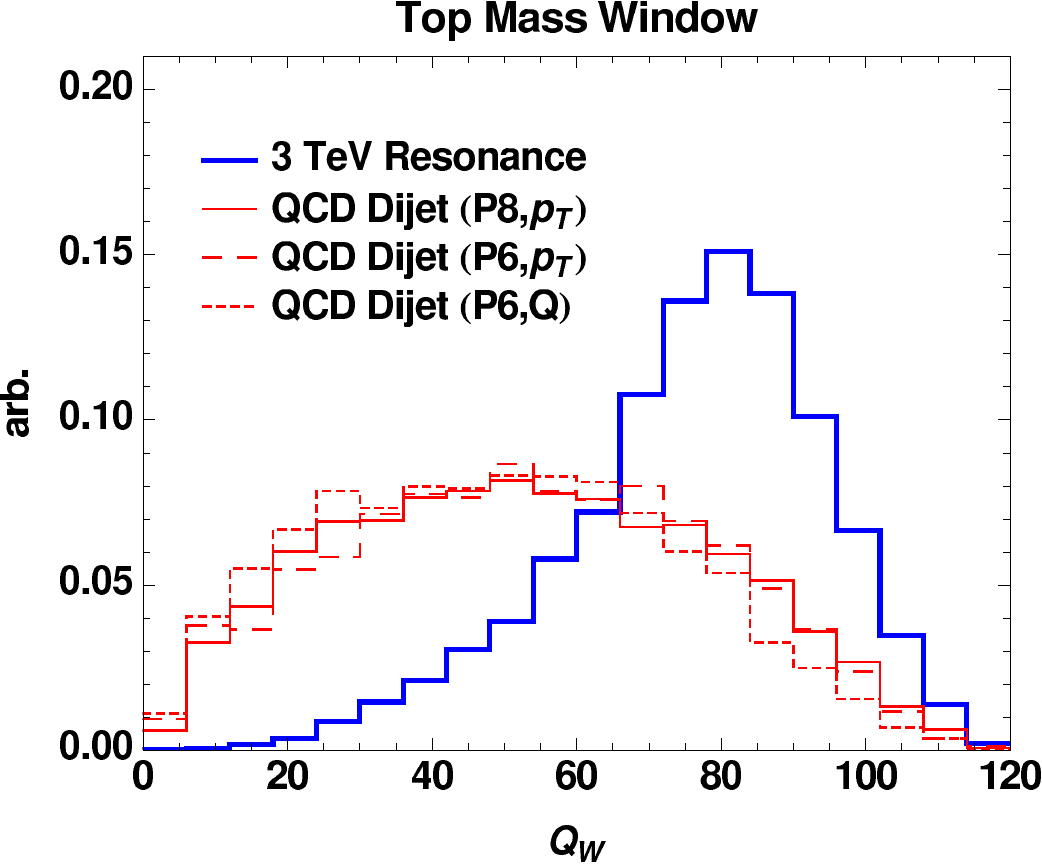}
  \includegraphics[width=0.32\textwidth]{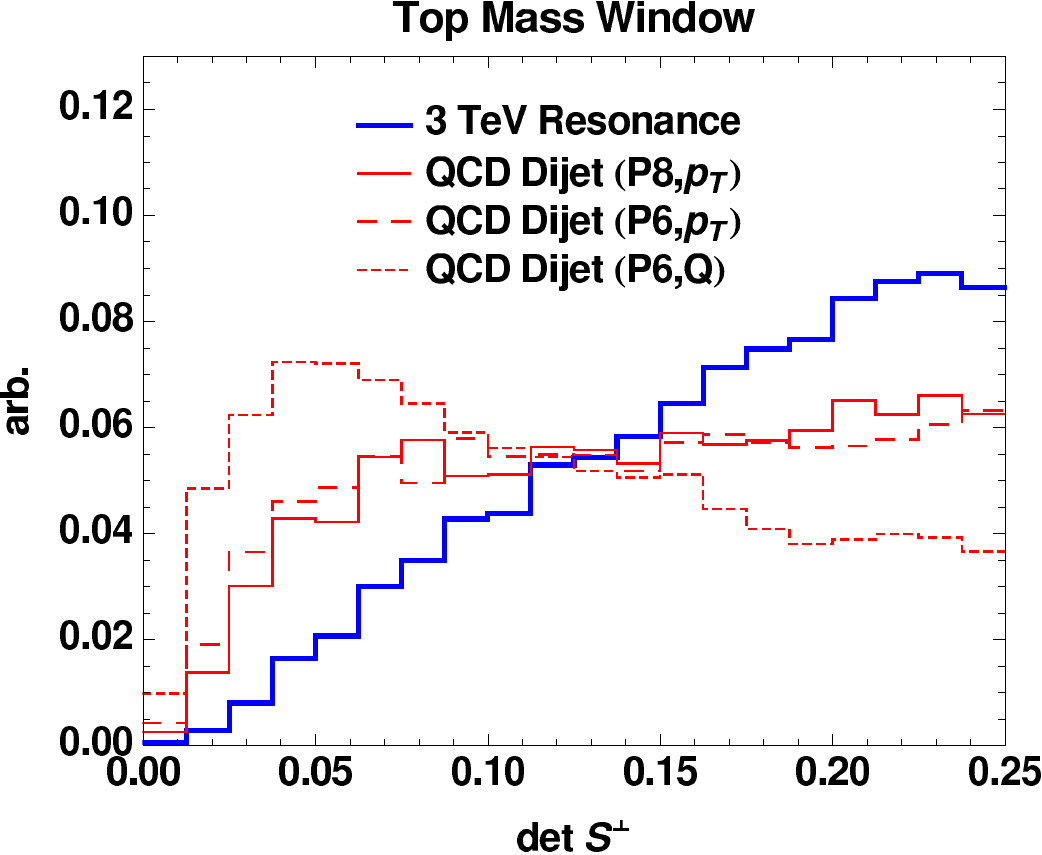}
\caption{Reconstructed kinematics variables for top jets and QCD jets
  in the top mass window of Eq.\eqref{eq:thaler_wang2} and after an
  additional cut of $p_T > 1200$~GeV on the fat jet: original energy
  ratio $z$ (left), $m_{jj}$ for the $W$ candidate subjets (center),
  and the determinant of the transverse sphericity (right).  Figures
  from Ref.~\cite{thaler_wang}.}
\label{fig:thaler_wang}
\end{figure}

The numerical values in the tagging criteria are optimized
for highly boosted top quarks with $p_{T,t} > 800$ at least, where we
require jet mass windows and large energy ratios for example using the
first definition in Eq.\eqref{eq:thaler_wang1}:
\begin{alignat}{5}
  m_{jjj} = 160...200~\gev
  \qqquad 
  m_{jj} = 60...100~\gev \; ,
  \qqquad 
  z > 0.1
\label{eq:thaler_wang2}
\end{alignat}
where the $W$ mass constraint has to be fulfilled by one subjet
combination and the $z$ value is extracted from the $t \to Wb$ decay
step. Both observables are shown in Fig.~\ref{fig:thaler_wang} and
show a clear difference for signal and backgrounds.

To this stage the Thaler-Wang tagger does not yet include a
distinctive feature of a three-body decay. Therefore, it is combined
with the classical sphericity event shape. The sphericity
tensor~\cite{sphericity} defined on the two-dimensional plane
transverse to the boost direction is defined in terms of calorimeter
objects
\begin{alignat}{5}
S^{\perp kl} 
= \dfrac{1}{\sum_{\alpha \in \text{jet}}|\vec{p}^\perp_\alpha|} \; 
  \sum_{\alpha \in \text{jet}}
          \dfrac{\vec{p}^{\perp k}_\alpha \vec{p}^{\perp l}_\alpha}
                {|\vec{p}^\perp_\alpha|} \; .
\label{eq:sphericity}
\end{alignat}
It is constructed out of the transverse momentum components
$\vec{p}_\alpha^\perp$ of all energy depositions $\alpha$,
perpendicular to the jet momentum. To avoid constructing the tensor
with explicit coordinates the actual observable is its determinant.
For two-body kinematics $\det S^{\perp kl}$ is zero. For three-body
decays it corresponds to two finite tensor eigenvalues summing to
unity and hence ranges within $0...0.25$.  In the right panel of
Fig.~\ref{fig:thaler_wang} we see that top jets show a clear bias
towards large $\det S^{\perp kl}$ values, but that the background
uncertainties are significant. In addition, event shapes will even
more than jet masses be affected by pile-up and the way we remove
it~\cite{boost2011}.

\subsection{N-Subjettiness}
\label{sec:scet}

$N$-Jettiness~\cite{scet0} is an event shape which describes the
number of isolated jets in an event. It can be adapted as
$N$-subjettiness to count subjets inside a fat
jet~\cite{Kim:2010uj,scet1,scet2}. Relative to $N$ subjet directions $\hat{n}_j$
it is defined as
\begin{alignat}{5}
\tau_N = 
\frac{1}{\sum_{\alpha \in \text{jet}} p_{T,\alpha} R_0^\beta} \;
\sum_{\alpha \in \text{jet}} p_{T,\alpha} \; 
\min_{k=1,...,N} (\Delta R_{k,\alpha})^\beta
\label{eq:nsubjet}
\end{alignat}
with an arbitrary weighting exponent $\beta >0$, to ensure infrared
safety. The normalization factor limits $\tau_N$ to the interval
$0...1$. In the first version of the tagging algorithm~\cite{scet1}
these $N$ axes are defined through a subjet algorithm. In a modified
version~\cite{scet2} they are defined in analogy to the thrust event
shape, namely as a minimization of the numerical value for $\tau_N$.

Fat jets with large values $\tau_N \to 1$ have many calorimeter
clusters far away from the $N$ main axes, which means they consist of
at least $N+1$ well separated subjets. In the ratio
$\tau_{N+1}/\tau_N$ typical QCD effects will drop out, and the ratio
will develop a dip for events which have $N+1$ subjets.\bigskip

Because of the largely unknown QCD effects the value of $\tau_3$, \ie
the quality of the three-subjet hypothesis is not the best
discriminator of top jets as compared to QCD jets. The $\tau_N$
distributions using the minimization criterion and $\beta=1$ tend to
peak in the $\tau_1 = 0.2...0.25$, $\tau_2 = 0.07...0.1$, or $\tau_3 =
0.04...0.05$ regimes, where the lower values are given for QCD jets
and the upper values are reached by top jets~\cite{scet2}. However,
the QCD-induced widths of the distributions are consistently larger
than the peak differences.

In Fig.~\ref{fig:scet} we instead show two ratios of $N$-subjettiness
values. The two constructions of the $N$ reference directions give very
similar results, with a little bias towards smaller ratios for the
explicit minimization condition.  For top decays producing three
separated subjets the ratio $\tau_3/\tau_2$ is expected to drop,
compared to the QCD case. Indeed, we see a significantly lower signal
peak than background peak in $\tau_3/\tau_2$, even though this is at
least as much due to an increase of the background peak as a decrease
of the signal peak compared to $\tau_2/\tau_1$.\bigskip

\begin{figure}[t]
  \includegraphics[width=0.40\textwidth]{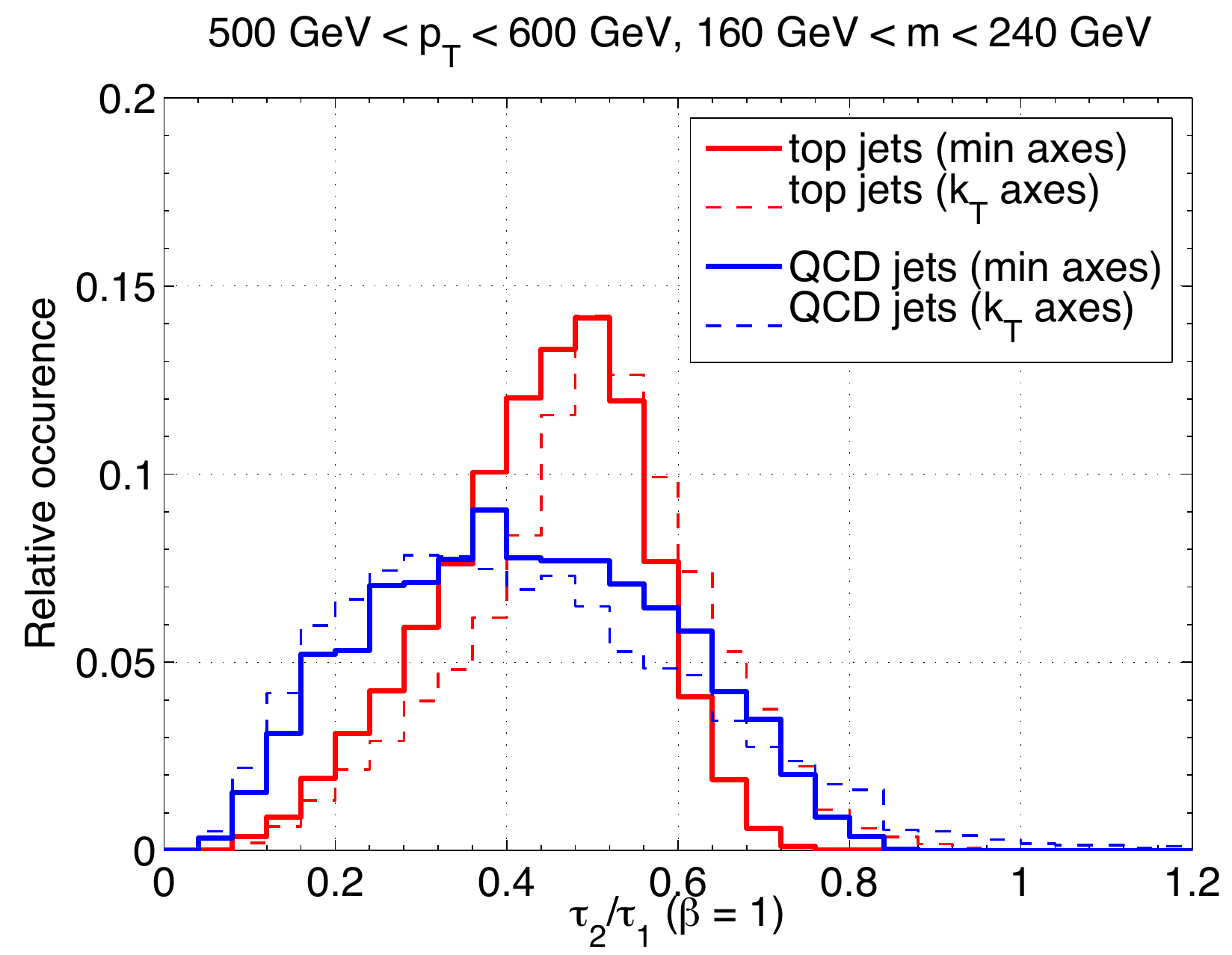}
  \hspace*{0.1\textwidth}
  \includegraphics[width=0.40\textwidth]{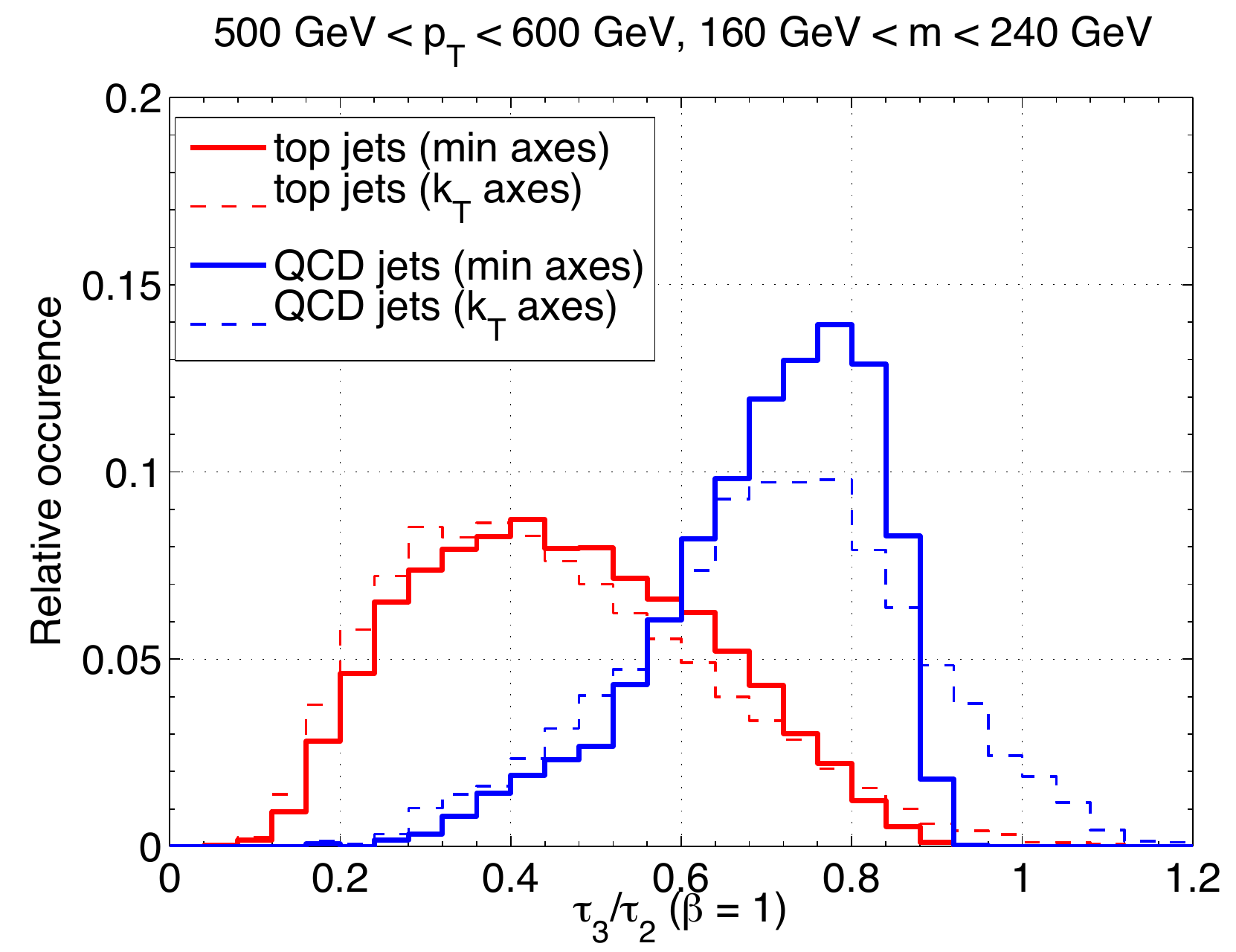}
\caption{$N$-subjettiness distributions for signal an background. Both
  methods for extracting the $N$ reference momenta are shown. Both
  panels use $\beta=1$ in the definition of $\tau$, which turns out to
  give the best tagging performance. Figures from Ref.~\cite{scet2}.}
\label{fig:scet}
\end{figure}

The associated top tagging algorithm is based on an anti-$k_T$ jet of
size $R=1.0$ and with $p_T > 200$~GeV. The choice of jet algorithm
reflects the fact that the clustering history will not be part of the
top selection criteria. Instead, it uses two basic jet shape
requirements on the top jet mass and on the ratio of subjettiness
values
\begin{alignat}{5}
m_\text{fat jet} = 160...240~\gev
\qqqquad 
\frac{\tau_3}{\tau_2} < 0.6 \; ,
\label{eq:scet}
\end{alignat}
Because the fat jet mass is not corrected for soft QCD and pile-up its
upper limit is larger than usual. The efficiencies obtained for
different methods of reconstructing the $N$ reference directions and
for $\beta = 1...2.5$ only slightly differ, likely within the
uncertainties induced by QCD and detector effects.

An obvious extension of the tagging criteria Eq.\eqref{eq:scet} would
be including all $\tau_N$ and $\tau_N/\tau_{N-1}$ measures for
$N=1,2,3$ and $\beta=1,2$. For fixed efficiencies this reduces the
mis-tag rate by roughly 20\%~\cite{scet2}.

\subsection{Alternative jet shapes}
\label{sec:alternatives}

After discussing a set of specialized top tagger which are currently
being tested by ATLAS and CMS we have to add a few more general
approaches. For example, the template method based on jet shapes or
the pure counting based tree-less approach are likely not going to be
the leading top tagging tools used at the LHC. However, their ideas might
well prove useful when the experimental task at the LHC goes beyond
identifying known Standard Model particles and features.\bigskip

The template method for top tagging~\cite{leandro3} relies on
anti-$k_T$-jets of size $R=0.5$ and a jet energy in the 1~TeV
range. In a similar ansatz~\cite{leandro1} this is replaced by a cut
on the transverse momentum of the leading jet of at least $p_{T,j} >
1$~TeV.  In addition, the fat jet mass has to lie in the
$160...190$~GeV range. Relevant additional observables are then
included as an overlap of measured correlations on the calorimeter
level and different parton-level templates, weighted by the geometric
energy deposition. 

Possible additional observables used in this top tagging study are jet
shapes.  Event shapes like thrust or the eigenvalues of the
sphericity tensor Eq.\eqref{eq:sphericity} can be used on the
content of geometrically large jets and their constituents. In that
framework they are often referred to as jet shapes. A jet shape which
is essential for all top tagging algorithms is the jet mass.  A major
theoretical issue is if jet shapes are infrared safe, which we will
skip in this discussion~\cite{jet_mass}. Obviously, this question
also includes the underlying jet algorithms.

An essentially equivalent alternative to the sphericity is
the planar flow. It is derived from the tensor $I_w$ and its two
eigenvalues $\lambda_{1,2}$~\cite{leandro0}
\begin{alignat}{5}
I_w^{kl} = \frac{1}{m_\text{jet}} 
          \sum_{\alpha \in \text{jet}} 
\frac{\vec{p}^{\perp k}_\alpha \vec{p}^{\perp l}_\alpha}{E_\alpha} 
\qqqquad
P = \frac{4 \det I_w}{(\text{tr} I_w)^2}
  = \frac{4 \lambda_1 \lambda_2}{(\lambda_1 + \lambda_2)^2} \; .
\label{eq:planar_flow}
\end{alignat}
For only two constituents $P$ again vanishes, as it does for any kind
of linear geometry. For a generic three body decay it can assume any
value between zero and one.  For example requiring $P > 0.5$ enhances
the number of top jets over the QCD background. In practice, the
template tagger uses a correlated cut in the template overlap vs
planar flow plane. Given that the overlap measure includes the full
kinematic event information it might be possible to further improve it
in the direction of the so-called matrix element method of
log-likelihood ratios.

Yet another class of jet shapes which we can use to describe two-body
as well as three-body configurations are
angularities~\cite{angularity,Chen:2011vq}. In the template method
they are only included for Higgs tagging, but they can also be used to
improve top tagging.  For different weights $a$ the angularity is
defined as
\begin{alignat}{5}
\tau_a = \frac{1}{m_\text{jet}}
        \sum_{\alpha \in \text{jet}}
E_\alpha \, 
\sin^a \frac{\pi \theta_i}{2 R} \, 
\left( 1 - \cos \frac{\pi \theta_i}{2R} \right)^{1-a} \; ,
\label{eq:angularity}
\end{alignat}
in terms of the angle $\theta_i$ with respect to the main axis. 
The correction factor $\pi/(2 R)$ includes the jet size $R$
and ensures that for the maximum value $\theta_i = R$ the argument of
the trigonometric functions does not exceed the hemisphere limit
$\pi/2$ from earlier $e^+ e^-$ applications. Infrared safety limits
the range of angularities to $ -\infty < a < 2$.  For $a=0$ we find
that $1 - \tau_0$ turns into thrust~\cite{thrust}, while for $a=1$
is becomes jet broadening~\cite{broadening}.  Because for each value
$a$ the angularity is a simple number we can correlate it with other
observables, like for example the azimuthal angle between the $W$
decay subjets and search for structures in such distributions.\bigskip

A second alternative approach to top tagging, explicitly not based on the
clustering history, is the tree-less substructure
analysis~\cite{treeless}. Unlike for example the $N$-subjettiness it
includes angular correlations. From the JADE distance measure
Eq.\eqref{eq:jade} we know that angular separation can be closely
linked to invariant masses of subjet combinations.

The geometric correlations between all possible pairs of subjets can
be analyzed in terms of the angular structure function and its
numerical derivative
\begin{alignat}{5}
\mathcal{G} (R) &= 
\frac{\sum_{j_1 \ne j_2} \; d_{j_1 j_2}^\text{(JADE)} \; \Theta(R-\Delta R_{j_1 j_2})}
     {\sum_{j_1 \ne j_2} \; d_{j_1 j_2}^\text{(JADE)}}
\qqquad
\Delta \mathcal{G} (R) &= R \; 
\dfrac{\sum_{j_1 \ne j_2} \; d_{j_1 j_2}^\text{(JADE)} \; K(R-\Delta R_{j_1 j_2})}
      {\sum_{j_1 \ne j_2} \; d_{j_1 j_2}^\text{(JADE)} \; \Theta(R-\Delta R_{j_1 j_2})} \; .
\label{eq:treeless1}
\end{alignat}
The function $K$ is nothing but a finite delta distribution, \eg $K(x)
= e^{-x^2/R_0^2}/\sqrt{\pi R_0^2}$ with $R_0 = 0.6$.  It fixes a
typical $R$ distance between two subjets. For values $R = R_*$
corresponding to observed subjet pairs inside the fat jet the function
$\mathcal{G} (R)$ makes a step and $\Delta \mathcal{G} (R)$ develops a
peak. Top decays with three hard decay subjets will show three such
peak values $R_{k*}$ with $k=1,2,3$, each corresponding to one side of
the triangle defined by three subjets. The number of observed peaks we
call $n_p$. For each of the peaks we define a mass value
\begin{alignat}{5}
m_*^2 = \frac{1}{R_*} \;
\sum_{j_1 \ne j_2} \; d_{j_1 j_2}^\text{(JADE)} \; 
\sqrt{\pi R_0^2} \; K(R_*-\Delta R_{j_1 j_2})
\label{eq:treeless2}
\end{alignat}
%
where the JADE distance is defined in Eq.\eqref{eq:jade}.  For massive
particle decays this mass variable $m_*$ scales with the invariant
mass of the parent subjet.

\begin{figure}[t]
  \includegraphics[width=0.60\textwidth]{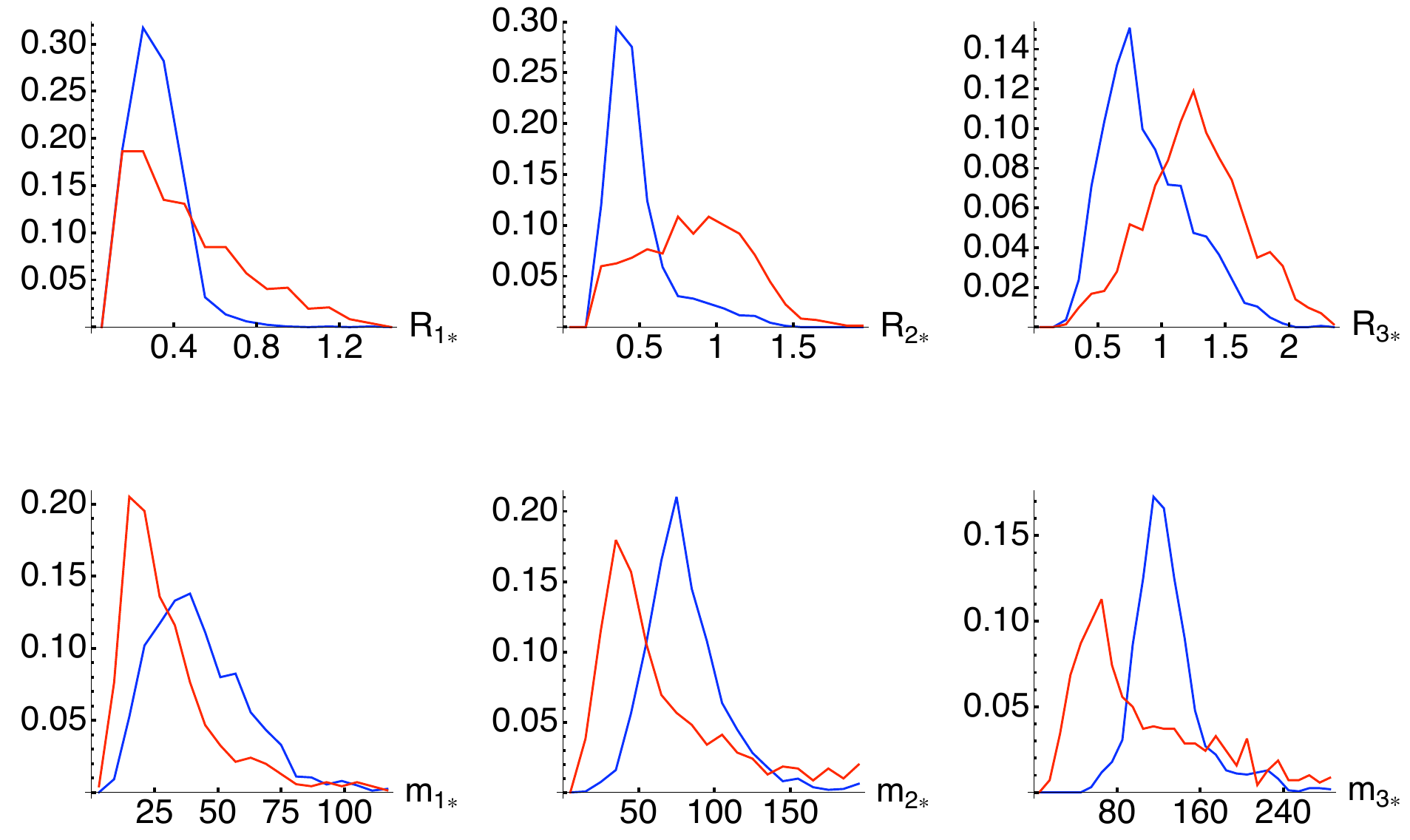}
  \caption{Peak positions and associated masses for $n_p = 3$ and fat
    jets with $p_T = 500...600$~GeV.  Shown are normalized
    distributions for the top signal (blue) and QCD backgrounds (red).
    Figure from Ref.~\cite{treeless}.}
  \label{fig:no_trees}
\end{figure}

In Fig.~\ref{fig:no_trees} we show the peak positions and their
associated mass values for three-subjet signal and background
configurations.  For QCD backgrounds the $R_*$ distributions are broad
and essentially scale invariant. The $m_*$ distributions points
towards small values, even though their typical values increase
typically by a factor two for increasing points. In contrast, for top
jets the $R_*$ distributions are peaked. Their mass scales correspond
to the given decay kinematics, as for example discussed in
Sec.~\ref{sec:heptop}.\bigskip

The associated tree-less top tagging algorithm starts with a fat C/A
jet of size $R=1.5$. From the peaks in the $\Delta \mathcal{G}$
spectrum we then extract one, two or three hard subjets. There exist
different sets of cuts, depending on the transverse momentum of the
fat jet and the number of peaks.  We quote the cuts applied to events
with three subjet structures and $p_T = 300...400$~GeV.  The original
uncorrected fat jet mass and two peak-associated mass values $m_*$
have to fulfill
\begin{alignat}{5}
m_\text{fat jet} > 102~\gev 
\qqquad 
m_{2*} > 26~\gev
\qqquad 
m_{2*} > 79~\gev \; ,
\end{alignat}
In addition, the angular correlations have to satisfy
\begin{alignat}{5}
R_{1*} < 0.81 
\qqquad
R_{2*} < 1.03 
\qqquad
R_{3*} < 2.11  \; .
\end{alignat}
\bigskip

While the two taggers discussed above might not give the best
efficiency for the usual signatures, they have the advantage of being
much more general than some of the established taggers. If jet shapes
should indeed turn out powerful QCD analysis tools at the LHC, these
approaches will allow us to efficiently utilize jet shapes in searches
for new physics.\bigskip

QCD observables which are not linked to traditional event shapes might
also help distinguishing massive electroweak splittings from QCD
backgrounds. The radiation of QCD jets possesses characteristic
features which we can use to discriminate a color octet gluon from a
decaying color singlet resonance. Angular ordering of soft gluon
radiation implies that most gluons are emitted in between color
connected partners~\cite{Marchesini:1983bm,qcd}. In the decay of a
color singlet, \eg $H \to b \bar{b}$, the two decay products are
always color connected. In leading color approximation this is not
true for a gluon which splits to $b \bar{b}$. Its gluon radiation is
therefore more likely to be outside the $b \bar{b}$ cone.

Two observables might exploit this feature in the top tagging
framework. The pull vector~\cite{Gallicchio:2010sw} can be defined for
each individual jet in an event
\begin{alignat}{5}
\vec{t} = \sum_{\alpha \in \text{jet}} \frac{p_{T,\alpha}}{p_{T, \text{jet}}} \;
           |\vec{r}_\alpha| \; \vec{r}_\alpha \; .
\end{alignat}
Here, $\vec{r}_\alpha$ is the constituent position relative to the jet
and $p_{T,\alpha}$ is the transverse momentum of this constituent. The
angle between the pull vectors of different jets can be used to decide
if two $b$-tagged jets come from a color singlet resonance or a color
octet gluon. Pull has been tested on $W$ bosons from top decays by D0
~\cite{Abazov:2011vh}. According to this measurement the fraction of
uncolored $W$ bosons is $0.56 \pm 0.42$ (stat+syst), indicating that
pull is a challenging observable already in the relatively clean
Tevatron environment.

As a second observable dipolarity~\cite{Hook:2011cq} can help
selecting the correct $W$ decay products in a boosted top
decay. Compared to the pull angle, its definition is modified such
that all radiation off the dipole is captured in one (sub-)jet. For a
jet splitting into two subjets $j_1$ and $j_2$ dipolarity is defined
on all calorimeter objects $\alpha$ as
\begin{alignat}{5}
\mathcal{D} = \frac{1}{R^2_{j_1 j_2}} \; 
              \sum_{\alpha \in \text{jet}} \; 
              \frac{p_{T,\alpha}}{p_{T,\text{jet}}} R^2_\alpha \; ,
\end{alignat}
where $R_\alpha$ is the distance between the $i$th constituent and the line
segment that runs from $j_1$ to $j_2$. Using the HEPTopTagger
framework is was shown that dipolarity might be able to reduce the mistag
rate significantly.

\section{QCD Effects}
\label{sec:qcd}

Hadronic final states of hard interactions resulting form proton-bunch
crossings at the LHC are subject to many sources of QCD radiation.
Final state radiation are soft and collinear jets radiated off the
produced particles, in our case the top quark. It can be described
well using the parton shower, and radiation off heavy states is
suppressed.  Initial state radiation are soft and collinear jets from
initial state radiation, arising because the incoming partons have to
bridge the gap in scale between the proton and the hard process. In
the collinear limit they are also well described by the parton shower,
in the harder regime they require matrix element
corrections~\cite{qcd}.

Underlying event is additional soft QCD activity arising from a given
proton-proton interaction and surrounding the hard event.  It is
caused by semi- or non-perturbative interactions between the proton
remnants. The soft continuous underlying event radiation can have a
large effect on the jet mass and critically depends on the size $R$ of
the fat jet~\cite{arXiv:0712.3014}
\begin{alignat}{5}
\left< \delta m_j^2 \right>  \simeq 
\Lambda_\text{UE} \; p_{T,j} \; 
\left( \frac{R^4 }{4} + \frac{R^8}{4608} + \mathcal{O}(R^{12}) \right) \; .
\label{eq:ue_scaling}
\end{alignat}
At the LHC, the amount of transverse momentum of the underlying event
radiation per unit rapidity, $\Lambda_\text{UE}$, is roughly
$\mathcal{O} (10)$ GeV~\cite{Cacciari:2009dp}.  

Finally, pile-up is the effect of multiple proton-proton collisions in
one beam crossing. Its effects are already observed now and are
expected to become even harder to deal with once the LHC runs at
design energy and design luminosity. Pile-up can add up to 100 GeV of
soft radiation per unit rapidity~\cite{Cacciari:2007fd}.\bigskip

As discussed in Sec.~\ref{sec:algos} the $k_T$ and C/A algorithms, for
a virtuality and an angular ordered shower, aim to reverse the shower
evolution. Approximately, they preserve the physical picture of the
jet evolution from the hard scale to the hadronization scale in the
recombination sequence. Initial state radiation, underlying event and
pile-up spoil this picture and add noise to the jet clustering.
Jet-mass-based algorithms using subjets as part of the
reverse-engineered cluster history are sensitive to a distortion by
uncorrelated soft radiation.

An additional complication in identifying events with hadronically
decaying electroweak resonances is that splittings of quarks and
gluons can geometrically induce a large jet mass,
\begin{alignat}{5}
\left< m_j^2 \right> \simeq C_i \, \alpha_s \, p_{T,j}^2 \, \Delta R_{j_1 j_2}^2 \; ,
\label{eq:light_geometry}
\end{alignat} 
where $C_i=3~(4/3)$ are the color factors for gluon (quark) induced
jets~\cite{ellisjets}.  For very hard jets this value can become of
the order of the electroweak scale. This makes initial state radiation
associated with heavy particle production dangerous, in particular in
events with generically large jet multiplicity. For the top tagger it
also means that while $p_{T,j}$ and $R$ are required to be large to
capture all decay products, they should not become too large.

To discriminate a hadronically decaying heavy resonance from a QCD
jet, \eg using its invariant mass, all final state radiation has to be
properly recombined.  This implies that we can separate it from
initial state radiation, underlying event and pile-up. While
underlying event and pile-up tend to be soft compared to the decay
products of a boosted resonance, initial state radiation is
not~\cite{skands}.  Its typical transverse momentum can be of the same
order as a $W$ decay jet, in particular for moderately boosted top
quarks. Therefore, different substructure approaches are needed to
cope with underlying event/pile-up and with initial state
radiation.\bigskip

Jet grooming methods, like filtering (Sec.~\ref{sec:filtering}),
trimming (\ref{sec:trimming}) and pruning (Sec.~\ref{sec:pruning}),
remove soft uncorrelated radiation from a fat jet while retaining
final state radiation off the resonance. For QCD jets grooming methods
reduce the upper end of the jet mass distribution, whereas for signal
events they yield a sharper peak near the true resonance mass
$m_j=m_\text{res}$. To keep these methods generic it is implicitly
assumed that for boosted heavy particles $p_{T,\text{FSR}} >
p_{T,(\text{ISR,UE,PU})}$. Thus, the transverse momentum of the
subjets is an important criterion to discriminate between final state
radiation and other radiation. Using soft-collinear effective theory
it has recently been shown that under certain conditions grooming
techniques factorize~\cite{walshzuberi}.

As a matter of fact, the problem of QCD effects inside geometrically
large jets was early on noticed by the authors of
Ref.~\cite{gerbush}. This is why their `top tagger' is based on narrow
$k_T$ jets for the top decay products which are then combined in the
spirit of the C/A-algorithm.

\subsection{Filtering}
\label{sec:filtering}

\begin{figure}[b]
  \includegraphics[width=0.22\textwidth]{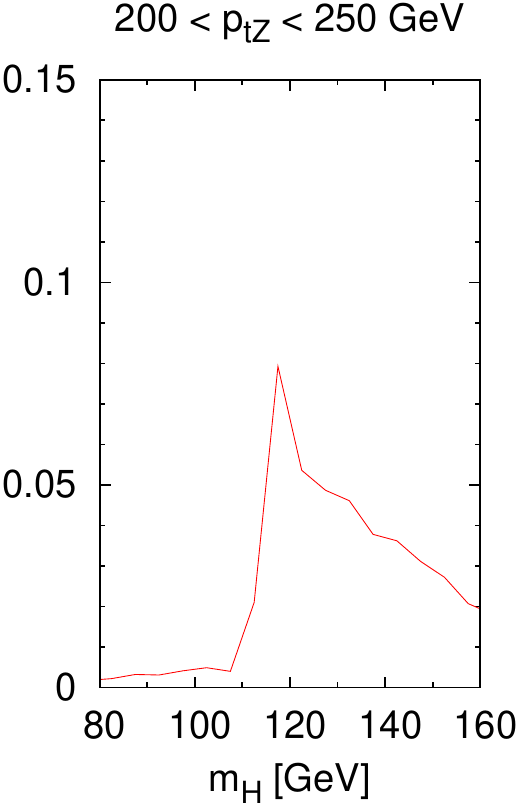}
  \includegraphics[width=0.22\textwidth]{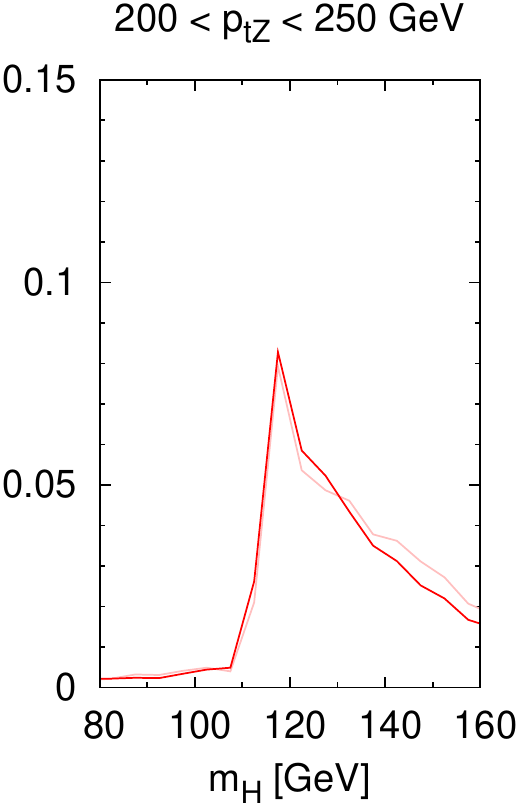}
  \includegraphics[width=0.22\textwidth]{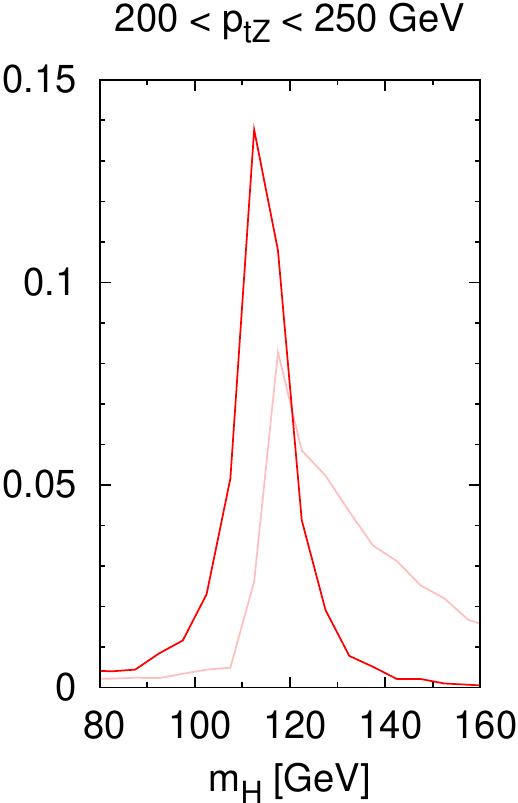}
  \includegraphics[width=0.22\textwidth]{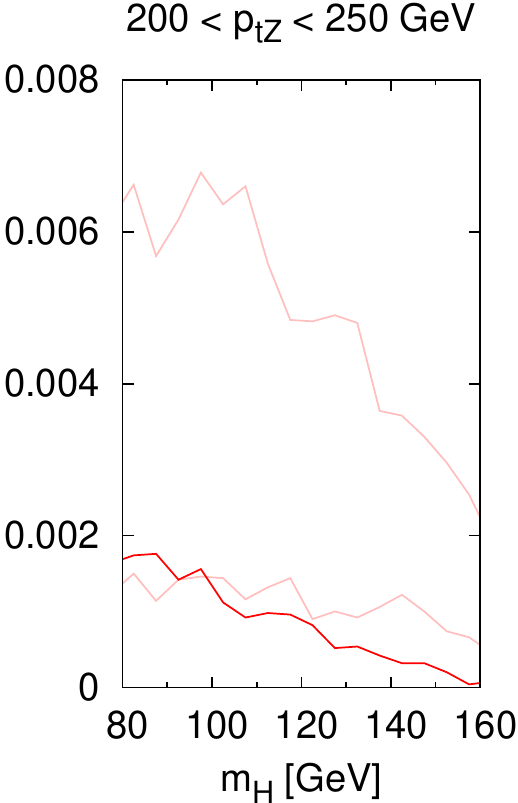}
  \caption{Mass of the hardest jet in the $pp \to H_{b\bar{b}}Z_\ell$
    final state before mass drop, after mass drop, and after filtering
    (from the left). The very right panel shows all three curves (top
    to bottom) for the $Zb\bar{b}$ continuum background.  In the BDRS
    analysis the C/A fat jet has size $R=1.2$ and the leptonic $Z$ is
    required to have $p_{T,Z} = 200...250$~GeV.  Figure from
    Ref.~\cite{gavinfiltering}.}
  \label{fig:higgsbdrs}
\end{figure}

Filtering, the first proposed jet grooming method~\cite{bdrs}, was
introduced as part of the BDRS Higgs tagger. Its target application is
$HW$ and $HZ$ production with a leptonic decay of the gauge bosons,
\ie events with relatively low jet multiplicity. For the fat Higgs
jets it applies the mass drop algorithm described in
Sec.~\ref{sec:early} to extract the relevant $b\bar{b}$ subjets and
their geometric structure. The size of the relevant two subjets is
large enough to contain most of the QCD radiation from the Higgs decay
\ie soft-collinear enhancement forces gluon radiation off the
$b$-quarks to be almost entirely emitted into the two subjets. 

To further improve the $m_{\bar b b}$ resolution the constituents of
the two $b$-tagged subjets can be recombined into smaller C/A-subjets
of size
\begin{alignat}{5}
R_\text{filter} = \min \left(0.3, \frac{\Delta R_{b\bar{b}}}{2} \right) \; .
\label{eq:filter}
\end{alignat}
This zooming-in obviously reduces the effective area of the fat jet
considered for mass reconstruction and this way tames any QCD effects
scaling with $R$ \eg as shown for the underlying event in
Eq.\eqref{eq:ue_scaling}.

For the Higgs boson the best mass resolution is achieved by
reconstructing the Higgs mass from the $n_\text{filter}=3$ hardest
filtered subjets. This means we include two $b$-jets and the hardest
wide-angle gluon radiation. Two free parameters, $R_\text{filter}$ and
$n_\text{filter}$ control the filtering performance.\bigskip

The effect of the different steps of the BDRS algorithm, including
filtering, we show in Fig.~\ref{fig:higgsbdrs}. The original object is
a C/A fat jet of size $R=1.2$, selected as the hardest jet in the
$H_{\bar{b}b}Z_\ell$ signal and background events. To ensure that the
Higgs decay products are contained inside the fat jet without
including many events where the $b$ quarks cannot be resolved, the
analysis requires $p_{T,Z} = 200...250$~GeV. We see that for the
signal the mass drop has no big effect on the fat jet
mass. Essentially all jets pass the mass drop criterion and only
little soft radiation is removed. The additional filtering clearly
sharpens the peak of the reconstructed Higgs mass and moved it towards
the input Higgs mass $m_H=115$~GeV. 

For the QCD background most events do not pass the mass drop
condition, which can be seen in the normalization of the upper two
curves in the right panel of Fig.~\ref{fig:higgsbdrs}. The
corresponding jet mass distribution after the mass drop becomes less
steep, indicating that following Eq.\eqref{eq:light_geometry} more
energetic jets are more likely to pass the mass drop criterion. The
additional filtering only has a mild effect on the continuum
reconstructed mass. This feature is very helpful because it means that
filtering does not sculpt the backgrounds, in spite of the carefully
chosen parameters in Eq.\eqref{eq:filter}. It allows us to analyze
hadronic Higgs decays including side bands to determine the
backgrounds.\bigskip

Applying filtering inside a top tagger is a straightforward
generalization. Starting with the HEPTopTagger algorithm described in
Sec~\ref{sec:heptop} we include a filtering stage and reconstruct all
jet masses from filtered subjets. The effect is the same as shown for
the Higgs tagger in Fig.~\ref{fig:higgsbdrs}: both $W$ and top mass
peaks are sharpened and moved towards the input values. Unlike for the
Higgs tagger, all mass windows are part of the tagging algorithm, so
there are no side bands in the usual sense. Still, choosing different
top masses for the MC simulation and inside the tagger we can check
that the tagger does not sculpt the backgrounds~\cite{heptop1}.

As mentioned above, the two filtering parameters $R_\text{filter}$ and
$n_\text{filter}$ have to be tuned including realistic simulations of
pile-up and detector effects. We find that using the $R_\text{filter}$
value from Eq.\eqref{eq:filter} and $n_\text{filter} = 5$ gives the
best performance for the more intensively radiating hadronic top
quark~\cite{gregor_private}.

\subsection{Trimming}
\label{sec:trimming}

Trimming~\cite{reftrimming} targets very similar effects as
filtering. In the first step we reconstruct a fat jet which will be
heavily impacted by QCD radiation. Its subjets we re-combine with a
higher resolution $R_\text{trim}$, defining a larger number of smaller
subjets. These subjets can be separated into two categories: hard and
soft. This discrimination is based on the transverse momentum, so hard
subjets obey
\begin{alignat}{5}
p_{T,j} > f_\text{trim} \; \Lambda_\text{trim} \; ,
\end{alignat}
where $f_\text{trim}$ is an adjustable parameter and
$\Lambda_\text{trim}$ is an intrinsic scale of the fat jet. It can for
example be chosen as its jet mass or its transverse momentum.  While
we discard all soft subjets the re-combined hard subjets define a
trimmed (fat) jet. Just like filtering this reduces the effective size of
the fat jet entering any kind of jet mass measurement.\bigskip

Because $\Lambda_\text{trim}$ can be different for each fat jet the trimming
procedure is self-adaptive: for a fat jet with large transverse
momentum and/or mass the subjets need to have a larger transverse
momentum to stay inside the trimmed jet. Just as the filtering
procedure, trimming requires two input parameters. However, because it
is self-adaptive the results are less sensitive to the origin of the
fat jet and trimming can be used as a generic tagging tool in a
multi-jet environment.

\subsection{Pruning}
\label{sec:pruning}

Unlike filtering or trimming, pruning removes underlying event and
pile-up while building the jet, \ie as part of the jet algorithm. In a
first step it defines a fat jet which can be based on any of the
recombination algorithms described in Sec.\ref{sec:algos}. In a second
step its constituents are pruned by checking in every recombination
step
\begin{alignat}{5}
\frac{ \min p_{T,j_i}} {p_{T,j}} < z_\text{prune} 
\qqquad 
\text{and}
\qqquad
\Delta R_{j_1 j_2} > R_\text{prune} \ .
\label{eq:prunecut}
\end{alignat}
If both conditions are met, the merging $j_1,j_2 \to j$ is
vetoed. Just as filtering and trimming, pruning depends on two
parameters: $z_\text{prune}$ and $R_\text{prune}$.  A global value for
$z_\text{prune}$ ensures that recombined well separated subjets are
not very asymmetric in $p_T$.  In contrast, $R_\text{prune}$ can be
determined on a jet-by-jet basis. Thus, pruning is, similar to
trimming, a self-adaptive procedure, applicable to a multi-jet final
state in an unbiased resonance search. \bigskip

\begin{figure}[t]
  \includegraphics[width=0.30\textwidth]{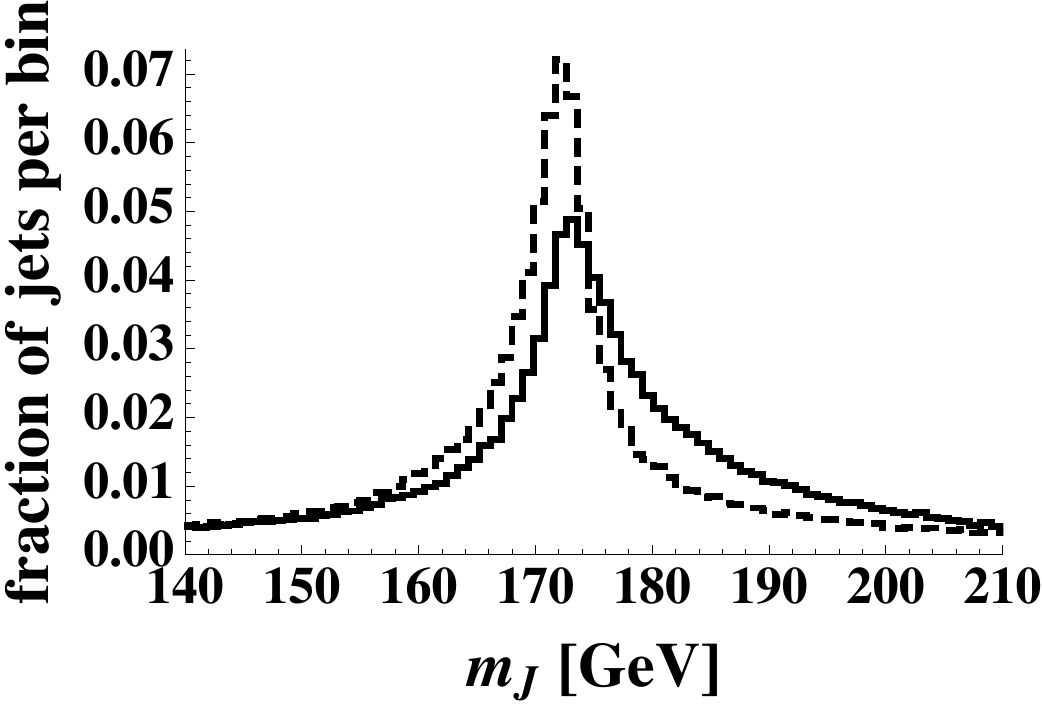}
  \hspace{0.1\textwidth}
  \includegraphics[width=0.30\textwidth]{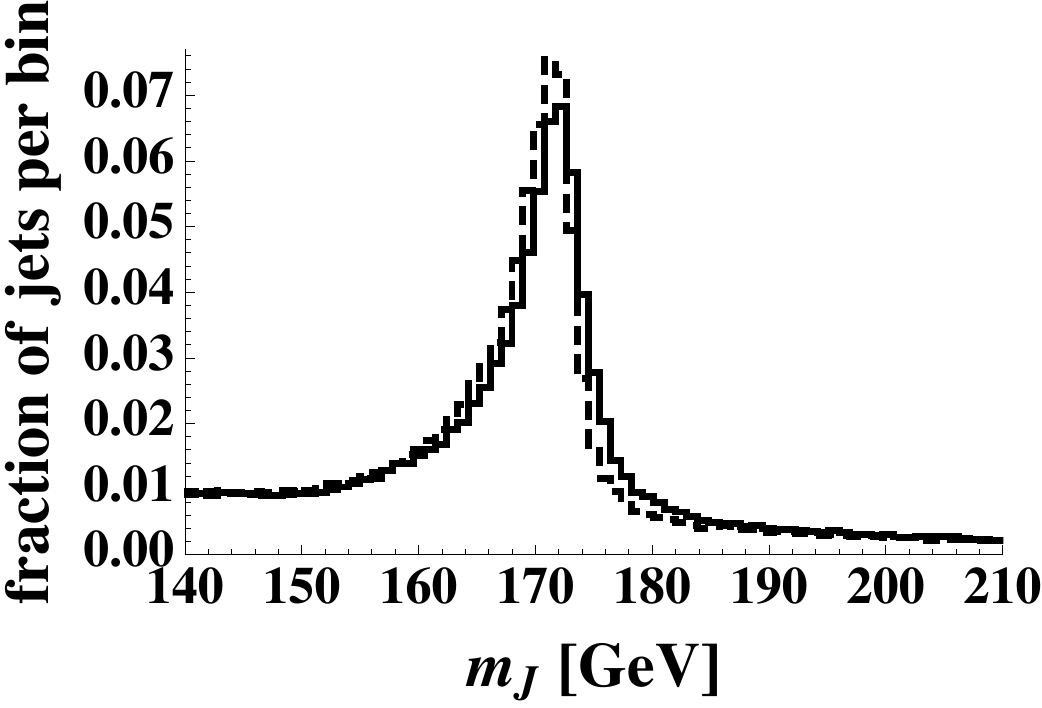} \\
  \includegraphics[width=0.30\textwidth]{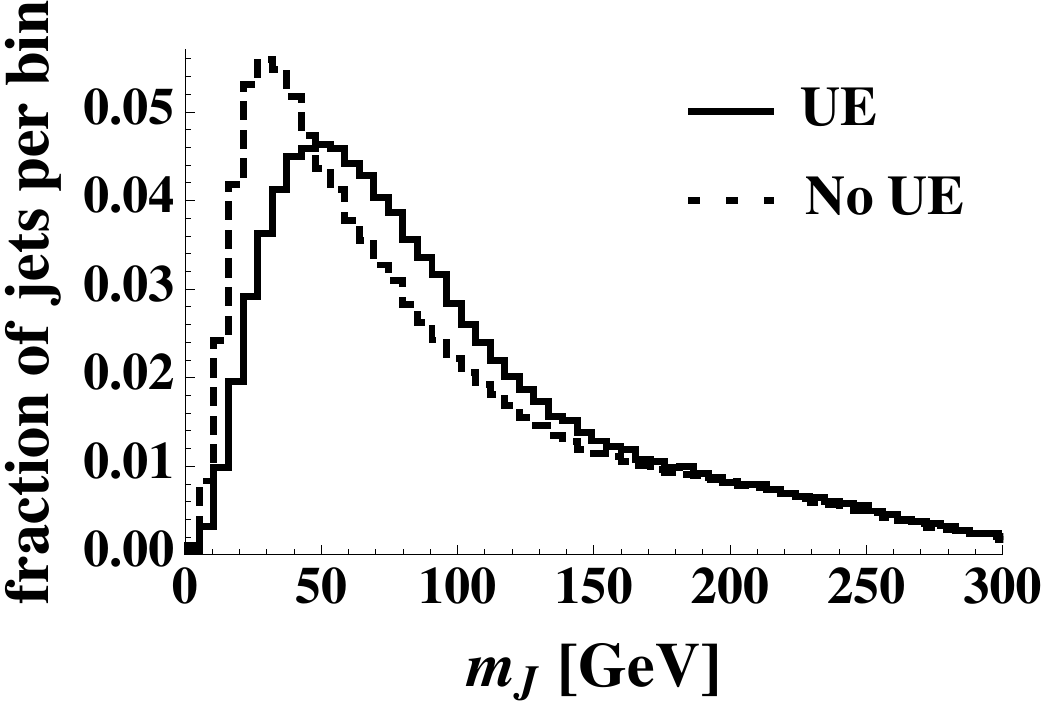} 
  \hspace{0.1\textwidth}
  \includegraphics[width=0.30\textwidth]{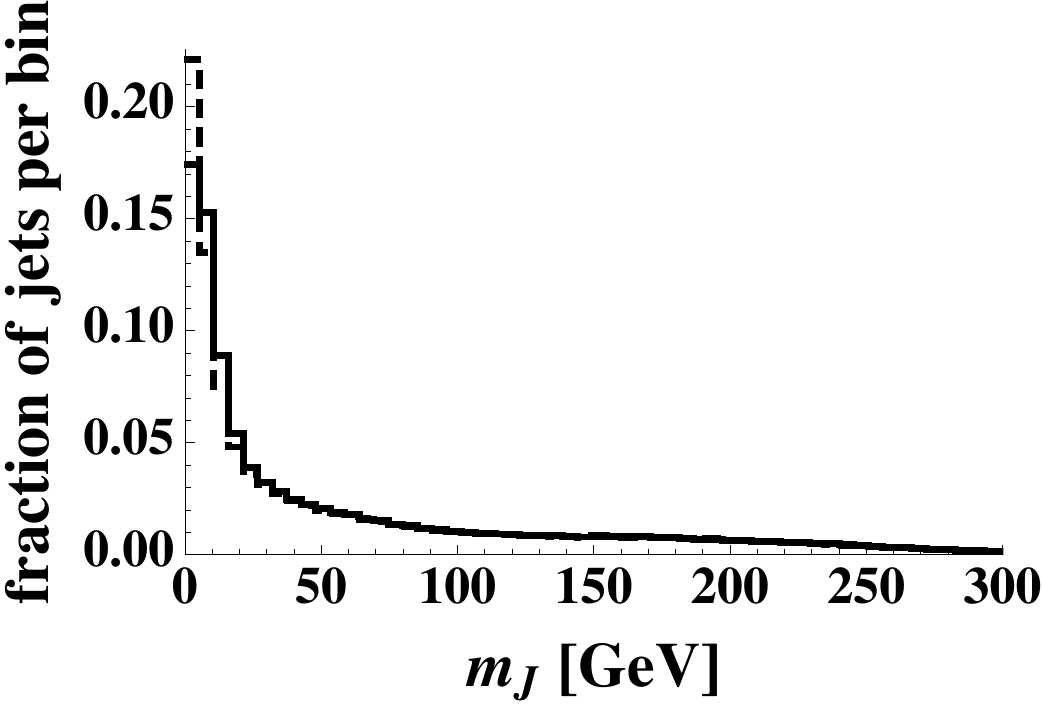} 
  \caption{Jet mass with and without underlying event before (left)
    and after pruning (right). Signal top jets are shown in the top
    row, background QCD jets on the bottom. Figures from
    Ref.~\cite{pruning2}.}
  \label{fig:pruning}
\end{figure}

Consequently, for the C/A-algorithm all subjets are merged unless
Eq.\eqref{eq:prunecut} holds and the minimal distance between the
subjets is $R_\text{prune}$. Once the $\Delta R$ condition is true
it automatically holds true for higher-level subsequent mergings, so
only the $z_\text{prune}$ condition needs to be considered. Thus, in
trimming the transverse momentum of the subjets is a parameter which
changes on a jet-to-jet basis, whereas in pruning the size of the
subjets changes from jet to jet. \bigskip 

Unlike filtering and trimming, pruning can be considered a full
self-adaptive tagging algorithm, as discussed in
Sec.~\ref{sec:hopkins}~\cite{pruning1,pruning2,Vermilion:2011nm}.
Fig.~\ref{fig:pruning} shows how pruning extracts a sharp peak in the
jet mass spectrum by removing soft radiation in the recombination
procedure. For QCD background events it moves the typical jet masses
from values in the $30...50$~GeV range to the expected values of $m_j
\lesssim 10$~GeV, even in the presence of underlying event.

\subsection{Filtripruning}
\label{sec:filtripru}

In Secs.~\ref{sec:filtering},~\ref{sec:trimming} and~\ref{sec:pruning}
we describe three grooming techniques, focusing on their similarities
and differences. At this stage they mainly act as tools to remove
unwanted QCD activity from fat jets, which means they are
indispensable tools for any substructure analysis. Filtering and
trimming target a very similar problem with an equally similar
approach. Both are mainly targeted at properly defining the objects
for the tagging algorithms.  Pruning works very differently in the
sense that it can be used to actively contribute to the top tagging
criteria; comparing Secs.~\ref{sec:seattle} and~\ref{sec:pruning}
indeed shows that the pruning and top tagging algorithms are not
easily separated.  The general question therefore becomes at what
level we can utilize observables inside the grooming procedure to
improve top tagging.\bigskip

\begin{figure}[t]
  \includegraphics[width=0.40\textwidth]{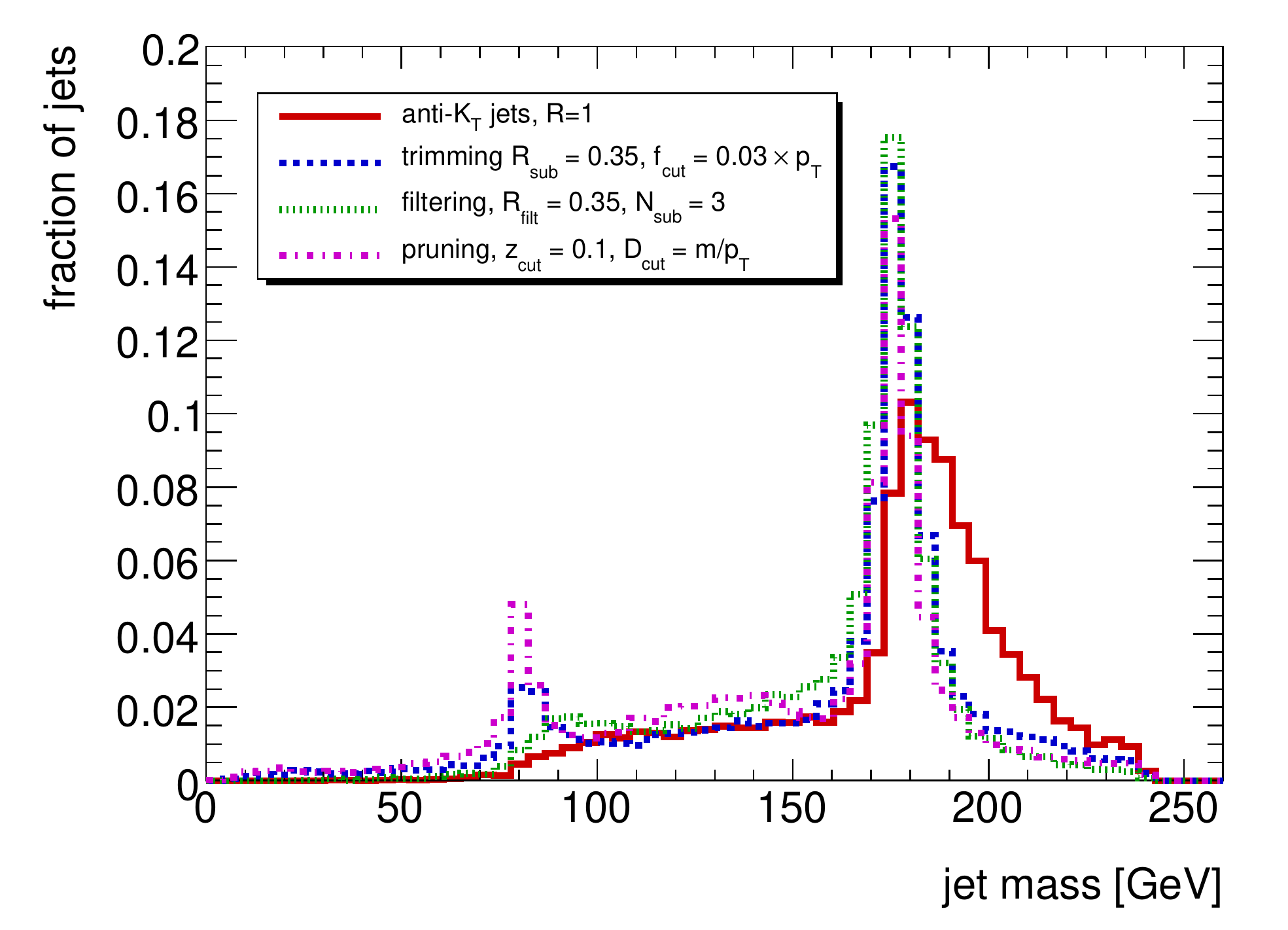}
  \hspace{0.1\textwidth}
  \includegraphics[width=0.40\textwidth]{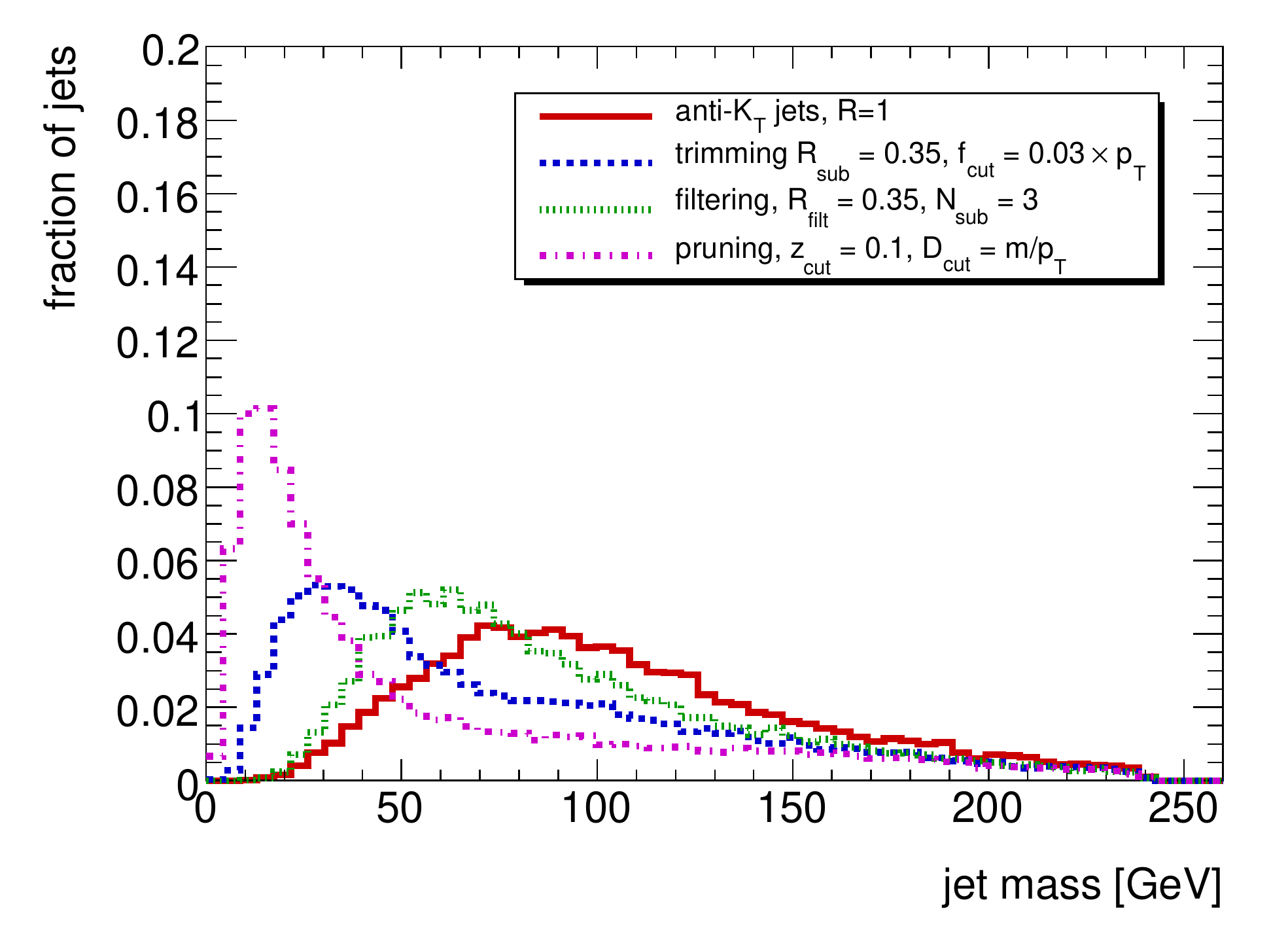}
  \caption{Jet mass for $t\bar{t}$ events (left) and QCD backgrounds
    (right) for all three jet grooming techniques. The solid curve
    (red) represents the uncorrected anti-$k_T$ jet mass for $R=1.0$
    and $500 \leq p_{T,j} \leq 600$~GeV. Figures from Ref.~\cite{boost2010}.}
  \label{fig:allgroom}
\end{figure}

In Fig.~\ref{fig:allgroom} we show the performance of pruning,
trimming and filtering on dijet and $t\bar{t}$ events for one
transverse momentum slice of the fat jet, $p_{T,j} =
500...600$~GeV~\cite{boost2010}. It is based on reasonable,
but not optimized parameters for trimming ($R_\text{sub}=0.35,
f_\text{prune}=0.03$, $\Lambda_\text{trim} = p_{T,j}$), pruning
($z_\text{prune}=0.1, R_\text{prune}=m_j/p_{T,j}$), and filtering
($R_\text{filter}=0.35, n_\text{filter}=3$). We see that the top mass
is well reconstructed by all three grooming methods. The second peak
appears because in some events the fat jet only captures the $W$ decay
subjets or the bottom quark is too soft to obey $p_{T,b} >
p_{T,\text{(UE,PU)}}$. In those cases pruned and trimmed subjets
nicely reconstruct the $W$ mass.  Filtered subjets using
$n_\text{filter}=3$ do not reconstruct the $W$ boson because they are
not adapted automatically.

For the QCD background all algorithms reduce the QCD-jet mass by
removing soft radiation as compared to the uncorrected values up to
$m_j = 150$~GeV. Again, this figure illustrates how crucial jet
grooming is before we can in any way rely on a measured jet mass.  In
contrast to the signal reconstruction the three approaches perform
differently. Pruning with its most sophisticated modelling of the QCD
splittings best closest to the expected QCD range $m_j \lesssim
10$~GeV. For the chosen set of parameters trimming and in particular
filtering perform much worse. While pruning internally vetoes mergings
which are wide angle and asymmetric in $p_T$ filtering always
recombines three subjets irrespective of their $p_T$ and relative
position inside the jet. QCD radiation is enhanced in the soft and
collinear limits, but the jet mass is most sensitive to wide angle
radiation. In the detector a QCD jet will on average only show few
isolated spots of energy accompanied by more continuous radiation from
pile-up and underlying event. Thus, with the fix choice of 3 subjets of
size R=0.35 filtering is revoked of most of its ability to reject the
dijet background. \bigskip

\begin{figure}[b]
  \includegraphics[width=0.35\textwidth]{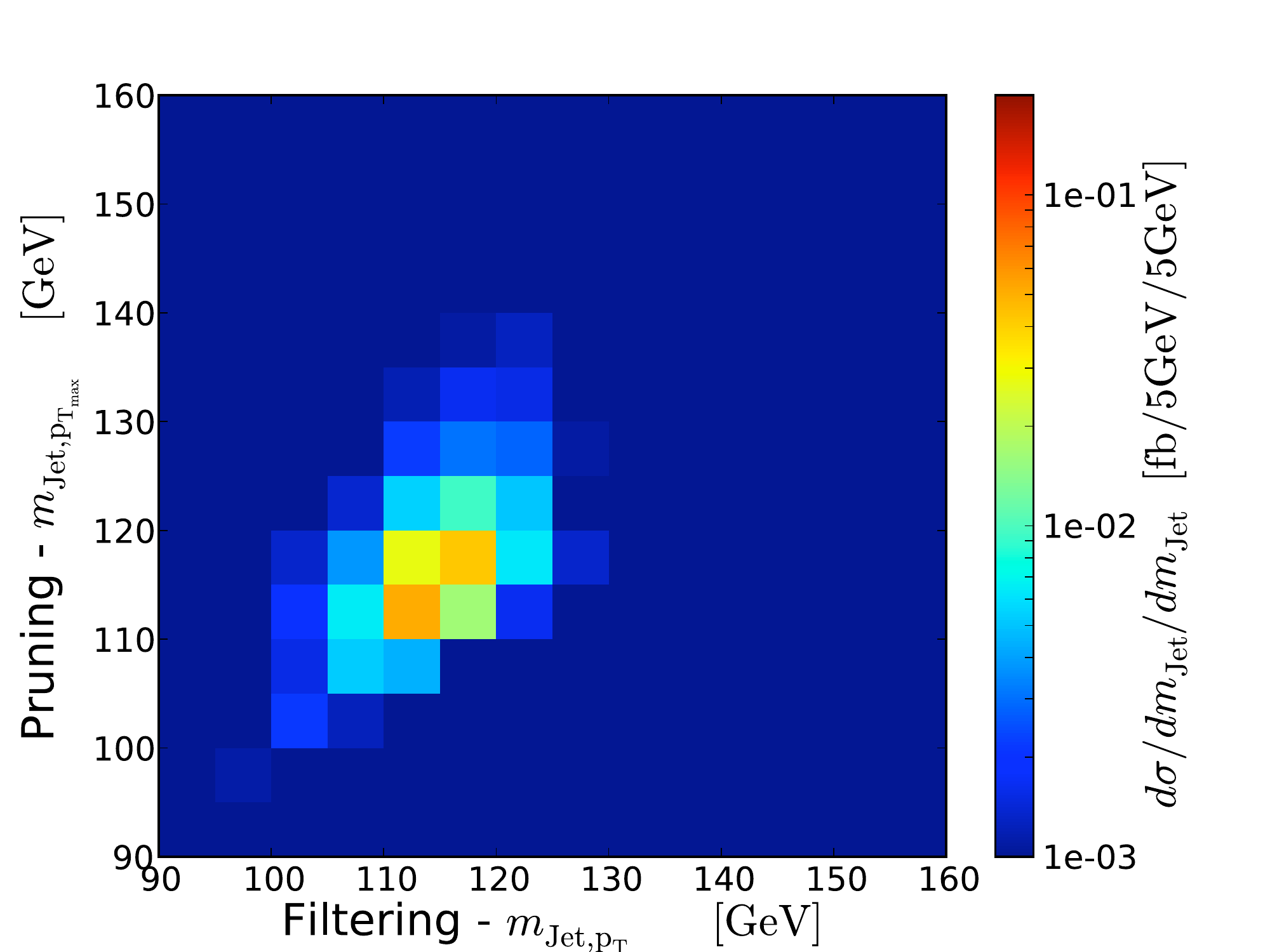}
  \hspace*{0.1\textwidth}
  \includegraphics[width=0.35\textwidth]{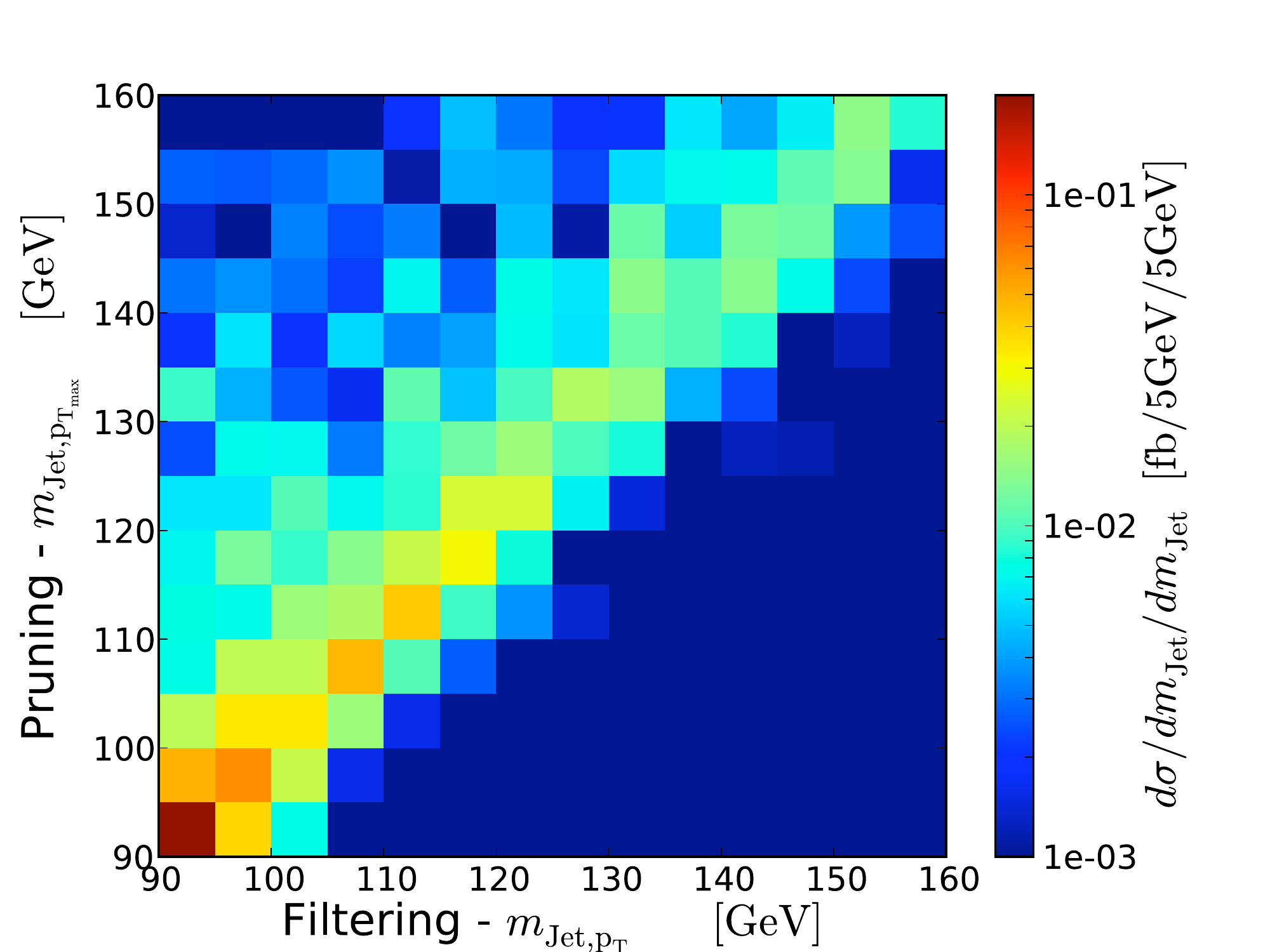} 
  \caption{Groomed jet masses in $HZ$ (left) and $Z$+jets (right)
    events. For example, pruning is compared to filtering.  All three
    algorithms reconstruct the Higgs mass equally well, while filtered
    or trimmed QCD-jet masses are smaller than pruned jet
    masses. Figures from Ref.~\cite{spanno_soper}.}
  \label{fig:pruntrim}
\end{figure}

In Ref.~\cite{spanno_soper} it was first shown that because grooming
techniques treat the QCD effects in particular in background events
differently they can in combination improve the tagging algorithm.
Following the BDRS approach, see Sec.~\ref{sec:early}, a C/A fat jet
of size $R=1.2$ is selected with $p_{T,j} \geq 200$~GeV and $|y| <
2.5$. After passing the mass drop requirement of Eq.\eqref{eq:bdrs}
the jet is either filtered, trimmed, or pruned. All grooming methods
reconstruct identical jet masses for the signal events, as we see in
the left panel of Fig.~\ref{fig:pruntrim}. However, while trimming and
filtering perform similarly on the $Z$+jet background, pruning tends
to reconstruct larger jet masses than filtering or trimming, as shown
in the right panel of Fig.~\ref{fig:pruntrim}. Thus, counting the
number of events in the two-dimensional jet mass region around the
input Higgs mass $(115 \pm 5)$~GeV $S/B$ can be improved by up to a
factor of two~\cite{spanno_soper}. The same approach can be used in to
reconstruct $Z$ bosons in $H \to ZZ$
decays~\cite{Hackstein:2010wk,Englert:2010ud}. In a similar spirit
many different jet substructure observables can be combined to tag
boosted $W$ bosons~\cite{Cui:2010km}.\bigskip

Applying different grooming techniques also significantly improves top
tagging, for example using filtered as well as pruned jet masses in
the HEPTopTagger framework~\cite{heptop2}.  After the mass drop
identification all possible three-subjet combinations with filtered
jet masses of $m_{jjj} = 160...200$~GeV are kept. In this step
filtering can simply be replaced by pruning, such that eventually the
filtered and pruned mass values can be compared.  Without any
optimization of the grooming parameters an improvement of $S/B$ by a
factor two seems realistic.  A generalization to other taggers is
straightforward.

\section{Performance}
\label{sec:performance}

Above all, top taggers are tools to reconstruct top quarks as
precisely and unambiguously as possible. They are designed to support
searches for new physics in otherwise inaccessible final states and
phase space regions. Because of the many sources of radiation in a
hadronic final states we cannot infer their performance from data
driven methods only. Instead, to evaluate the reliability of taggers
in a hadronic environment we need to also estimate their tagging
efficiencies, background fake rates and top-momentum reconstruction
using Monte Carlo programs. However, it is difficult to reproduce the
full complexity of high-$p_T$ events at the LHC using Monte Carlo
programs. Therefore, after estimating tagging efficiencies using event
generators we have to validated the predictions on data.

Although the concepts of the taggers are quite different, all of them
make use of information contained in the structure of the fat
jet. Therefore, all taggers are subject to the question to what extent
we can make use of this information at the LHC. Thus, already by
studying some of the taggers in detail the experimental collaborations
can obtain insights valuable for all taggers and jet substructure
methods.

Consequently, the interplay between the evaluation of top taggers on
Monte Carlo samples (Sec.~\ref{sec:comparisons}), the application of
taggers on data (Sec.~\ref{sec:data}) and the identification of
interesting new physics scenarios (Sec.~\ref{sec:applications}) is of
great importance for the optimal use of top taggers at the LHC.

\subsection{Comparisons}
\label{sec:comparisons}

Most of the top taggers described in Sec.~\ref{sec:algos} are publicly
available and included in collections of jet tools~\cite{spartyjet,
  fastjet}. Due to the different approaches in the reconstruction of
the top quark it can be instructive to study how they perform on
simulated event samples. For the BOOST~2010 proceedings
Herwig~\cite{herwigfortran} and Pythia~\cite{pythia} signal and
background samples have been prepared for such an
analysis~\cite{websamples}. All samples are divided in equally-sized
sub-samples with parton $p_T$ ranges from $200-300$ GeV, $300-400$
GeV, ... , $1.5-1.6$ TeV, thus covering the full range from topologies
with moderate boost to extremely energetic events. For each $p_T$ bin
10.000 events were generated. Combining all samples yields an
approximately flat $p_T$ distribution. These samples were used to
compare a subset of the taggers described in Sec.~\ref{sec:algos}
in~\cite{boost2010}. For each event, jets were clustered with the
anti-$k_T$ algorithm with an R-parameter of $1.0$. In this study, for
each top-tagging algorithm, the input parameters were optimized for
each efficiency by minimizing the mistag rate. Because the top quark
decays in a three body decay and the finite jet size can result in
loosing one of the decay products different definitions of tagging
efficiency come to mind, \eg the correct reconstruction of the top
quark mass or the reconstruction of the top quark momentum. In
Ref.~\cite{boost2010} the tagging efficiency and mistag rate was
defined as the number of top-tags divided by the total number of
anti-$k_T$ jets in the background and signal sample, respectively.

All tagging efficiencies are flat, particularly for the Thaler/Wang
and ATLAS top tagger. Both have for small efficiency working points a
large fake rate, but have a comparably small fake rate at working
points with large tagging efficiencies. The CMS and Hopkins top tagger
are very similar in design. Unsurprisingly, they perform very
similarly over the whole efficiency range.

\begin{figure}[t]
  \includegraphics[width=0.45\textwidth]{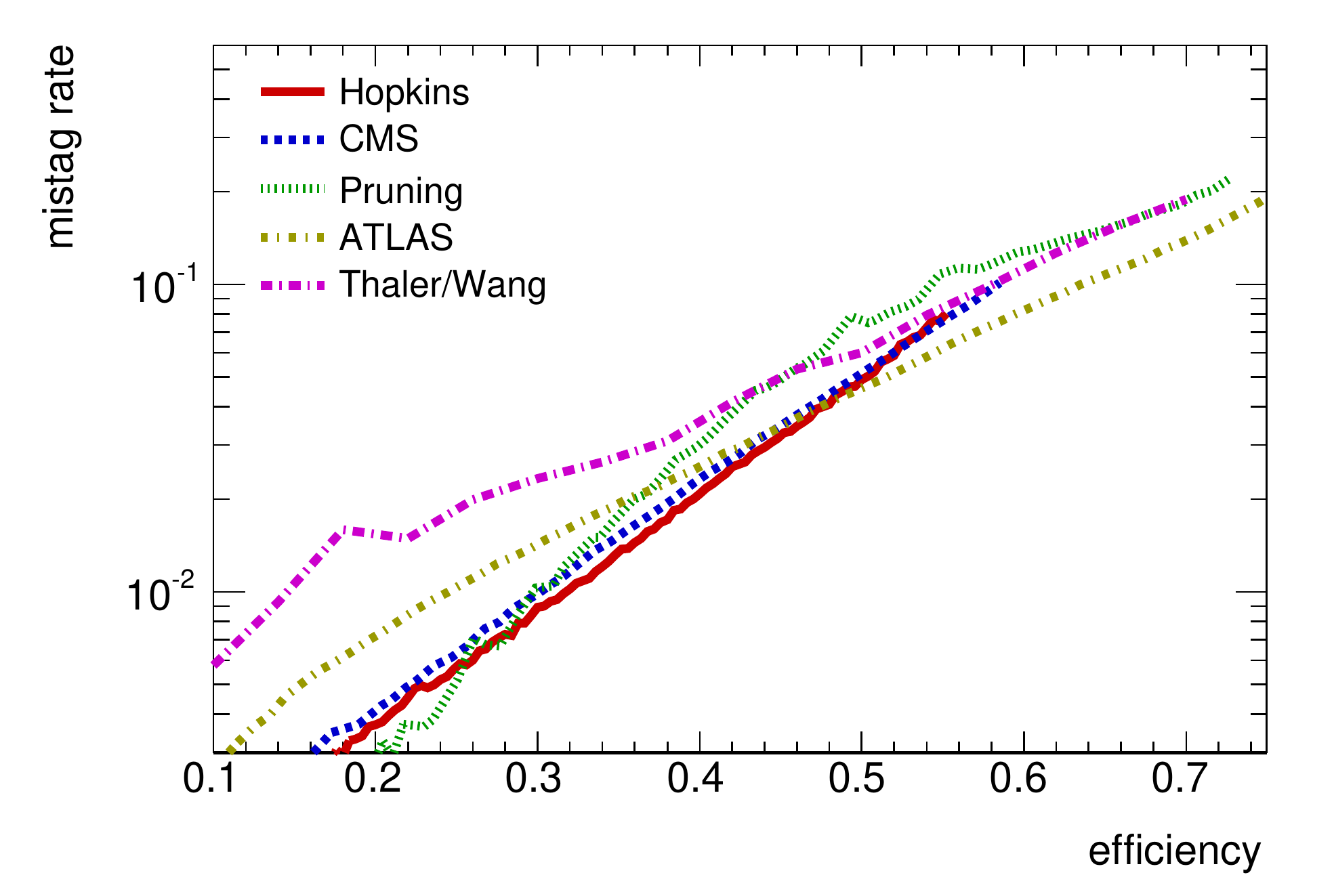}
  \includegraphics[width=0.4\textwidth]{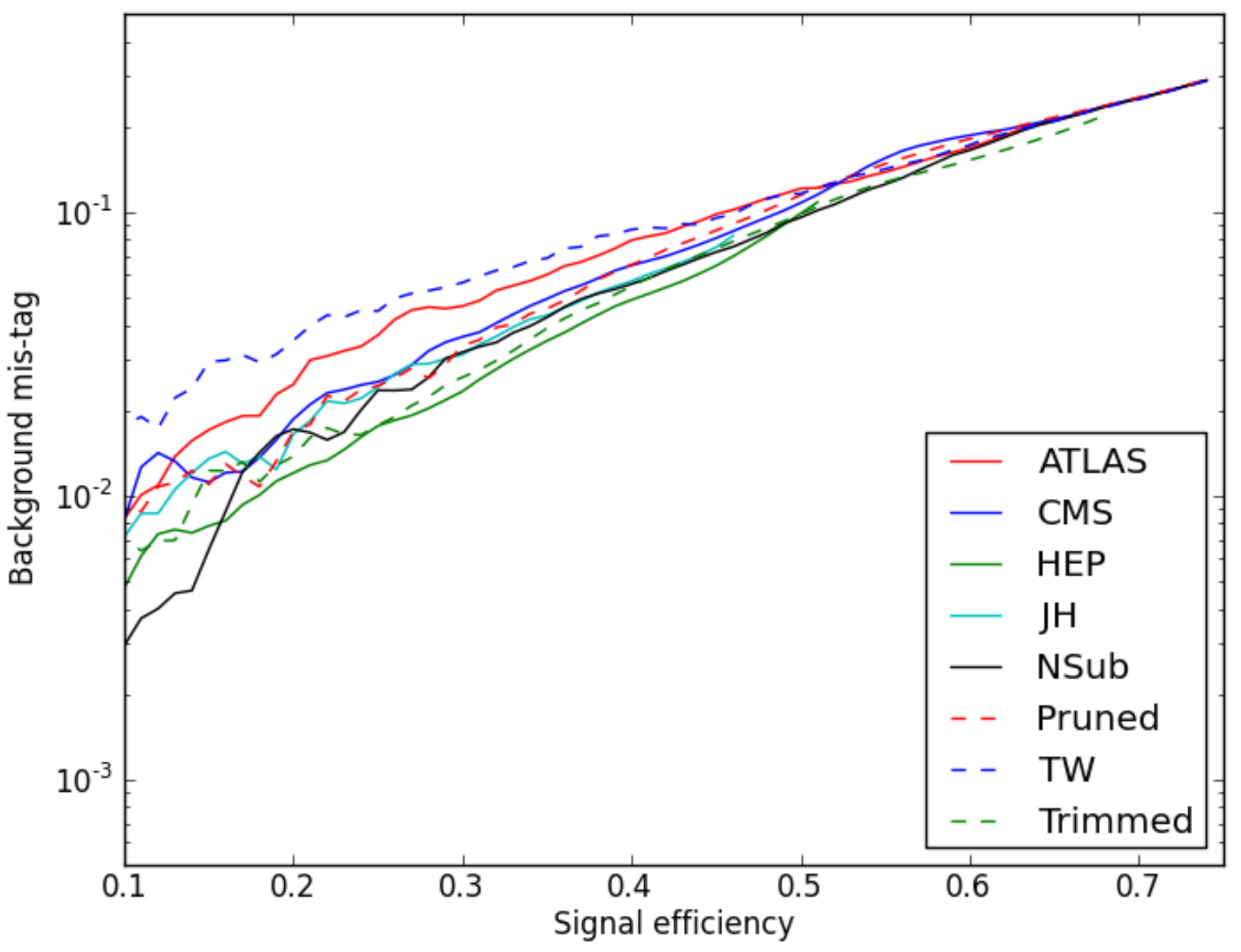}
  \caption{Mistag rate versus efficiency after optimization for the
    studied top-taggers. Left: tagging rates averaging over all Herwig
    generated $p_T$ subsamples of Ref.~\cite{websamples}. Right:
    tagging rates for Sherpa generated CKKW samples with $p_T$ $500$-$600$ GeV, as described in Ref.~\cite{boost2011}. It includes a simple detector
    simulation~\cite{lochtool}. Figures from Refs.~\cite{boost2010} 
    and \cite{boost2011}.}
  \label{fig:tagger2010}
\end{figure}

For the BOOST 2011 proceedings~\cite{boost2011} the list of taggers
and the samples were extended. New Herwig++~\cite{Bahr:2008pv} and
Sherpa~\cite{refsherpa} samples were generated with 100,000 unweighted
events for each $p_T$ slice. The Sherpa samples allow to examine the
effects of higher-order matrix elements~\cite{ckkw}. The list of
taggers was extended by a trimming-based tagger, N-subjettiness and
the HEP tagger. The input parameters for the taggers were optimized
for the individual efficiency working points. To make the study as
realistic as possible a simple calorimeter simulation was
included~\cite{lochtool}. This simulation smears energy according to a
radial profile based on performance of the ATLAS detector.

In Fig.~\ref{fig:tagger2010} (right) we include results for matched
Sherpa samples with unoptimized input. Comparing the results of the
unmatched Herwig samples of Fig.~\ref{fig:tagger2010} (left) with the
matched Sherpa samples (right), the efficiencies of all taggers
decline when applied to matched samples. In both comparisons the
Thaler/Wang and ATLAS tagger perform slightly worse than the other
taggers. The mass drop and jet shape based taggers however perform
very similarly.

Only further detailed studies on data can provide a
final assessment which tagger performs favorable compared to
others. Most likely tagging performance is subject to the fat jet cone
size, the transverse momentum of the top and the amount of additional
uncorrelated radiation in the event. Thus, it is possible that not one
tagger alone will be able to cover the whole spectrum of final states
in an optimal way.

\subsection{Data}
\label{sec:data}

Early jet substructure measurements are already available from
HERA~\cite{Chekanov:2002ux, Chekanov:2004kz, Chekanov:2009bc}. The
ZEUS collaboration studied the average number of subjets in $ep$
collisions and the distribution of energy flow within jets. These
measurements are well described by QCD and present Monte Carlo
simulations.

At the Tevatron both experiments, D0 and CDF, measured the
substructure of QCD jets.  D0 studied the $k_T$-subjet multiplicity
for central jets of size $R=0.5$ and $p_T = 55,...100$~GeV at
$\sqrt{s} = 0.63$ and 1.8~TeV. It turns out that the average number of
subjets in jets which can be linked to hard gluons is significantly
larger than in quark-related jets, as expected from the color charge
and the form of the splitting kernels~\cite{Abazov:2001yp}. It is
worth noting, however, that any link of an observed jet to hard quarks
or gluons is somewhat at odds with our QCD splitting picture and at
least scale dependent.

Based on $0.17~\ifb$ of Run~II data CDF measured jet shapes in
inclusive jet production and compared them with different underlying event
tunes~\cite{Acosta:2005ix}. Using $5.95~\ifb$ of data CDF was the
first experiment to look at substructure of massive jets with $p_T >
400$~GeV~\cite{Aaltonen:2011pg}. Both Midpoint and anti-$k_T$ jet
algorithms were used with jet sizes of $R = 0.4$, 0.7, and 1.0. The
measured jet shapes are the jet mass, angularity
(Eq.\eqref{eq:angularity}), and planar flow
(Eq.\eqref{eq:planar_flow}). For the jet mass the theory predictions
are in good agreement with the data. In contrast, for low angularity
and planar flow the Pythia predictions show some level of
disagreement. Using the same data sample CDF finds an upper cross
section limit of 38~fb (at $95 \%$~CL) for boosted top quarks. This is
approximately one order of magnitude higher than the estimated
Standard Model rate, and is limited by the QCD background
rates. However, it is the most stringent limit on boosted top quark
production to date.\bigskip

Spurred by very positive Monte Carlo results for new physics
searches~\cite{ATLASHV}, ATLAS and CMS both initiated studies on jet
substructure.  An early ATLAS study on differential and integrated jet
shapes~\cite{Aad:2011kq} uses $3~\ipb$ of data to look at anti-$k_T$
jet with $p_T = 30...600$~GeV. The data shows sensitivity to the
details of the parton shower, fragmentation, and underlying event
models in the Monte Carlo generators. However, for an appropriate
choice of the parameters the agreement between theory and experiment
is good.\bigskip

Pile-up is considered to be the biggest risk for the applicability of
more elaborate jet substructure methods. As a test ATLAS measures the
jet substructure including mass drops and filtering using $35~\ipb$ of
data at $\sqrt{s}=7$~TeV~\cite{ATLASpileup}. Jets are selected using
the anti-$k_T$ algorithm with size $R=1.0$ and the C/A-algorithm with
$R=1.2$. They are required to be central, $|y|<2$, and hard,
$p_{T,j}>300$~GeV. The jet mass distributions for both jet definitions
are then compared with Monte Carlo predictions from Pythia and
Herwig++. In addition, a set of substructure observables are
included. For the anti-$k_T$ jets this is the first $k_T$-splitting
scale $\sqrt{d_{12}}$ as defined in Eq.\eqref{eq:jetalgos}; for
C/A-jets it is the mass distribution after requiring a mass drop and
filtering the subjets.

\begin{figure}[t]
  \includegraphics[width=0.32\textwidth]{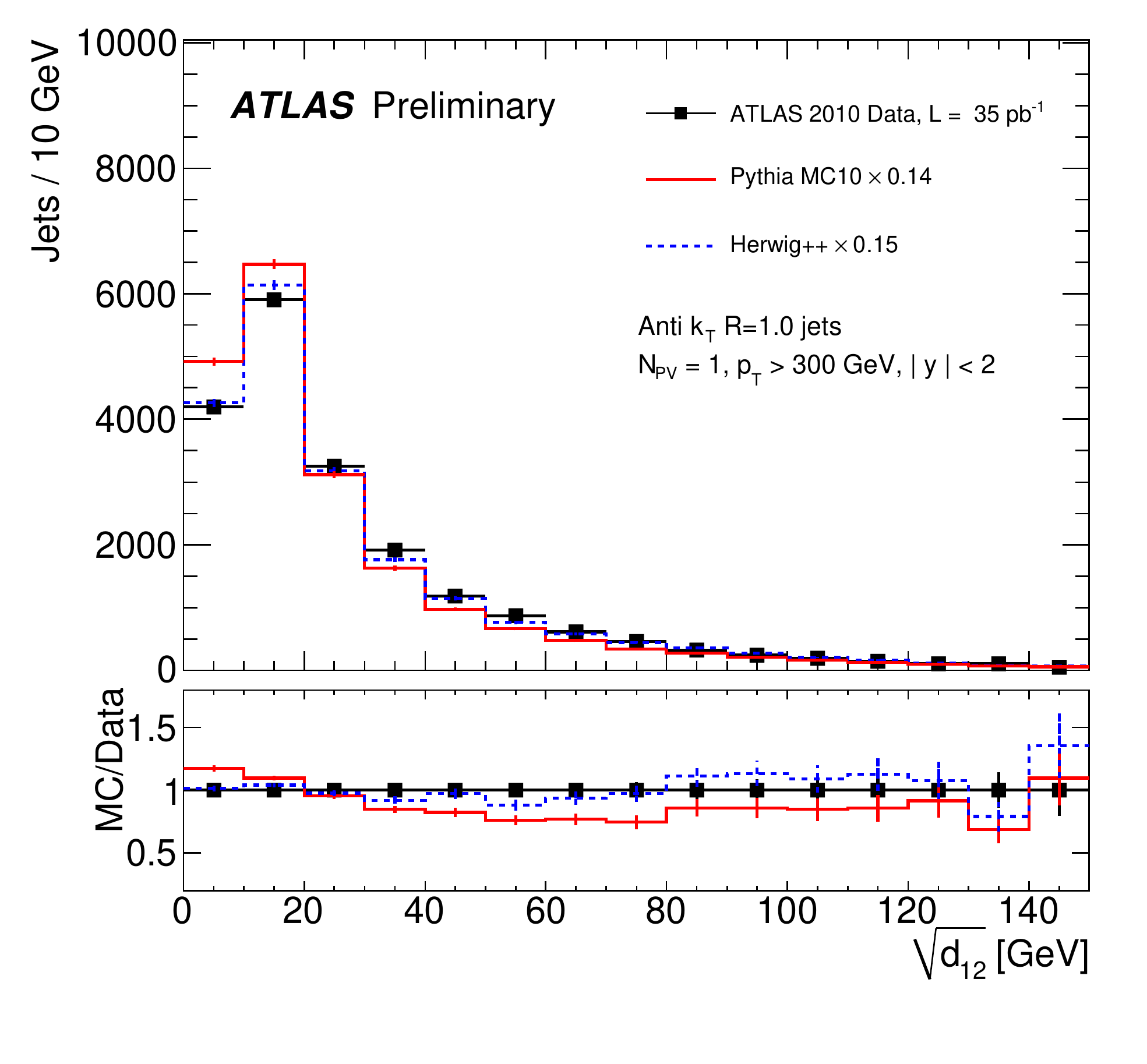}
  \includegraphics[width=0.32\textwidth]{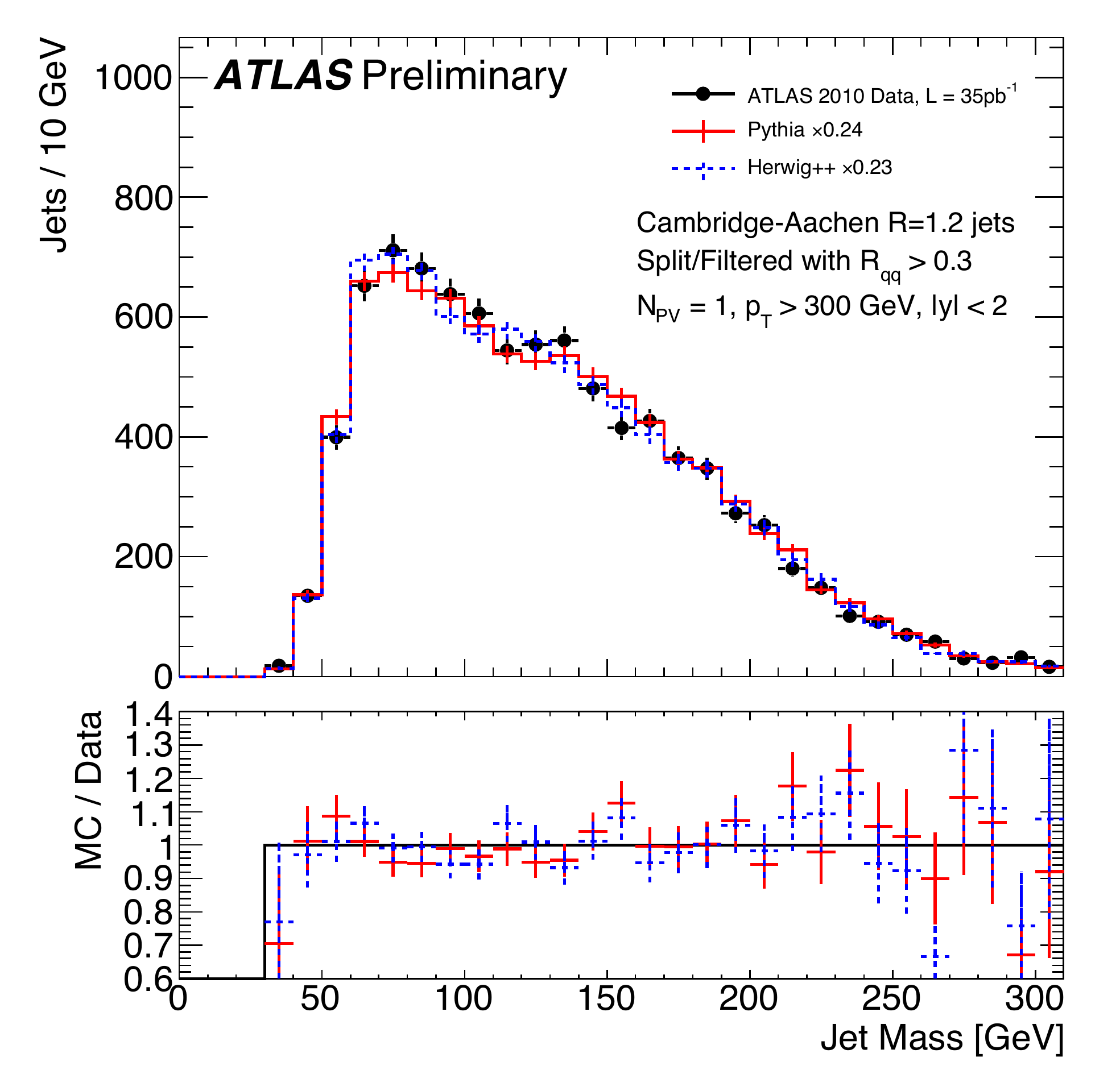}
  \includegraphics[width=0.32\textwidth]{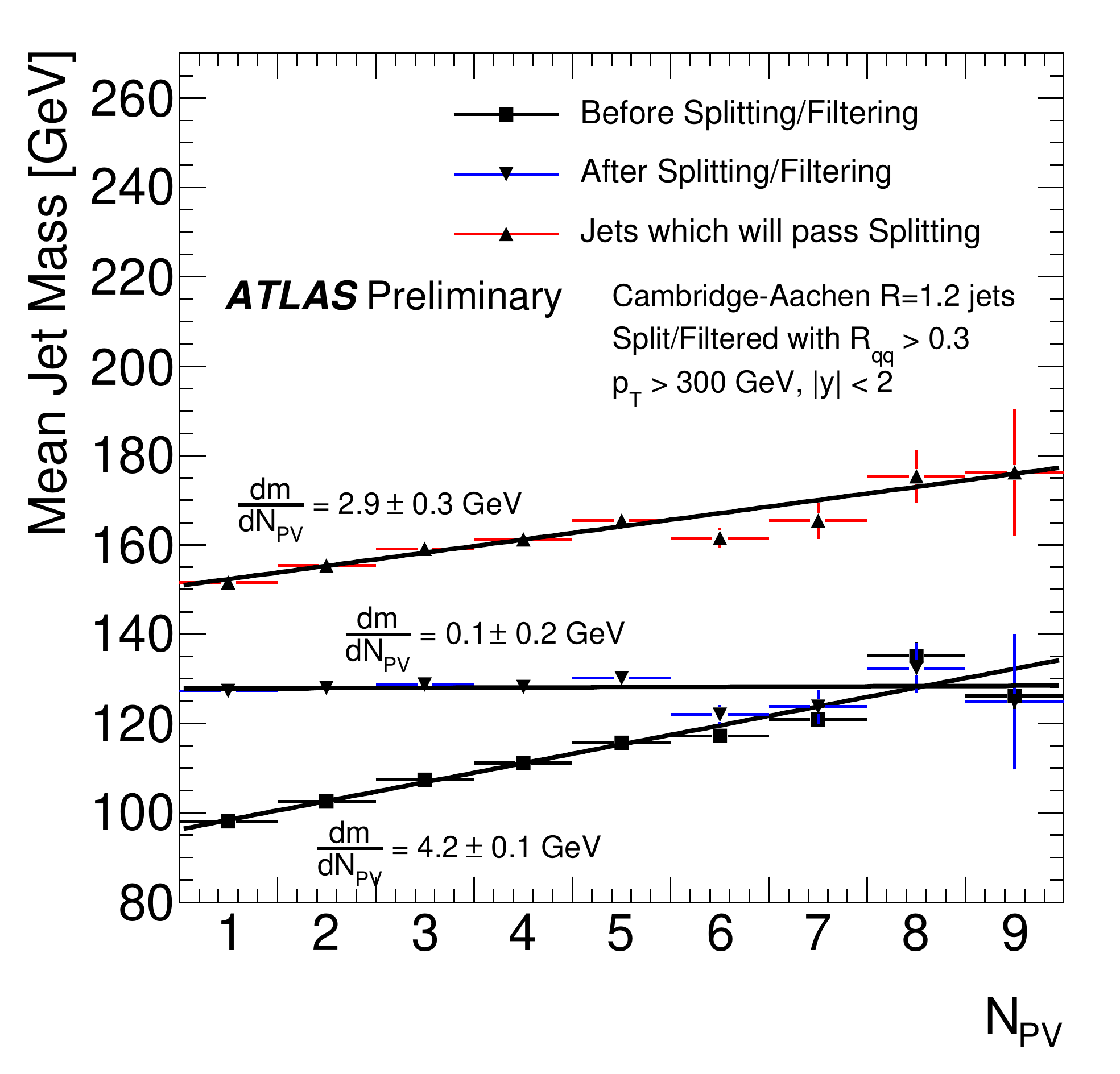}
  \caption{$k_T$ splitting scale (left), jet mass after mass drop and
    filtering (middle) and dependence of the jet mass on the number of
    primary vertices in the event (right). Figures from
    Ref.~\cite{ATLASpileup}.}
  \label{fig:atlaspileup}
\end{figure}

The left panel of Fig.~\ref{fig:atlaspileup} shows that the
measurement of the $k_T$ splitting scale agrees within $10-20$\% of
the Monte Carlo predictions over the whole analyzed range.  We know
that the un-groomed jet mass for large jets suffers from pile-up
contributions. For this uncorrected jet mass different Monte Carlo
predictions from Pythia and Herwig++ show sizable deviations from the
data.  However, after applying a mass drop and subjet filtering, as
described in Sec.~\ref{sec:early}, the jet mass distributions agree
well between theory and experiment, see Fig.~\ref{fig:atlaspileup}.

Finally, the number of primary vertices is a measure for pile-up. The
right panel of Fig.~\ref{fig:atlaspileup} shows the mean jet mass
before and after splitting and filtering as a function of the number
of primary vertices.  After requiring an additional large angular
subjet separation $R_{j_1,j_2} > 0.3$ the splitting naturally select
more massive jets. The tilted lowest line shows the mean mass of jets
which pass the splitting but before filtering. The filtering step then
significantly reduces the impact of pile-up and the slope shown in
Fig.~\ref{fig:atlaspileup} is consistent with zero.\bigskip

Already with the 2011 data CMS performed a search for heavy $Z'$ and
color octets decaying to a $t\bar{t}$ final state~\cite{cmssearch}. It
relies on the CMS top tagger, a variant of the Johns Hopkins tagger
introduced in Sec.~\ref{sec:hopkins}. All jets are reconstructed by
the C/A-algorithm with $R = 0.8$. One set of events, called `type 1+1'
at least two jets with $p_{T,j}>350$~GeV have to be present and allow
for a top reconstruction. The `type 1+2' set requires a leading jet
with $p_{T,j}>350$~GeV, a second with $p_{T,j}>200$~GeV, and the third
$p_{T,j}>30$~GeV. The hardest jet needs to correspond to a hadronic
top whereas the other two are reconstructed using pruning.  Combining
both topologies, a Randall-Sundrum KK gluon with mass between 1.0 and
1.5~TeV is excluded at 95\% CL, based on $886~\ipb$ of integrated
luminosity, see Fig.~\ref{fig:cmssearch}.

\begin{figure}[b]
  \includegraphics[width=0.45\textwidth]{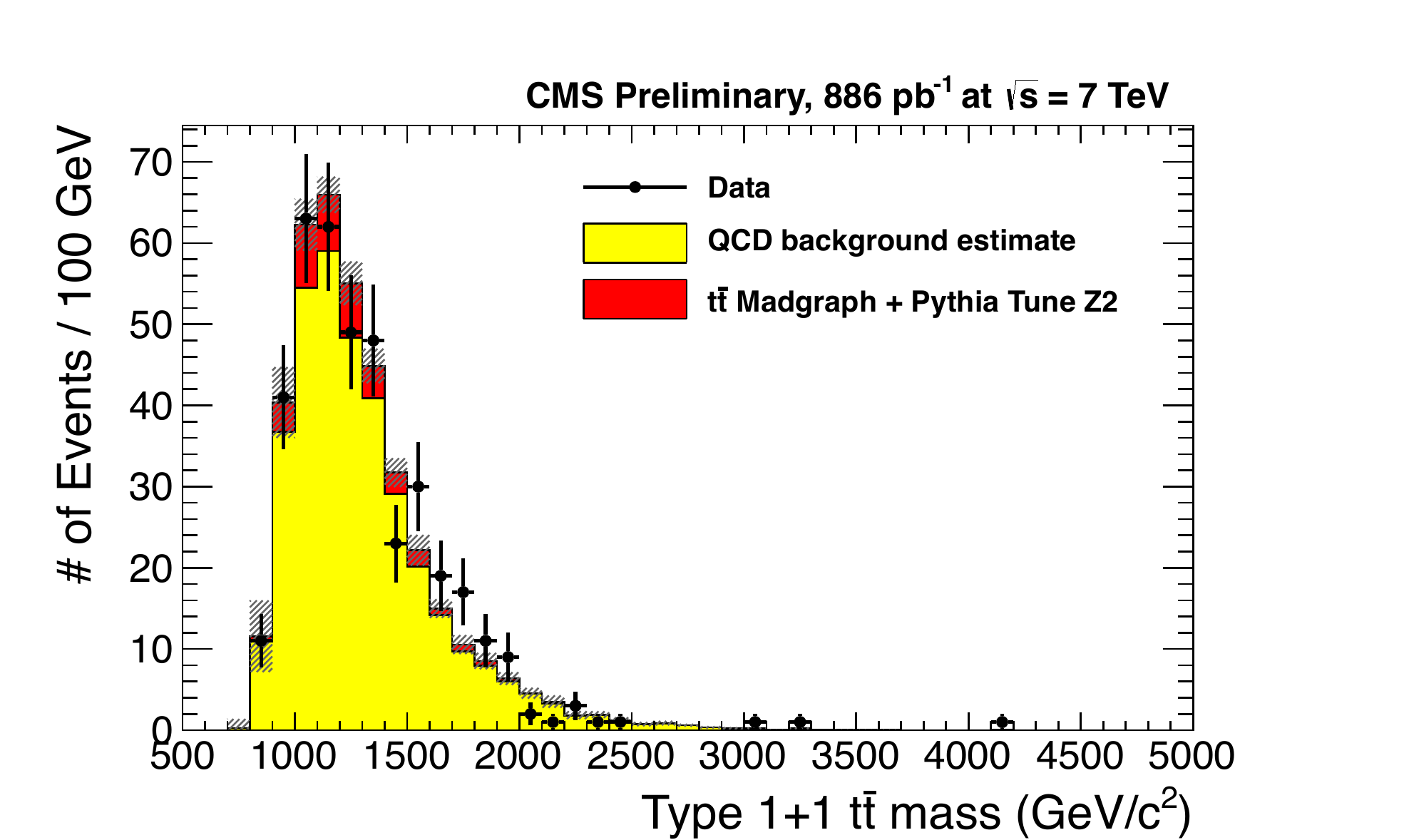}
  \hspace*{0.05\textwidth}
    \includegraphics[width=0.45\textwidth]{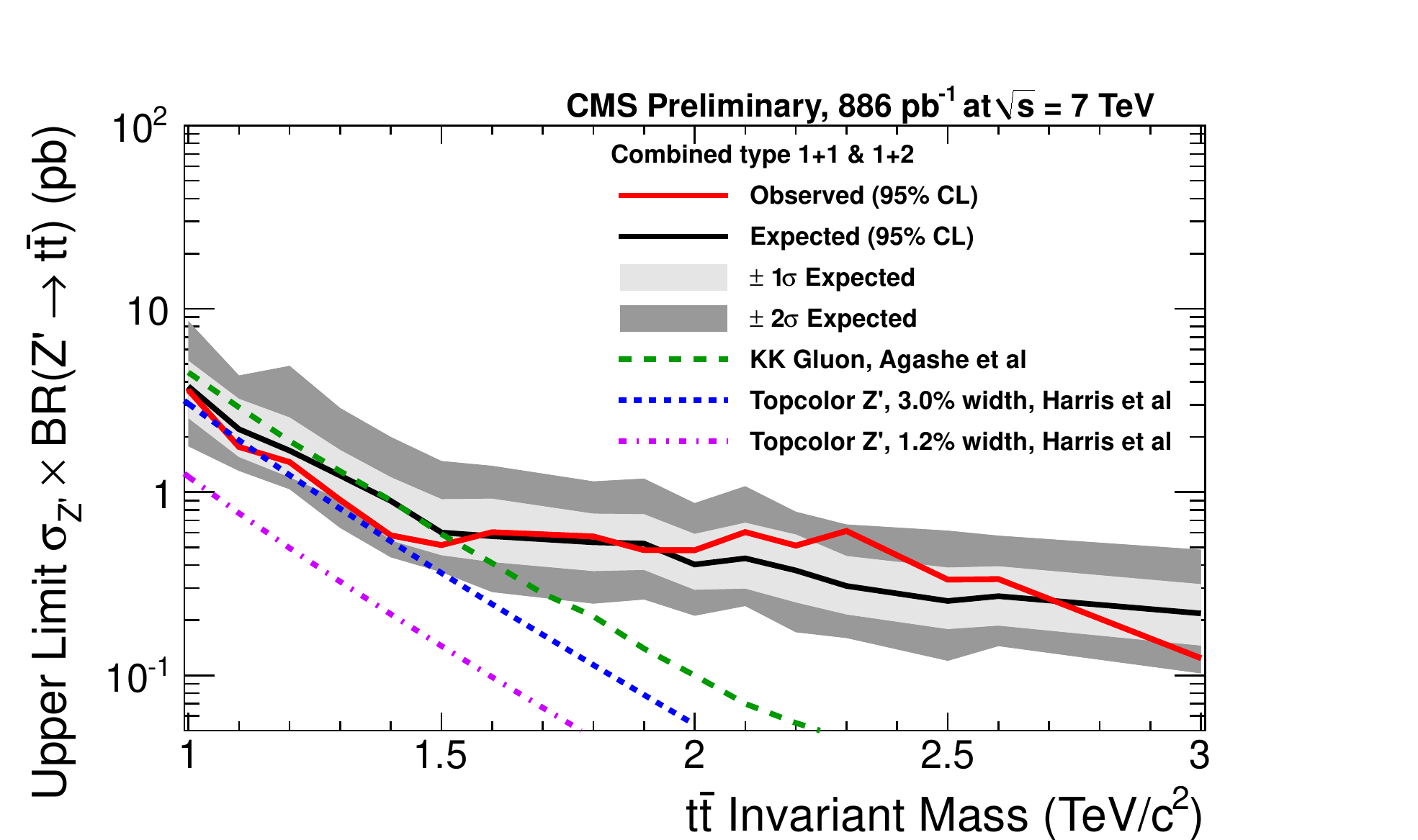}
  \caption{Observed and predicted $t\bar{t}$ invariant mass
    distribution in the `1+1' topology (left). Limits on the possible
    cross section times branching ratio of $t \bar{t}$ resonances
    (right). Figure from Ref.~\cite{cmssearch}.}
  \label{fig:cmssearch}
\end{figure}

\subsection{Applications}
\label{sec:applications} 
As outlined in Sec.~\ref{sec:case} boosted tops arise in a plethora of
physics scenarios. Thus, the application of top taggers has become
increasingly popular in searches for new physics resonances or
effects. 

The need to reconstruct top quarks in decays of very heavy resonances
was the initial motivation to study boosted tops. If the $s$-channel
resonance is heavy and both tops decay hadronically, the decay
products are highly collimated. To overcome the large dijet background
it is unavoidable to study the internal structure of the jets. Because
the tagging efficiency of the tops does not depend on the spin or the
color charge of the heavy resonance, the only parameters which
determine if the resonance search is feasible are the resonance's mass
and the production cross section. The scenario of a heavy $s$-channel
resonance decaying to two top quarks has been discussed in context of
various $Z'$ and color octet models
models~\cite{atlasresonance,gerbush,Bhattacherjee:2010za,Bai:2011mr,Berger:2011af,Yang:2011jk}.

Alternatively, top
partners~\cite{Kribs:2010ii,Gregoire:2011ka,Bi:2011ha,Fan:2011jc,heptop1},
fourth generation~\cite{holdom} or vector quarks~\cite{early2} can be
sources of boosted top quarks. Pair-produced top partners or heavy
quarks can decay into tops. In these small cascade decays the tops are
often accompanied by missing transverse energy, Higgs bosons or gauge
bosons, all forming useful handles to suppress backgrounds.

Recently, CDF~\cite{Aaltonen:2011kc} and D0~\cite{Abazov:2011rq}
measured an unexpectedly large forward-backward asymmetry of the top
quarks. Measurements of this quantity are subtle at the LHC, due to
its proton-proton initial state. However, one can define a forward
central charge asymmetry which captures the physics. Unfortunately,
for the dominating $gg$ initial state at the LHC there is no asymmetry
at all. To enhance the subdominant $q \bar{q}$ and $q g$ production
processes it is beneficial to require a large invariant mass of the $t
\bar{t}$ system, \ie require boosted tops. By reconstructing the
momentum of the hadronic top and measuring the charge of the second
top's lepton, it is possible to count the number of tops and anti-tops
in the forward region. This allows to measure the forward-charge
asymmetry at the LHC~\cite{Afb}.

\section{Beyond hadronic tops}
\label{sec:beyond}

Boosted resonances are a natural product of any ultraviolet completion
of the Standard Model with new decaying particles around the TeV
scale. In addition, even for continuum production processes the LHC
will provide enough energy to probe a boosted regime if background
suppression requires it.  This means that the reconstruction of
boosted heavy objects will be useful in a wide range of LHC
searches. Particles which can be tagged using their hadronic decay
channels include $W$ and $Z$ bosons, a light Higgs boson, or the top
quark.  Search strategies which directly use jet substructure in the
identification and reconstruction of those Standard Model particles
can indeed be a superior way of discovering new physics. As an analogy
we can remind ourselves that for many decades we have not considered
the decay products of a $B$ meson the appropriate analysis objects in
high-$p_T$ searches.\bigskip

Top quarks are predominantly produced in pairs at hadron colliders. As
discussed in Sec.~\ref{sec:case} non-boosted top quarks decaying to
$bl\nu$ provide three good handles for reducing QCD backgrounds: a
charged lepton suitable for triggering, large missing transverse
energy and a taggable $b$ jet. Following the logic of hadronic top
tagging we can ask if boosted leptonic top quarks have useful
properties for LHC searches.

Apart from heavy $s$-channel resonances, boosted top quarks naturally
arise in decays of top partners, \eg supersymmetric stops. The large
$t\bar{t}$ backgrounds make the reconstruction of such top partners
challenging~\cite{meade}. Using a top tagger on the purely hadronic
decays we know how to extract and reconstruct such top
partners~\cite{heptop1}. 

For the semi-leptonic sample we start with one hadronic tagged top
quark. Next, we need to identify and approximately reconstruct the
leptonically decaying top quark. First, QCD jets again pose a
dangerous background, because the rejection through $b$-tagging
degrades at large boost and the lepton-$b$ isolation becomes
marginal~\cite{thaler_wang}.  To ameliorate this problem, we can
introduce a tracker-level mini-isolation cut for the
lepton~\cite{Rehermann:2010vq}.  This results in a high background
rejection rate. To reconstruct the leptonic top momentum without
relying on the measured missing transverse momentum we define an
appropriate coordinate system for the top decay products. In those
coordinates the neutrino momentum component either in or orthogonal to
the $\ell$-$b$ decay plane can be approximately
neglected~\cite{Plehn:2011tf} . With this assumption the $W$ and top
mass constraints are sufficient to reconstruct the neutrino and top
momenta and in turn use the measured $\vec{p}_T$ value for background
rejection. This again allows us to extract top partner signals out of
large top pair backgrounds~\cite{Plehn:2011tf}.\bigskip

As described in Sec.~\ref{sec:early} the first applications of jet
substructure were the $W$ reconstruction from a heavy Higgs
decay~\cite{mike} or from $WW \to WW$ scattering~\cite{butterbrot}.
The latter can provide important insights into the nature of
electroweak symmetry breaking, especially if no Higgs boson is
found. Adding information on the polarization of the $W$ bosons to
this analysis should enhance its sensitivity~\cite{Han:2009em}. 

If a heavy Higgs boson decays mostly into gauge bosons the branching
ratio into the cleaner leptonic final state is either small (for $Z$
bosons) or difficult to reconstruct (for $W$ bosons). New substructure
methods can help to disentangle the hadronically decaying gauge bosons
from the QCD backgrounds, allowing for a reconstruction of the Higgs
mass~\cite{Hackstein:2010wk,Cui:2010km} and spin
properties~\cite{Englert:2010ud}.

In the busy environment of the LHC the identification of hadronic $W$
and $Z$ decays is hard and it can be beneficial to combine several
observables to discriminate them from QCD backgrounds. In
Ref.~\cite{Cui:2010km} a large number of observables is combined in a
multivariate approach. For highly boosted $W$ bosons this technique
might be able to improve on the BDRS method significantly.\bigskip

As a third massive Standard Model particle a light Higgs boson with a
hadronic decay $H \to b\bar{b}$ is a natural candidate for tagging. As
a matter of fact, this is precisely the channel where the BDRS tagger
re-started the broad tagging effort for the LHC.  In the Standard
Model the Higgs tagger can be applied to the $WH$ and $ZH$ production
processes~\cite{bdrs,Soper:2011cr,Almeida:2011ki} or to $t\bar{t}H$
production~\cite{tth}.  Higgs bosons as decay products of new
particles can for example appear from heavy $s$-channel $Z'$
decays~\cite{Katz:2010mr, Katz:2010iq} or supersymmetric cascade
decays~\cite{Butterworth:2007ke,Butterworth:2009qa,Kribs:2009yh,Kribs:2010hp,Kribs:2010ii,Bellazzini:2010uk}.\bigskip

Finally, new particles themselves could be tagged. Light weakly
interacting states for example arise in extended Higgs sectors. A
light additional CP-odd scalar which couples to the Higgs potential
can become an intermediate step in a four-body Higgs decay. If this
scalar is light enough it will be boosted even in the decays of a
SM-like Higgs boson with $m_H < 200$~GeV. Such a decay is a perfect
scenario for tagging methods and essentially impossible to detect
using standard
methods~\cite{Chen:2010wk,Falkowski:2010hi,Englert:2011iz}.

\section{Outlook}
\label{sec:outlook}

In this paper we have given an overview of different top taggers and
related aspects. Top tagging as one of the major applications of
subjet methods has in the past years been a rapidly developing
field. While the first paper on top tagging was written by Mike
Seymour in 1994, the developments relevant to top taggers used at
ATLAS and CMS only started around 2008. Since then, whole conferences
on the topic have been organized, with proceeding collections
effectively setting the standard in the field. As we are writing this
review we are waiting for the first tagged top quarks being announced
by ATLAS and CMS. Obviously, we will not be able to give an
appropriate summary of top tagging at this point in time.\bigskip

On the other hand, top tagging tools have been developed and tested at
an impressive level of sophistication. We have found that
\begin{itemize}
\item in practice, hadronic top quark identification, at least in the
  boosted regime, will be as easy as a $b$-tag or any other particle
  identification.
\item top taggers always include the jet mass as one jet shape
  observable, but two fundamentally different approaches rely on
  either the jet clustering history or additional jet shapes for the
  top tagging algorithm.
\item soft QCD, underlying event, and pile-up are a major challenge to
  top taggers, so eventually any top tagger will need to include an
  efficient way of removing these effects.
\item comparing top taggers will tell us much more than which tool to
  use for the identification of hadronic top decays. It will be a
  first glimpse on what physics concepts to use for QCD studies at the
  LHC.
\end{itemize}
All of these aspects have been developed as the first truly new
analysis concept for the LHC experiments. There still are many open
questions which can only be answered in close collaboration between
theory and experiment and which will make the coming years exciting
times in top physics:
\begin{itemize}
\item Can we actually tag tops in Standard Model events?
\item What are the efficiencies we can achieve, compared to Monte Carlo truth?
\item How well can we reconstruct the top momentum?
\item Most importantly: will we discover or measure anything new using top taggers?
\end{itemize}
We are only now about to see the first comparisons with LHC
data. Therefore, this is where our review of this rapidly moving field
ends and the future begins.

\begin{acknowledgments}
First, we would like to thanks our collaborators Gavin Salam,
Michihisa Takeuchi, and Dirk Zerwas for almost three years of exciting
research.  Next, we would like to thank all our experimental
colleagues we had the pleasure of working with on developing top
taggers. This most importantly includes Gregor Kasieczka, Sebastian
Sch\"atzel, and Andre Sch\"oning in Heidelberg. Chris Vermilion and
the other coordinators of Ref.~\cite{boost2011} we would like to thank
for giving us the right panel of Fig.~\ref{fig:tagger2010} before it
was published.  Finally, we would like to thank all the colleagues
working on these developments for the open, inspiring and exciting
atmosphere which all of us should not take for granted. A perfect
example was the outstanding performance of `the old guys' during the
BOOST~2011 meeting in Princeton~\cite{video}.
\end{acknowledgments}

\bigskip


\end{document}

%% file: declare.tex
\def\tablename{Table}
\def\figurename{Figure}

\newcommand{\heptop}{{\sc HEPTopTagger \,}}

\newcommand\one{\leavevmode\hbox{\small1\normalsize\kern-.33em1}}
\newcommand{\Mpl}{M_\mathrm{Pl}}
\newcommand{\p}{\partial}
\newcommand{\lag}{\mathcal{L}}
\newcommand{\qqquad}{\qquad \qquad}
\newcommand{\qqqquad}{\qquad \qquad \qquad}

\newcommand{\qb}{\bar{q}}
\newcommand{\matx}{|\mathcal{M}|^2}
\newcommand{\really}{\stackrel{!}{=}}
\newcommand{\msbar}{\overline{\text{MS}}}
\newcommand{\qns}{f_q^\text{NS}}
\newcommand{\lqcd}{\Lambda_\text{QCD}}
\newcommand{\met}{\slashchar{p}_T}
\newcommand{\pmiss}{\slashchar{\vec{p}}_T}

\newcommand{\st}[1]{\tilde{t}_{#1}}
\newcommand{\stb}[1]{\tilde{t}_{#1}^*}
\newcommand{\nz}[1]{\tilde{\chi}_{#1}^0}
\newcommand{\cp}[1]{\tilde{\chi}_{#1}^+}
\newcommand{\cm}[1]{\tilde{\chi}_{#1}^-}

\providecommand{\mg}{m_{\tilde{g}}}
\providecommand{\mst}{m_{\tilde{t}}}
\newcommand{\msn}[1]{m_{\tilde{\nu}_{#1}}}
\newcommand{\mch}[1]{m_{\tilde{\chi}^+_{#1}}}
\newcommand{\mne}[1]{m_{\tilde{\chi}^0_{#1}}}
\newcommand{\msb}[1]{m_{\tilde{b}_{#1}}}

\newcommand{\mev}{{\ensuremath\text{MeV}}}
\newcommand{\gev}{{\ensuremath\text{GeV}}}
\newcommand{\tev}{{\ensuremath\text{TeV}}}
\newcommand{\fb}{{\ensuremath\text{fb}}}
\newcommand{\ab}{{\ensuremath\text{ab}}}
\newcommand{\pb}{{\ensuremath\text{pb}}}
\newcommand{\sign}{{\ensuremath\text{sign}}}
\newcommand{\ifb}{{\ensuremath\text{fb}^{-1}}}
\newcommand{\ipb}{{\ensuremath\text{pb}^{-1}}}

\def\slashchar#1{\setbox0=\hbox{$#1$}           
   \dimen0=\wd0                                 
   \setbox1=\hbox{/} \dimen1=\wd1               
   \ifdim\dimen0>\dimen1                        
      \rlap{\hbox to \dimen0{\hfil/\hfil}}      
      #1                                        
   \else                                        
      \rlap{\hbox to \dimen1{\hfil$#1$\hfil}}   
      /                                         
   \fi}
\newcommand{\dslash}{\slashchar{\partial}}
\newcommand{\Dslash}{\slashchar{D}}

\def\eg{{\sl e.g.} \,}
\def\ie{{\sl i.e.} \,}
\def\etal{{\sl et al} \,}